\documentclass[lettersize,journal]{IEEEtran}
\usepackage{amsmath,amsfonts}
\usepackage{algorithmic}
\usepackage{array}
\usepackage[caption=false,font=normalsize,labelfont=sf,textfont=sf]{subfig}
\usepackage{textcomp}
\usepackage{stfloats}
\usepackage{url}
\usepackage{verbatim}
\usepackage{graphicx}
\usepackage{cite}
\usepackage{multirow}
\usepackage{xcolor}
\usepackage{xurl} 
\usepackage{hyperref} 
\def\BibTeX{{\rm B\kern-.05em{\sc i\kern-.025em b}\kern-.08em
    T\kern-.1667em\lower.7ex\hbox{E}\kern-.125emX}}
\usepackage{balance}
\begin{document}
\title{A Survey of Blockchain, Artificial Intelligence, and Edge Computing for Web 3.0}
\author{Jianjun Zhu, Fan Li, and Jinyuan Chen
\thanks{Jianjun Zhu, Fan Li, and Jinyuan Chen are with Louisiana Tech University, Department of Electrical Engineering, Ruston, LA 71272 USA (e-mail: jzh013@latech.edu; fanli@latech.edu; jinyuan@latech.edu). }}

\markboth{}%
{How to Use the IEEEtran \LaTeX \ Templates}

\maketitle

\begin{abstract}
Web 3.0, as the third generation of the World Wide Web, aims to solve contemporary problems of trust, centralization, and data ownership. Driven by the latest advances in cutting-edge technologies, Web 3.0 is moving towards a more open, decentralized, intelligent, and interconnected network. 
However,  increasingly widespread data breaches have raised awareness of online privacy and security of personal data.
Additionally, since Web 3.0 is a sophisticated and complex convergence, the technical details behind it are not as clear as the characteristics it presents. In this survey, we conduct an in-depth exploration of Web 3.0 from the perspectives of blockchain, artificial intelligence, and edge computing.
Specifically, we begin with summarizing the evolution of the Internet and providing an overview of these three key technological factors. 
Afterward, we provide a thorough analysis of each technology separately, including its relevance to Web 3.0, key technology components, and practical applications. 
We also propose decentralized storage and computing solutions by exploring the integration of technologies.
Finally, we highlight the key challenges alongside potential research directions.
Through the combination and mutual complementation of multiple technologies, Web 3.0 is expected to return more control and ownership of data and digital assets back to users. 

\end{abstract}


\section{Introduction}
\IEEEPARstart{S}{ince}
the second half of 2021, Web 3.0 has become ubiquitous and gained widespread attention across various industries.  
Web 3.0 is a new version of the World Wide Web (also known as the Web), which is the world's dominant software platform.  
According to the well-known statistical site Statista,  as of October 2023, there were 5.3 billion web users worldwide, which accounts for 65.7\% of the global population. Notably, social media users account for 93.4\% of web users \cite{Statista:22}. However, the vast online world is controlled by tech giants who have absolute power compared to the general public. The centralized platforms developed by tech giants gather personal data from users and serve as both gatekeepers and arbiters in the delivery of their services, allowing them to shape and control what is available on the Web. Those platforms drive most traffic to online news and hence have a significant influence on the sources of information that the general public consumes on a regular basis. The primary beneficiaries are a small group of stakeholders. These platforms are unelected and difficult to audit. 
For the current dilemma, there is an urgent need for solutions to improve web environment while returning control to users. This has laid a solid foundation for the emergence of Web 3.0. 
Web 3.0 is a counter-proposal to the current technological monopoly, providing a more holistic view of how society should use technologies \cite{KS:2022}.

\subsection{Motivations and contributions}
While Web 3.0 is still in its infancy, it has made significant advances in creating an open, trustless, and permissionless web where users can share and exchange data without relying on centralized organizations.
However, there are still some major problems that need to be addressed. 

\emph{1)} True decentralization: The emerging industries supporting decentralized web are highly consolidated, which may undermine the realization of the decentralized vision of Web~3.0.

\emph{2)} Mass adoption: Technical barriers for using Web 3.0 remain high, making it challenging for users to migrate to Web 3.0 platforms. In addition, the differences between developing environments also pose challenges to Web 3.0 adoption.

\emph{3)} Storage/computation: 
Web 3.0 requires decentralized storage and computation.
However, building efficient decentralized solutions is challenging.

We are motivated by these problems to explore mitigation strategies from a technical perspective.
Fundamentally, Web 3.0 is a convergence of emerging technologies. 
Among them, blockchain, artificial intelligence (AI), and edge computing are essential in this revolution, enabling users to use the Web securely and intelligently. 
Blockchain will allow individuals, companies, and computers to exchange data in a decentralized manner.   
AI will empower computers with the ability to learn and reason to provide user-centered interactions.  
Edge computing allows faster processing and decision-making with low latency by bringing computing and storage closer to where users actually consume the data.

\begin{figure} []
	\centering
	\includegraphics[scale=0.055]{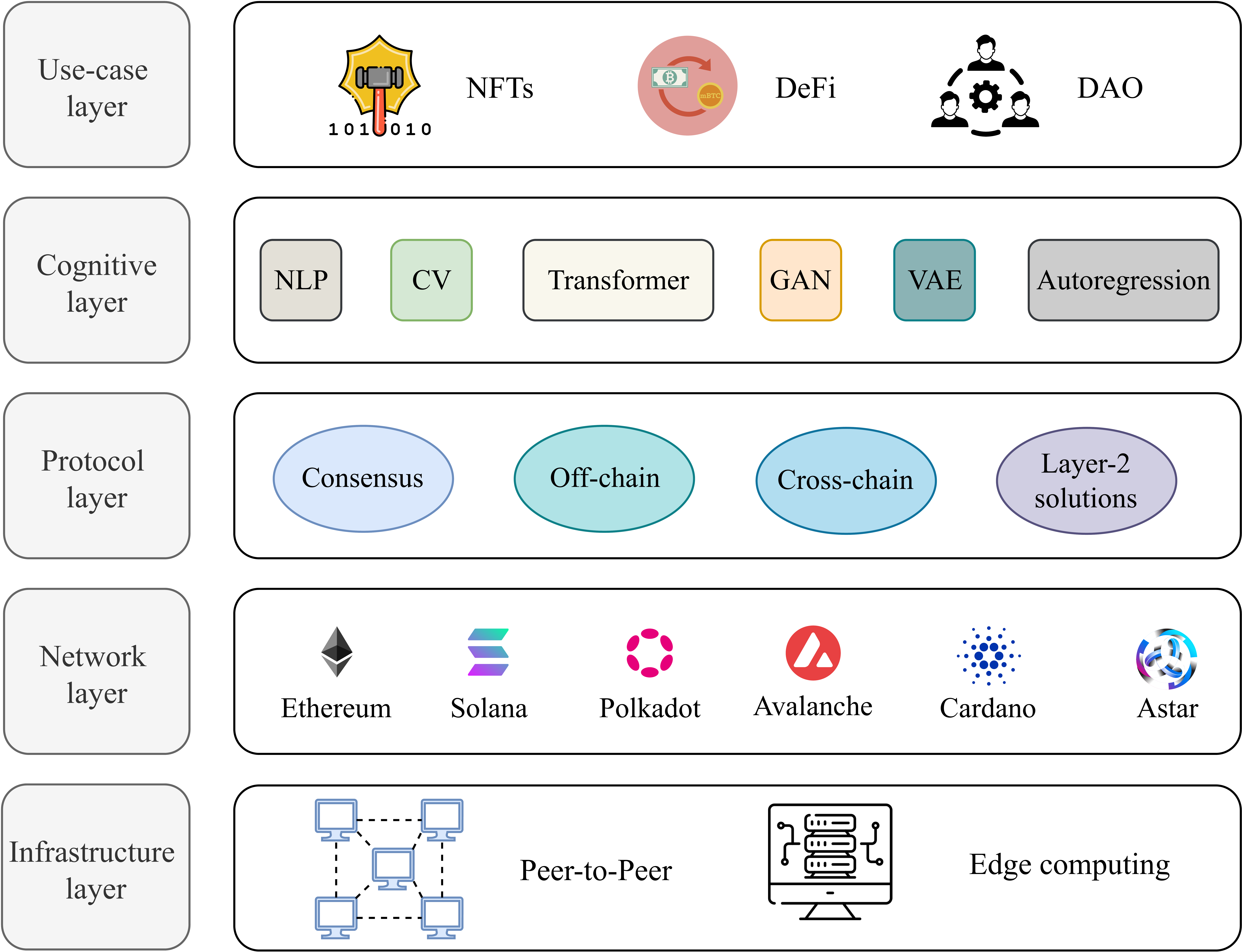}
	\caption{Web 3.0 stack architecture.}
        \label{fig:stack}
\end{figure}

Based on the concept that the architecture of Web 3.0 is typically constructed using technology stacks \cite{Web3Foundation}, 
we propose an alternative stack architecture with varying emphases as illustrated in Fig.~\ref{fig:stack}. 
In this paper, we will explore Web 3.0 in-depth from the perspectives of blockchain, AI, and edge computing.  
The corresponding technology enablers will be discussed in depth in each chapter. 
To the best of our knowledge, 
there has been no comprehensive survey on Web 3.0 from the perspective of these three technologies simultaneously.
To summarize, the main contributions of our work are discussed as follows:

\indent \emph{1)}
We discuss in detail the technical building blocks of Web 3.0 and their necessity from the perspectives of blockchain, AI, and edge computing.

\indent \emph{2)}
We survey state-of-the-art Web 3.0 practical applications to explore more convergence possibilities to improve the Web 3.0 ecosystem.  

\indent \emph{3)}
We propose and illustrate storage and computing solutions in decentralized scenarios by exploring the integration of blockchain, AI, and edge computing.

\indent \emph{4)}
We highlight critical challenges encountered in the development of Web 3.0 with concrete data and discuss future research directions accordingly.

\subsection{Comparison with other surveys}
To date, there have been several survey papers on different aspects of Web 3.0. Our survey is distinctive from all the other surveys as we conducted an in-depth analysis of the impact of blockchain, AI, and edge computing on the development of the Web 3.0 ecosystem.
A comparison of our survey with other works on Web 3.0 is provided in Table~\ref{tab:surveyComparison}.

\begin{table*}[]
\centering
\small 
\setlength{\tabcolsep}{6pt} 
\renewcommand{\arraystretch}{1.83} 
\caption{A comparison between our survey and other works on Web 3.0. }
\begin{tabular}{c|ll}
\hline \hline
\multicolumn{1}{c|}{\textbf{Reference}}              & \multicolumn{1}{c}{\textbf{Description}} \\ \hline \hline
\cite{VK:21}  & Providing an integrative literature review on the development of the Web.  \\ \hline
\cite{Zarrin:21}  & Conducting a review of decentralized Internet with a focus on consensus and other emerging technologies.  \\ \hline
\cite{Wang:22}  & Exploring blockchain-based Web 3.0 from an architecture identification and evaluation perspective.  \\ \hline
\cite{Ray:23}  & Analyzing key advances and potential impacts to emphasize the importance of Web 3.0.  \\ \hline
\cite{Gan:23}  & Introducing the evolution, technologies, and challenges of Web 3.0.  \\ \hline
\cite{Ren:23}  &  Surveying the impact of quantum technology on the development of blockchain-based Web 3.0. \\ \hline
\cite{Huang:23}  & Investigating Web 3.0 application categories, popularity, challenges, and opportunities. \\ \hline
\cite{Shen:23}  & Conducting a comprehensive overview of the latest advances of AI in Web 3.0.  \\ \hline
\multirow{3}{*}{\hfil This work }  & \multicolumn{1}{l}{Providing an in-depth analysis of Web 3.0 from the perspectives of blockchain, AI, and edge computing,} \\ [-1ex]  & \multicolumn{1}{l}{discussing practical applications through a significant amount of concrete research works, and proposing }  \\ [-1ex]  & \multicolumn{1}{l}{solutions for decentralized storage and computation through technological integration.}  \\ \hline \hline

\end{tabular}
\label{tab:surveyComparison}
\end{table*}

Comparatively, Voj\'{i}\v{r} {\textit{et al.}} in \cite{VK:21} provided an integrative literature review examining the development of the Web by analyzing its evolution from centralization to decentralization and the reactions it provoked. 
Zarrin  {\textit{et al.}} in \cite{Zarrin:21} showed the potential of blockchain technology to provide a robust and secure decentralized Internet by exploring consensus algorithms and how blockchain can be combined with emerging technologies.
Wang \textit{et al.} in \cite{Wang:22} conducted an in-depth exploration of Web 3.0 from the perspective of blockchain, identifying twelve types of architecture by decoupling existing systems into core components.
Ray in \cite{Ray:23} emphasized the importance of Web 3.0 in shaping a decentralized and democratized Internet by analyzing the key advances and impacts of Web 3.0 applications and their integration with emerging technologies.
Gan \textit{et al.} in \cite{Gan:23} provided an overview of Web 3.0 in terms of technologies, challenges, potential, and prospects.
Ren \textit{et al.} in \cite{Ren:23} explored the fusion of various quantum information technologies with blockchain to develop a resilient digital ecosystem based on blockchain.
Huang \textit{et al.} in \cite{Huang:23} empirically investigated the categories of Web 3.0 applications and their popularity, as well as the potential of this emerging field.
Shen \textit{et al.} in \cite{Shen:23} provided a comprehensive overview of the latest advances of AI in Web 3.0, proposing and investigating the main challenges at each layer of the Web 3.0 architecture.

\begin{figure}[!h]
    \centering
    \includegraphics[scale=0.066]{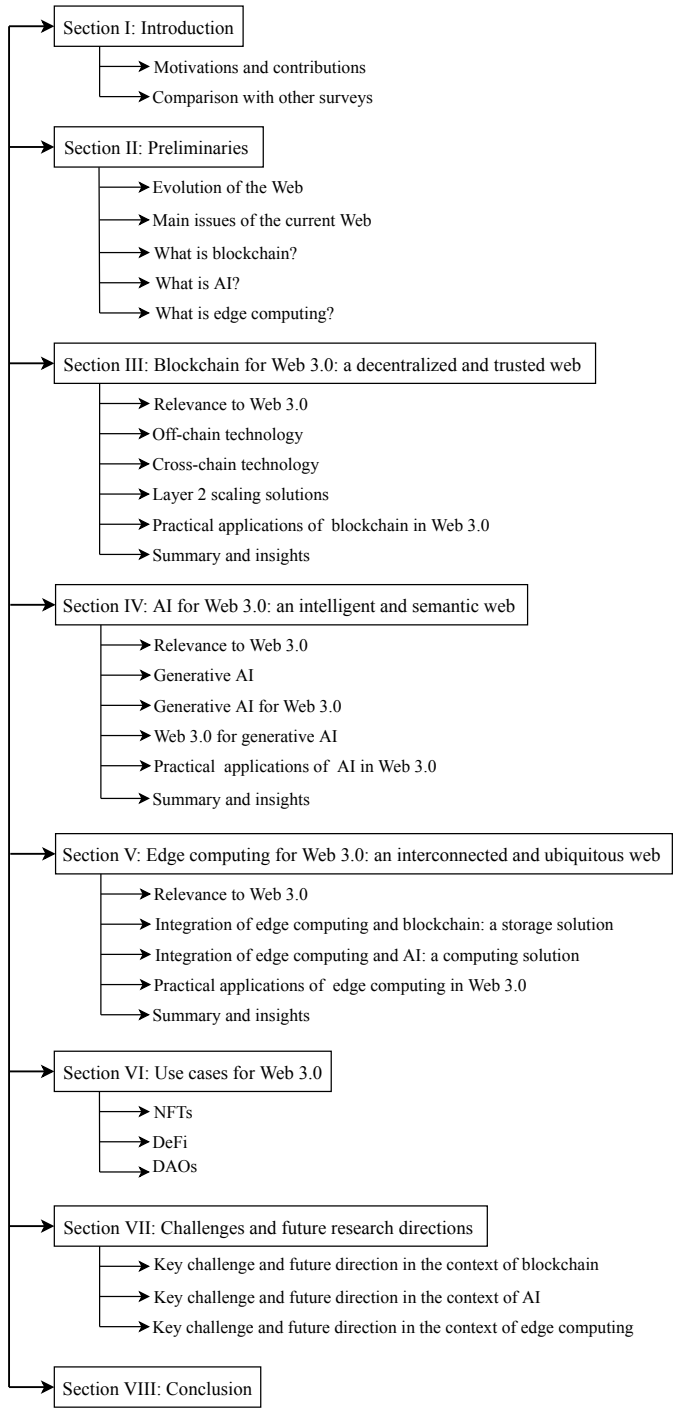}   
    \caption{An illustrative organizational structure of this paper.}
    \label{fig: paperstructure}
\end{figure}

The rest of this paper is organized as follows.
Section~\ref{sec:Preliminaries} provides relevant preliminaries such as the evolution of the Web, current web issues, and background on blockchain, AI, and edge computing.
Sections~\ref{sec:cornerstone}-\ref{sec:edgecomputing} provide an in-depth analysis of each technology (i.e., blockchain, AI, and edge computing) for Web 3.0 in terms of relevance, fundamental components, practical applications, and further insights.
Section~\ref{sec: applications} delves into the primary use cases of Web 3.0 and their practical applications. Additionally, major issues from both technical and non-technical aspects are discussed.
Section~\ref{sec:challenges} discusses the key challenges and further research directions.
Finally, the work is concluded in Section~\ref{sec:conclusion}.
An illustrative organizational structure of this survey is presented in Fig.~\ref{fig: paperstructure}.

\begin{figure*}[]
	\centering
	\includegraphics[scale=0.041]{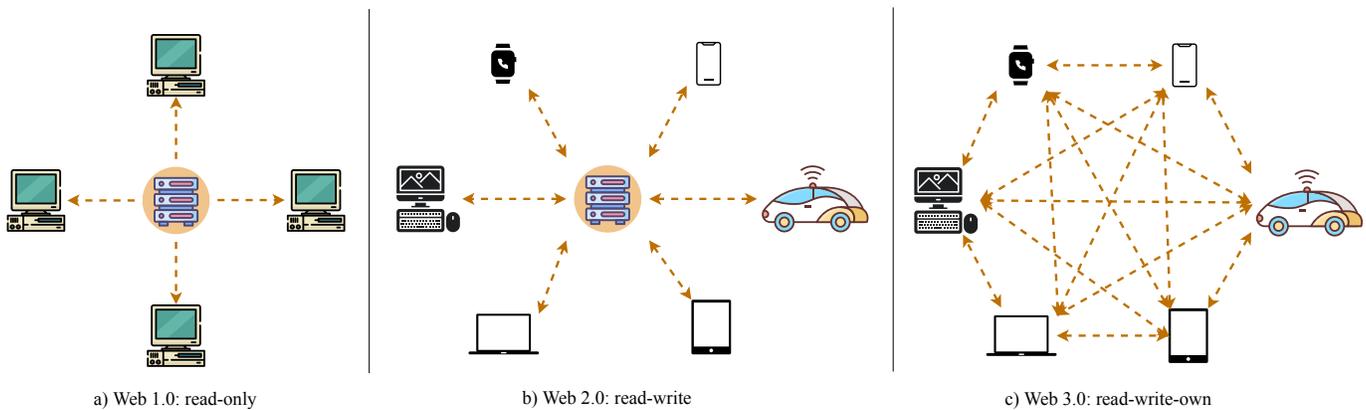}
	\caption{Evolution of the Web.} 
	\label{fig: evolution}
\end{figure*}

\section{Preliminaries} \label{sec:Preliminaries}
In this section, we introduce the fundamentals of Web 3.0 from the following aspects: evolution of the Web, main issues of current Web, and enabling technologies (i.e., blockchain, AI, and edge computing).

\subsection{Evolution of the current Web}
The Web is a hypertext document management system accessible via the Internet. People can use a web browser to access web pages hosted on web servers.  
In March 1989, Sir Tim Berners-Lee proposed a system for managing information for what would become the Web. He envisioned an intelligent, connected, and data-driven network in which computers could analyze all network data, including content, connections, and transactions between users and computers. He advocated for the European Council for Nuclear Research to provide the underlying code for free in April 1993. This decision led to today's Web \cite{Beniiche:2022, Park:2023}.

Web 1.0, also called the Static Web,  
is described as a web of interconnected information. Tim Berners-Lee coined it as ``read-only'' Web (see Fig.~\ref{fig: evolution}-a) 
since a massive majority of participants were content consumers. Content creators were basically  reporters, writers, and developers. Web 1.0 was not interactive as it did not provide the features to browse the Web, which made it extremely difficult for content users to find the needed information. The web pages were also proprietary, as many web browsers tried to attract users and stand out by providing proprietary content. This led to a series of incompatibility issues between web pages and web browsers. 

Web 2.0, also called the Social Web,  
is the most familiar and widely used Web today. The term ``Web 2.0" was first proposed by Darcy Dinucci in 1999 and later promoted by Tim O'Reilly and Dale Dougherty as a ``read-write" Web in late 2004 (see Fig.~\ref{fig: evolution}-b)\cite{O'Reilly:05}. 
The focus of Web 2.0 is on enabling users to interact with web content. Users are becoming more engaged in generating and sharing web content in addition to browsing. Users can communicate with each other over the Web through social media platforms. They can create content in their own blogs that other users can access and interact with in discussion forums.
The Web 2.0 period is also a time when mobile web access has boomed. People can utilize their phones, tablets, and almost any other web-connected devices to access web content at any time. Overall, Web 2.0 applications demonstrate a powerful front-end revolution with more opportunities to interact with end users.

Web 3.0 was first coined by Gavin Wood, co-founder of Ethereum, in 2014 as a way to minimize trust in a handful of private companies \cite{Web3Foundation, Ethereum:22}.
It is a decentralized and fair Internet reconstructed using distributed technology, where users can control their own data and identity. 
Up until 2021, Web 3.0 gradually entered the public consciousness and became mainstream. 
It is described as a ``read-write-own" Web (see Fig.~\ref{fig: evolution}-c)  that enables users to acquire, create, and execute. 
Web 3.0 does not rely on any third parties but rather allows users to interact directly with each other and with the web content they are accessing through a peer-to-peer (P2P) network. This makes privacy and security more assured because information is not stored in a data center that could be compromised. Additionally, power will be handed back to the end users, who no longer need permission from the tech giants to create, trade, and collaborate.  
A typical example is Solid proposed by Tim Berners-Lee in late 2020 \cite{Berners-Lee:20}. The purpose is to propose a specification that allows users to securely store data in a decentralized manner to achieve true data ownership and improve privacy.
Notably, Web 3.0 is more than just a new wave of innovation. It is an opportunity to reset and enable new benefits for ordinary users while solving some of the toughest challenges posed by disruptive technologies of the past.

\subsection{Main issues of the current Web}
The Web is among the most crucial innovations in the progression of human technologies. It aims to serve as an open platform that facilitates interaction, access, and information sharing across regional and cultural boundaries. From this viewpoint, the current Web has partially realized this vision since it has brought many revolutionary changes to the Internet. However, it has also raised some major issues \cite{WebFoundation:17}.

\begin{itemize}
\item \emph{Lack of ownership}:
The current business strategy of many data-driven companies is to provide free services in exchange for personal data. The collected data is invisible to users. In addition, users lack ways to inform third parties that they do not want their data to be disclosed.
According to a survey on personal information by Pew Research Center \cite{PEW:19}, 62\% of Americans believe that it is impossible for their data not to be collected in their daily lives; 79\% of Americans are very concerned about how companies use their personal data; 81\% of Americans believe that they have little control over their personal data. To make matters worse, the government is working with tech companies to track online behavior and create severe regulations that violate people's privacy.
\item \emph{Centralization}:
Despite the decades-long history of the Web, the network architecture is still based on the concept of stand-alone computers. Companies with significant control over large data platforms are progressively concentrating data in their hands. These tech giants provide users with online identities by holding their personal data and acting as gatekeepers of information. At the same time, they likely have control over who is able to use their digital products. All activities on the platform need to be carried out with the permission of the platform.   
This becomes a problem when people are denied the choices that should be rightfully theirs. Whenever people interact over the Web, data will be sent to the central server. It should be clear that the users will lose control over the data when that happens which will also cause a serious crisis of trust. 
\item \emph{Privacy}:
Web 2.0 applications collect vast amounts of personal and sensitive data by extensively tracking an individual's digital activities, social media posts, physical location, purchasing habits, and more to build highly detailed digital profiles. However, individuals have little knowledge or control over what information is collected and how it is used. This data may be accessed or disclosed by unauthorized parties, or misused by the service provider itself. According to Statista, the number of exposed data records detected since 2020 has exceeded 473 million data records\cite{Statista:23}. This points to serious issues with this pervasive data collection under opaque usage policies, which can lead to unfounded behavioral profiling, intrusive advertising practices, and inadvertent targeting of vulnerable audiences. 
\item \emph{Spread of misinformation}:
Data quality issues include data inaccuracy, data inconsistency, and data duplication. In Web 1.0, data quality was primarily determined by the reputations of publishers. However, Web 2.0 reduces data quality and leads to the spread of misinformation. The majority of individuals acquire news and information from a handful of social media platforms and search engines. These companies determine what to display based on the recommendation algorithms that learn from 
personal data so that more profit can be achieved when users click on links. This leads to these sites showing some poor-quality data or even fake news that may surprise and shock people and can spread like wildfire.  
Additionally, those with malicious intent may exploit the system to disseminate false information for monetary or political gain.
\item \emph{Unfair reward system}:
The reward system in Web 2.0 is unfair, with platforms reaping the vast majority of profits from the content created by content creators. Only a small percentage of the profits go to the individuals who actually create the content. There are two main reasons for this situation: The first reason is high concentration. All network resources are owned and controlled by the platform. The second reason is the unconsciousness of data value. Most users are not aware of data value they generate. The platform is also subtly telling people the uselessness of data. In addition, centralized recommendation algorithms prefer data that can be widely disseminated regardless of whether the data content itself has positive value. This type of content recommendation is unbalanced, which greatly enriches the interests of a small group of top influencers.
\end{itemize}

\subsection{What is blockchain?}

\begin{table*}[]
\caption{Blockchain consensus protocols.}
\centering
\small
\setlength{\tabcolsep}{2pt} 
\renewcommand{\arraystretch}{1.95} 
\begin{tabular}{c|l|l}
\hline \hline
\textbf{Types}             & \multicolumn{1}{c|}{\textbf{Reference}} & \multicolumn{1}{c}{\textbf{Description}} \\ \hline \hline
\multirow{6}{*}{\textbf{\begin{tabular}[c]{@{}c@{}}BFT\\ consensus\end{tabular}}} 
& BEAT \text{\cite{DRZ:18}}   &        A set of five asynchronous protocols focusing on practicality and efficiency \\ 
\cline{2-3} 
& HotStuff \text{\cite{YMRGA:19}}      &  A partially synchronous and leader-based protocol for linearity and responsiveness  \\
\cline{2-3} 
& HoneyBager \text{\cite{MXCSS:16}}  &  An asynchronous protocol constructed from an asynchronous common subset \\
\cline{2-3} 
& Proteus \text{\cite{JBR:19}}    &    A  protocol that  achieves consistent performance by  applying a root committee strategy    \\
\cline{2-3}      
& SBFT \text{\cite{GAGMPRSTT:19}}                &   A  protocol combining collectors, threshold signatures, a fast path, and redundant servers   \\ 
\cline{2-3} 
& Gosig \text{\cite{LWCLX:20}}           & A  protocol applying transmission pipelining and aggregated signature gossip  \\ 
\hline 
\multirow{8}{*}{\textbf{\begin{tabular}[c]{@{}c@{}}PoX\\ consensus\end{tabular}}} 
& Proof of Work \text{\cite{Nak:08}}            &  Participants compete to solve computational puzzles to validate transaction blocks \\
\cline{2-3} 
& Proof of Stake \text{\cite{Wood:16, KRDO:17, GHMVZ:17}}       &   Validators are selected to validate transaction blocks
based on their stake\\ 
\cline{2-3} 
& Proof of Activity \text{\cite{BLMR:14}}    &    Participants first solve cryptographic puzzles and then transit to a staking mechanism \\  
\cline{2-3} 
& Proof of History \text{\cite{Yakovenko:18}}  &  Verifiers collectively creat a ledger with verifiable passage of time between events \\
\cline{2-3} 
& Proof of Burn \text{\cite{KKZ:20}} &   Participants  burn some of their own tokens to gain the right to validate transactions \\ 
\cline{2-3} 
& Proof of Importance \text{\cite{NEM:18}}    &  Participants harvest new blocks depending on their importance to the network \\
\cline{2-3} 
& Proof of  Replication \text{\cite{Fisch:18}}       & Prover provides evidence that a unique copy of the data is stored \\
\cline{2-3} 
&  Proofs of Space \text{\cite{DFKP:15}}    & Participants prove their commitment by  allocating significant amounts of storage space\\
\hline \hline
\end{tabular}
\label{tab:consensus} 
\end{table*}

A blockchain is essentially a global network of interconnected nodes that serves as a distributed database or ledger. It contains blocks of transaction records that are shared by all participants in the network.
Each block carries the digest of the previous block, which is the output of the cryptographic hash function. This digest can be used to verify the validity of the previous block, such that it connects the blocks into a growing chain. Blocks cannot be changed backward without affecting all the following blocks. This assures data confidentiality and integrity, as well as the ability for  blockchain participants to verify and audit transactions.

The consensus protocol is the fundamental building block of blockchain networks as well as the Web 3.0 ecosystem. 
According to the latest research on the priority of Web 3.0 development factors, the most important factor in the development of Web 3.0 is the consensus mechanism, with a weight of 20.0\% \cite{Agbozo:23}.
Fundamentally, a consensus protocol was a fault-tolerant mechanism used between distributed processes to achieve a common agreement on a single data value. From the perspective of blockchain, the consensus protocol is used by each participant to agree on the state of a distributed ledger. 
Consensus protocols ensure the reliability of blockchain networks by fostering trust among anonymous peers in a decentralized setting and enacting regulatory economic incentives in this way.  
There are many consensus mechanisms powering the blockchain systems. Fundamentally, they can be broadly grouped into two categories: Byzantine fault-tolerant (BFT) consensus and Proof of Something (PoX) consensus (see Table~\ref{tab:consensus}). 
A comparison between some of the most commonly used consensus protocols can also be found in \cite{YC:18, FCHC:20, BMB:20}.

\subsection{What is AI?}

AI is an interdisciplinary science that uses computers and data to mimic the problem-solving and decision-making abilities of the human brain. It includes machine learning (ML) and deep learning (DL), which are creating a paradigm shift and making breakthroughs in all different fields. Typically, humans will play a role in supervised learning, providing positive feedback for good decisions while preventing bad ones. However, certain AI systems are designed for unsupervised learning, where they eventually figure out the rules through pattern recognition and learning using large amounts of data.

\begin{figure*}[]
    \centering
    \includegraphics[scale=0.073]{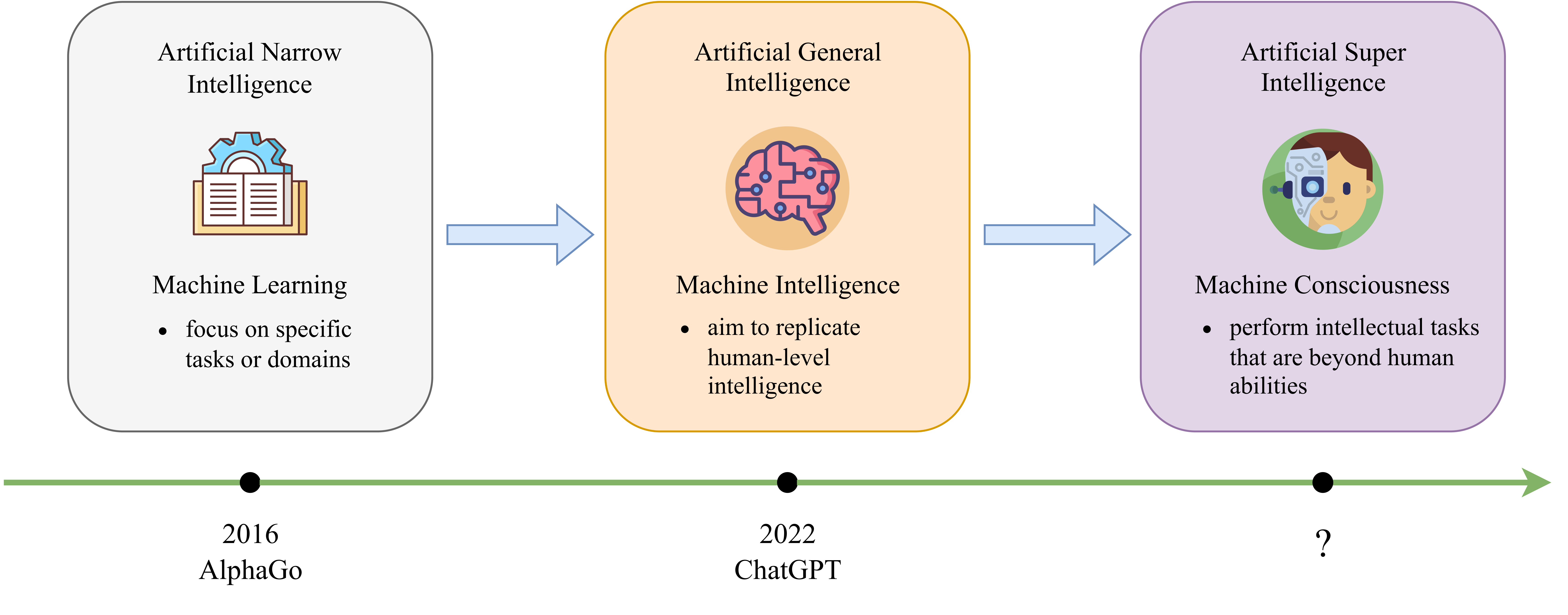}   
    \caption{Levels of AI sophistication.}
    \label{fig: aiLevel}
\end{figure*}
In terms of technical capabilities, there are three types of AI as shown in Fig.~\ref{fig: aiLevel} \cite{IBMAI:20}. 
One type is known as Artificial Narrow Intelligence (ANI), which has been integrated to improve people's daily lives. 
Both voice-controlled personal assistants and self-driving cars have benefited greatly from this type of AI. In particular, breakthroughs in healthcare are also dependent on it, as it can greatly reduce repetitive tasks that can lead to human errors, enhance the process of developing medical materials, and improve treatment outcomes. Another type of AI is called Artificial General Intelligence (AGI), in which the computer has greatly improved in intelligence compared to the level of Web 2.0. It would be self-aware, capable of problem-solving, learning, and long-term planning.  
If development continues, it will reach the third type of AI, i.e., Artificial Super Intelligence (ASI). With the advent of Web 3.0, the booming development of various edge technologies will propel AI into a more advanced stage.

\subsection{What is edge computing?}

\begin{figure*}[]
    \centering
    \includegraphics[scale=0.065]{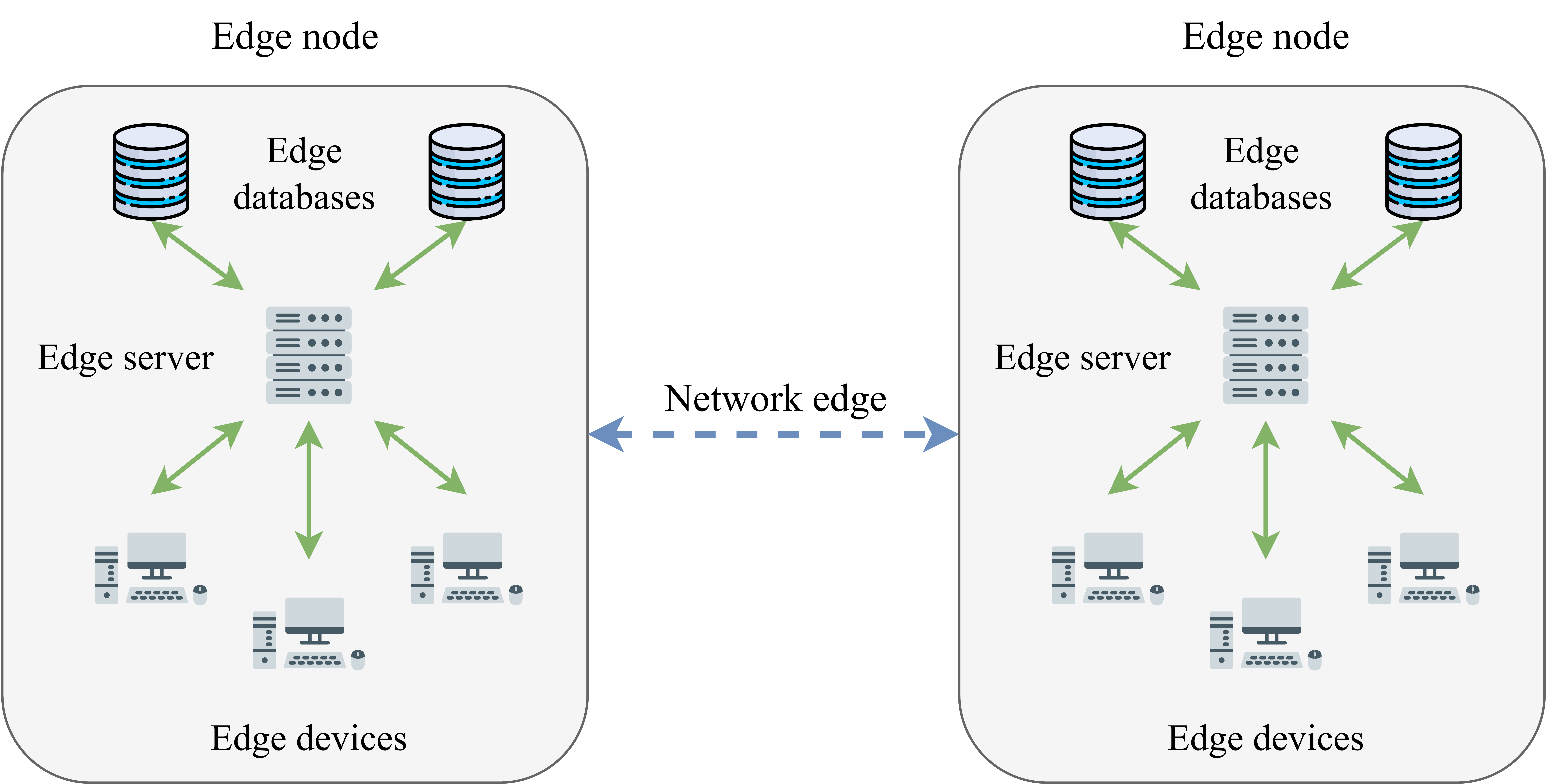}   
    \caption{Edge computing.}
    \label{fig: edge}
\end{figure*}

Edge computing refers to the offloading of data storage and data processing that were previously handled by centralized servers to the edge of the network close to end-user devices. This technique helps reduce data transfer times and device response latency while easing bandwidth congestion on the network. Costs associated with data transmission can be decreased by localizing processing at the edge. Decentralization is accomplished by dispersing computing away from central hubs.  
Additionally, shifting workloads to edge nodes with processing and storage capabilities helps effectively optimize resource usage.

The key components enabling edge computing are edge devices, edge servers, edge databases, edge nodes, and network edge as shown in Fig.~\ref{fig: edge}  \cite{HUA:23}.  
Edge devices refer to the devices that process data near the data source. For example, smartphones, laptops, sensors, and industrial robots.
Edge servers are information technology (IT) computing devices located near edge devices for computing IT workloads and resource management.
Edge databases are database systems that are deployed at the edge of the network. They are used to store the data generated by the edge devices locally instead of sending to a centralized data server.
Edge nodes are nodes that hold edge devices, edge servers, and edge databases. They are connected to the network and each other, acting as intermediaries to facilitate data exchange and resource sharing. 
The network edge is the network infrastructure such as 5G and high-speed satellite Internet that connects edge devices and edge servers with low latency.

\section{Blockchain for Web 3.0: a decentralized and trusted web} \label{sec:cornerstone}
\begin{figure}[!b]
    \centering
    \includegraphics[scale=0.43]{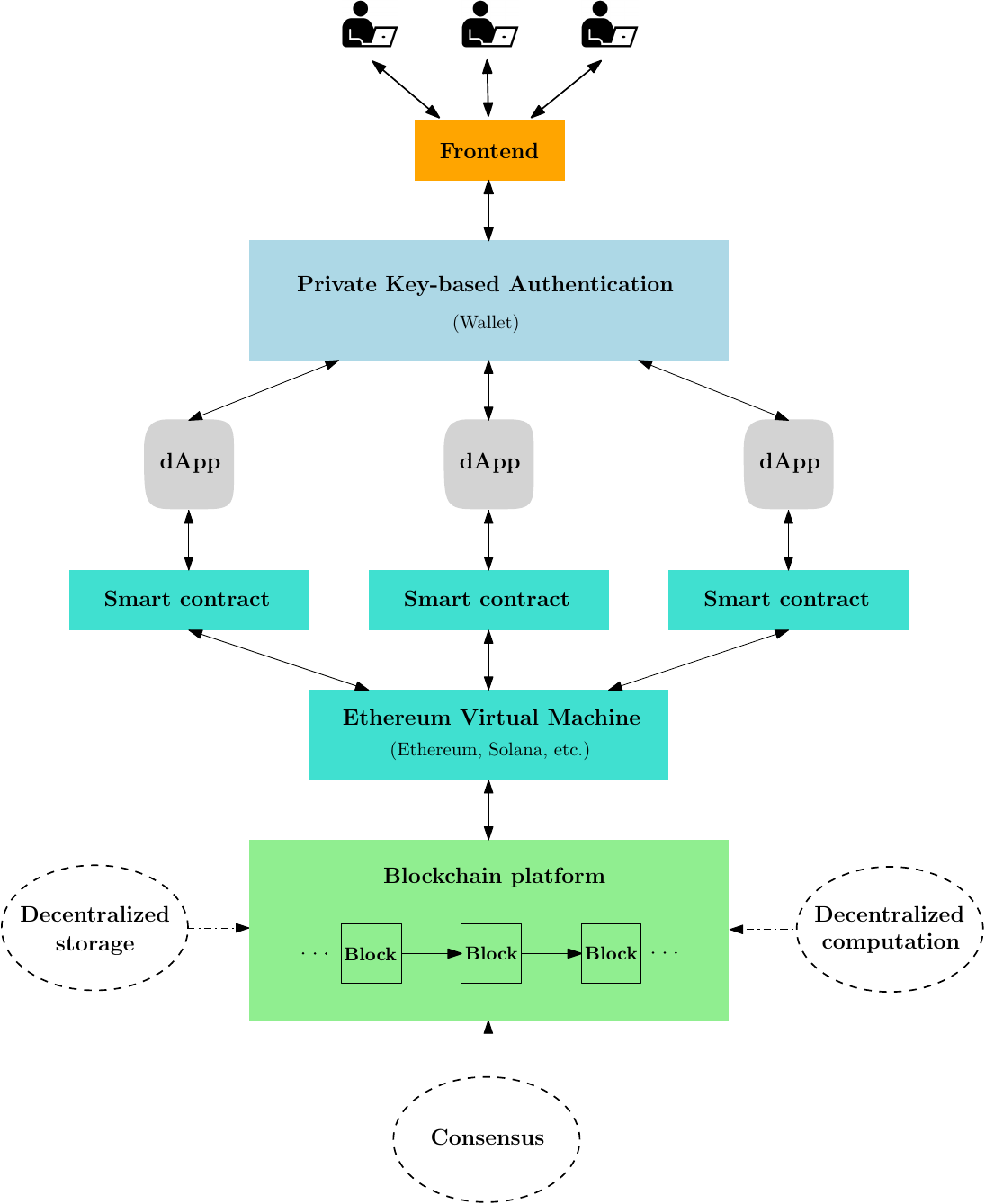}   
    \caption{Web 3.0 infrastructure from the perspective of blockchain.}
    \label{fig: web3architechture}
\end{figure}

Web 3.0 is a decentralized network based on blockchain technology, collectively maintained by nodes scattered around the globe. Blockchain redefines the way data is stored and managed. It uses cryptographic techniques to provide a unique set of states that can enable true P2P transactions without third parties. 
With blockchain, data will be stored on a decentralized network rather than on a centralized server so that privacy and ownership will be given back to individuals \cite{BP:22}. 
Notably, blockchain combined with privacy-preserving technologies has the potential to further enhance the protection of user privacy.
Typically, website fingerprinting is a technique widely used in web browser analysis to infer sensitive information about users by examining traffic patterns.
In response to such attacks, some significant improvements have been proposed for website fingerprinting-based methods. For example, a TCP/IP traffic-based defense mechanism was proposed in \cite{Huang:2023}. It is an efficient and low-overhead defense mechanism against attacks that can filter out the injected noise.
A traffic splitting-based defense mechanism was proposed to limit the data that can be observed by a single entry node in \cite{Cadena:2020}. 
By embedding these defense technologies, blockchain can provide additional layers of security on top of its inherent cryptographic protections and decentralized framework.

Due to its ability to store data in P2P networks, blockchain lays the foundation for Web 3.0. The protocol specifies the management rules, which are guaranteed by a majority vote of all members of the network. Participants are rewarded for their contributions to network security and maintenance. It enables the individuals to reach a consensus while the network collectively records previous user interactions or events. As a result, blockchain technology is certainly a powerful force that can make the network more decentralized \cite{Voshmgir:20}.
Web 3.0 is a backend revolution with a network architecture as shown in Fig.~\ref{fig: web3architechture} \cite{Kasireddy:21, deController:2023}.  
For average users, there will be no change to the interface of the Web. 
From a technical perspective, it is a set of blockchain-based protocols designed to transform the backend of the Web.
In this section, we will first show how blockchain technology is closely related to Web 3.0.  
Then, we will explain the criticality of blockchain for Web 3.0 from the perspective of specific technologies, i.e., off-chain technology, cross-chain technology, and layer 2 solutions. 
Afterward, we will introduce the practical applications of blockchain in Web 3.0.
Finally, the summary and insights are provided.

\subsection{Relevance to Web 3.0}
\begin{figure}[]
    \centering
    \includegraphics[scale=0.057]{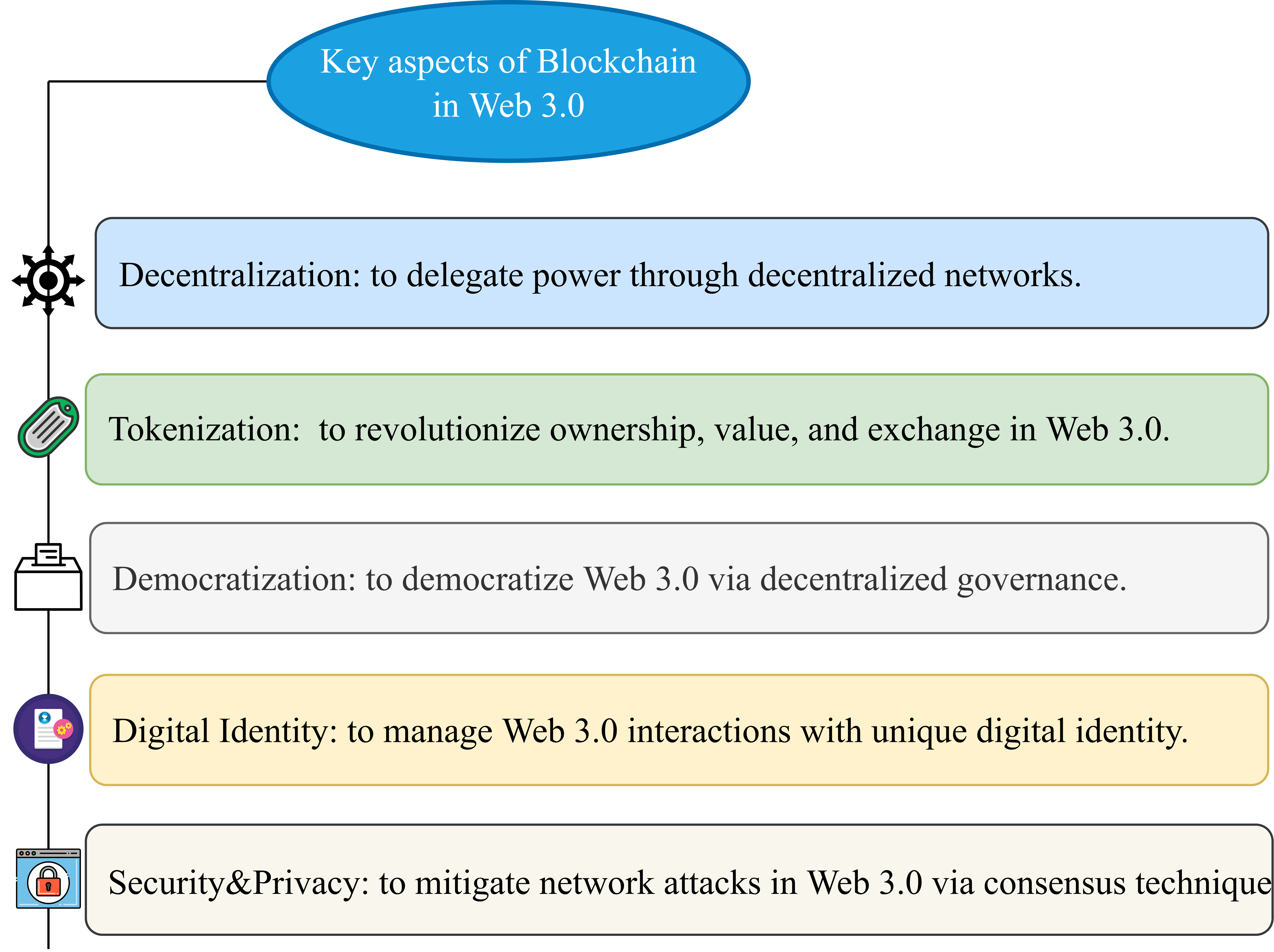}
    \caption{Relevance of blockchain to Web 3.0.}
    \label{fig: blockchainRelevance}
\end{figure}

Web 3.0 is a decentralized and trustworthy Web, aiming to transform a centralized network platform into a decentralized, secure, and user-centric platform. Blockchain technology plays a crucial role in realizing the vision of Web 3.0. This can be demonstrated in several key aspects as shown in Fig.~\ref{fig: blockchainRelevance}.

\subsubsection{Decentralization}
One of Web 3.0's visions is to build a decentralized system that does not depend on a single entity but operates on a distributed network of nodes. The core principle of decentralization is the delegation of power which requires the distribution of power and control outside of the central authority. The technological support comes from blockchain and its underlying P2P networks. Web 3.0 can help create more sustainable networks by decentralizing the energy consumption associated with centralized data centers. 
By distributing data and computing power among multiple nodes, blockchain-based systems can be more energy-efficient and environmentally friendly.
For example, according to measurements \cite{Bassi:21, Kohli:22, Digiconomist:23, Digiconomist:2023}, the energy consumption per Algorand transaction is approximately 0.000008 kWh, whereas each Visa transaction consumes around 0.0015 kWh.
In terms of carbon footprint, each Algorand transaction generates 0.0004 gCO2 compared to the 0.45 gCO2 of each Visa transaction.

\subsubsection{Tokenization}
Token is a digital scarcity in Web 3.0, which refers to the representation of real-world assets by a string of numbers on the blockchain network. The immutability and public verifiability of the blockchain network guarantee the uniqueness, scarcity, and security of this digital string. Tokens can be used to represent access rights, voting rights, or other types of ownership in applications. It also enables fractional ownership, which means assets can be divided into smaller parts that can be easily bought, sold, or traded, enabling more people to participate in investment opportunities. Tokenization has great potential to revolutionize ownership, value, and exchange in the Web 3.0 ecosystem.

\begin{figure*}[]
    \centering
    \includegraphics[scale=0.06]{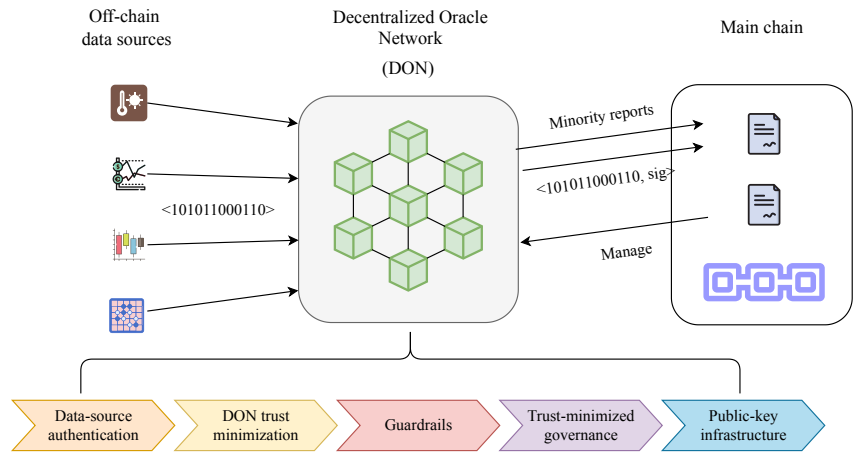}   
    \caption{Interactions between off-chain and on-chain resources via DON.}
    \label{fig: off-chain}
\end{figure*}

\subsubsection{Democratization}
One of the main features of Web 3.0 is the democratization enabled by blockchain technology. Web 3.0 promises to allow users to be rewarded based on their contributions to the Web. However, democratizing in an error-free and fair manner is a major challenge. Essentially, the main problem that the blockchain solves is how to transfer value and control from the platform to the community, while maintaining the prosperity of the platform. 
Decentralization and tokenization are two key aspects that are at the heart of the democratization of Web 3.0. Specifically, the decentralized nature of blockchain technology allows for trustless cooperation between stakeholders within the platform; then the tokenization of blockchain provides a fair way to incentivize different stakeholders to participate in the governance and use of the platform.

\subsubsection{Digital universal identity}
Digital identity is one of the core elements of Web 3.0. It plays a key role in managing Web 3.0 interactions. However, as data breaches and hacks expose the vulnerability of personal data, the utility and relevance of digital identities are becoming increasingly apparent. In particular, the proliferation of social media has resulted in individuals having different digital identities on different platforms. Online identity management faces serious challenges. Blockchain technology can provide a decentralized and interoperable identity system for Web 3.0 users. This ensures that each user has a secure, unique digital universal identity across multiple platforms, eliminating the need for multiple usernames and passwords.

\subsubsection{Security and privacy}
Ensuring the security and privacy of data and transactions is a crucial part of the rapid development of Web 3.0. Blockchain-based Web 3.0 can fundamentally eliminate the need for central institutions to manage data. Moreover, the integration of blockchain in Web 3.0 will greatly mitigate network attacks.   
Every data stream on the blockchain network has to be verified by different nodes. A successful network attack requires controlling the majority of nodes on the blockchain network in order to reach a consensus on its proposals. At the same time, it has to compete with potential new blocks that are added. This is impractical for hackers, as they would be trapped in an endless computational loop, making it harder for them to compromise the network and access users’ personal information and transaction records.

\subsection{Off-chain technology}
Off-chain technologies play a critical role in augmenting blockchains in the Web 3.0 ecosystem.
When blockchains are connected to off-chain resources, their capabilities are greatly enhanced, from incorporating real-world information into on-chain execution to reducing costs and increasing throughput by shifting computation off-chain. 
However, to securely and immutably connect blockchain to external resources, the oracle problem needs to be overcome.
This problem refers to the inherent limitation of blockchains being unable to directly access or use external data resources due to their isolated nature. 
A blockchain oracle is an entity that connects blockchain to external systems, allowing smart contract (SCs) to execute based on real-world inputs and outputs in the Web 3.0 ecosystem.

The decentralized oracle network (DON), proposed by Chainlink in \cite{Breidenbach:2021}, acts as a secure middleware to facilitate communication between on-chain and off-chain resources as shown in Fig.~\ref{fig: off-chain}. 
The DON obtains data from off-chain resources and then forwards the data to the SC deployed on the main chain.
Additionally, the DON incorporates a separate SC for node management. 
Notably, one of the goals of Web 3.0 is to minimize trust.
The implementation of a DON involves integrating a variety of trust-minimized technologies.

\subsubsection{Data-source authentication}
An important component of trust minimization involves strengthening data-source authentication through support for data signing tools and standards. 
By cryptographically signing the data they provide to SCs, a DON enables users to identify which nodes sent data 
and track the behaviors to determine the quality of their performance.
In this way, the end-to-end integrity of the data can be guaranteed.

\subsubsection{DON trust minimization}
To minimize trust in DON, there are two main methods.  
 The first is the failover clients, which are backup clients for nodes in case of unexpected events.  
 They do not increase the number of potential attacks and can reduce reliance on individual client security as they are not deployed on the mainline.
The second is the minority report, which is a parallel report that is forwarded to SCs on the main chain. This is an important mechanism that operates in a threshold manner to maintain the integrity and reliability of the data sources provided by the DON.

\begin{figure*}[]
    \centering
    \includegraphics[scale=0.031]{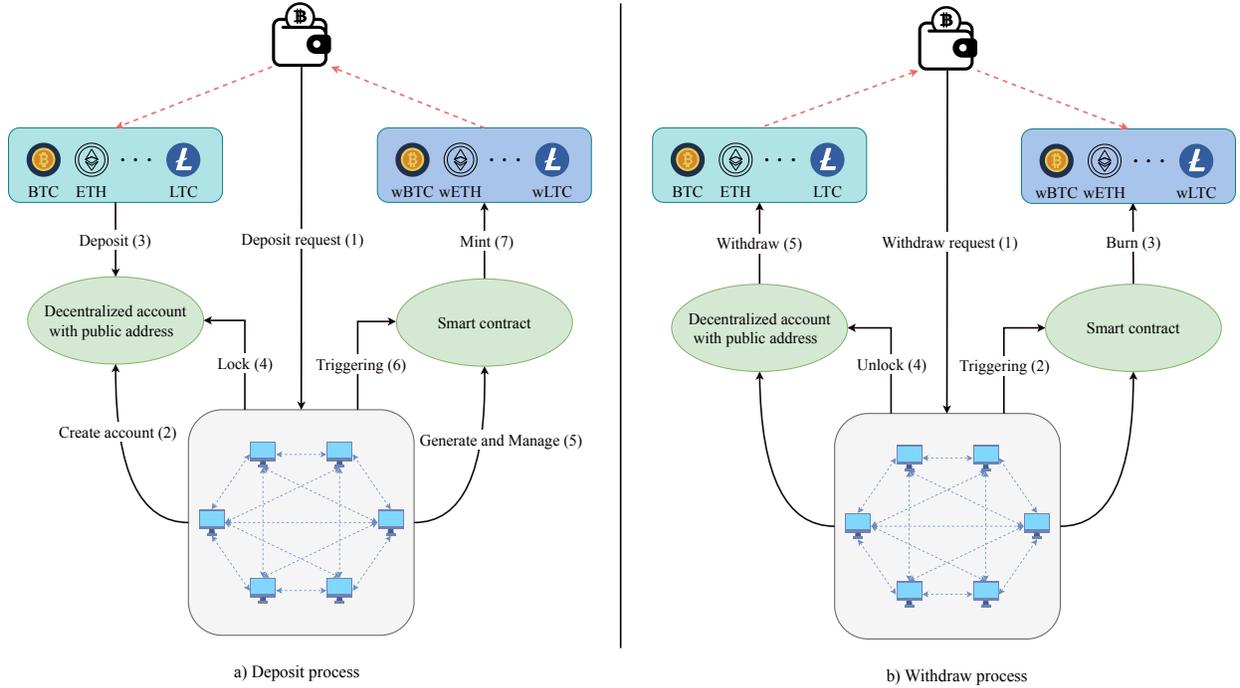}   
    \caption{Cross-chain assets deposit/withdraw flowchart.}
    \label{fig: cross-chain}
\end{figure*}

\subsubsection{Guardrails}
Guardrails are a collection of trust minimization mechanisms involving the implementation of monitoring and fail-safety in SCs. 
A circuit breaker is a guardrail where a SC may control state updates based on inputs. For example, it might be triggered if minority reports change significantly over time.
Escape hatches are emergency facilities that SCs can invoke to terminate pending transactions and future transactions in adverse circumstances. 
Failover means that SCs can provide a failover mechanism to ensure the continuity of services even in the event of DON failure.

\subsubsection{Trust-minimized governance}
Evolutionary governance and emergent governance are two types of trust-minimizing governance mechanisms. Evolutionary governance is about deploying changes gradually, ensuring the community has the opportunity to respond. Emergency governance refers to vulnerabilities in SC that require immediate intervention to avoid catastrophic consequences. Specifically, emergency governance uses a multi-signature intervention mechanism to ensure that signers dispersed across organizations are always available to authorize emergency changes.

\subsubsection{Public-key infrastructure}
As decentralization continues to advance, a strong public key infrastructure (PKI) is required to reliably identify network participants, including DON nodes. The foundation of PKI in DON is the Ethereum Name Service (ENS). ENS allows human-readable Ethereum names to be mapped to blockchain addresses.  
Tampering with the name is inherently as difficult as tampering with the SC that governs it unless the keys are compromised.

\subsection{Cross-chain technology}

A blockchain is a decentralized system powered by distributed ledger technology. However, it is not a cumulative ecosystem since each blockchain is designed for a particular use. They have specific advantages, limitations, and varying levels of decentralization. For example, if a blockchain aims to achieve high transaction throughput, it may be less decentralized and less secure. Since each blockchain is isolated from the others, leveraging the functionality of one blockchain cannot compensate for another.  
To fully take advantage of blockchain technology, cross-chain technology was developed to address interoperability, which can greatly boost productivity in the Web 3.0 ecosystem \cite{Buterin:2016, BVGC:21, WWC:23}. Everyone involved will benefit from the increased flexibility, as users will be able to easily transfer assets and data between blockchains \cite{BMRS:18}. 
Given the wide variety of bridge designs, the most widely adopted bridge design is based on the lock-mint-burn method as shown in Fig.~\ref{fig: cross-chain} \cite{Multichain}. Basically, this type of bridge designates a public address on the source chain for users to deposit their tokens.
On the destination chain, a SC will mint the wrapped tokens 1:1 with the tokens held in the decentralized managed account and send them to the user's wallet.

Cross-chain technology enables trade-offs between two or more blockchains in terms of efficiency, decentralization, and security. Additionally, cross-chain technology can improve chain efficiency, reduce fragmentation, and enable a free flow of users and features between multiple blockchains.  
In recent years, cross-chain platforms with different goals have been developed. For example, 
tBTC  is an Ethereum-like token that is linked to the value of Bitcoin \cite{Threshold}. It enables Bitcoin holders to gain access to the Ethereum ecosystem and decentralized finance (DeFi) applications. 
Parity Bridge is a cross-chain solution to connect fast and cheap Proof of Authority chains with the Ethereum public network and any other Ethereum-like PoW chain \cite{Parity}.
Wormhole is a communication bridge between Solana and other mainstream blockchain networks \cite{Solana:20}. Existing projects, platforms, and communities are able to seamlessly transfer digital assets across blockchains by utilizing Solana's high-speed and low-cost features.

\subsection{Layer 2 scaling solutions}
\begin{figure} []
	\centering
	\includegraphics[scale=0.061]{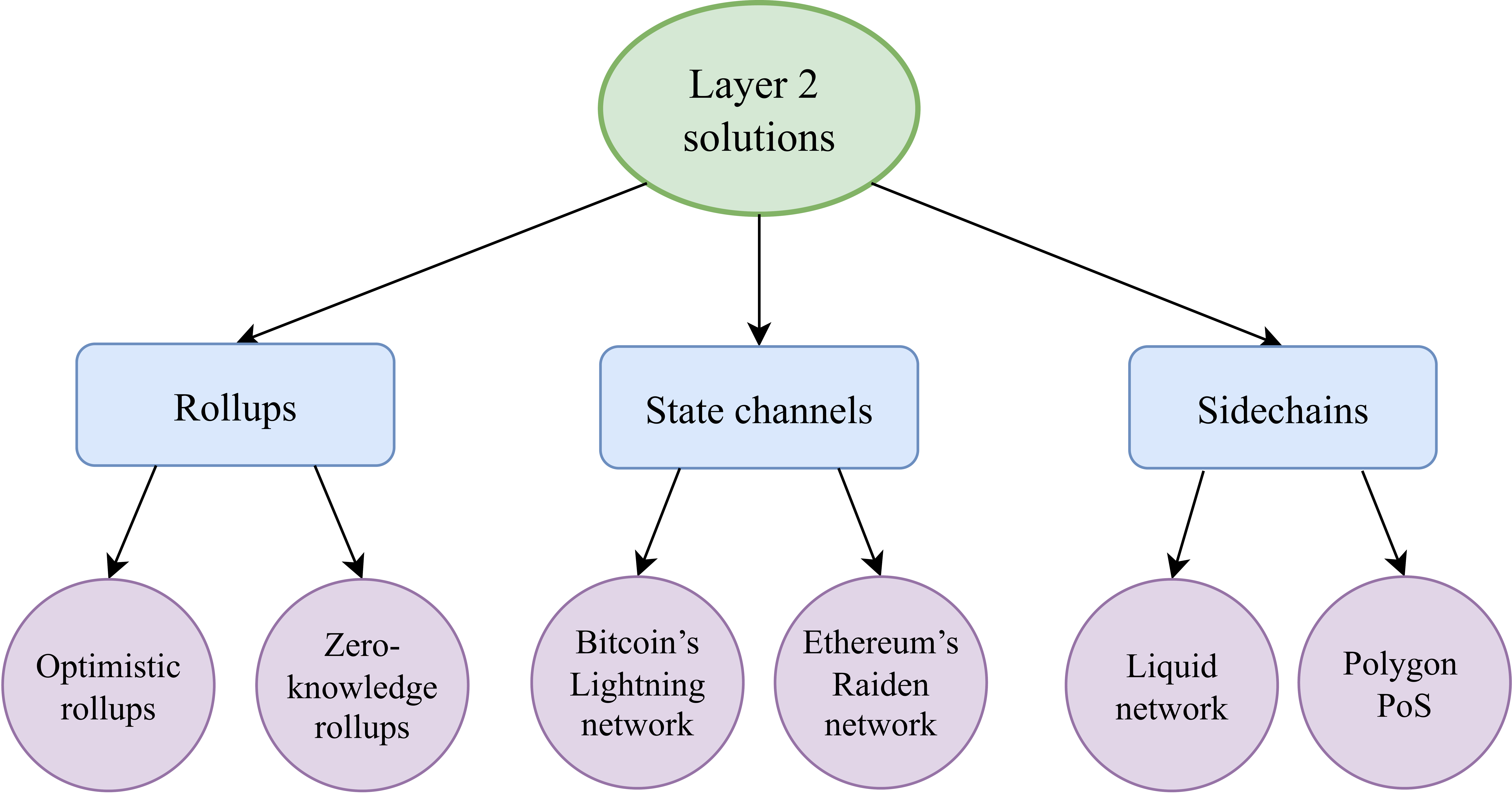}
	\caption{Layer 2 solutions with typical projects.}
        \label{fig:layer2}
\end{figure}

Blockchain-based Web 3.0 will revolutionize the way people transact and transfer value with each other. However, with its rapid development, scalability has become a bottleneck.   
To solve this problem, scaling solutions developed on top of Layer 1 blockchain networks, also called Layer 2 solutions, are widely adopted.
The purpose is to improve the scalability, efficiency, privacy, and other characteristics of the underlying blockchain network. 
There are several types of Layer 2 solutions. For example, rollups, state channels, and sidechains. 
Below is an overview of Layer 2 solutions as shown in Fig.~\ref{fig:layer2}.

\subsubsection{Rollups}
Rollups allow transactions to be executed outside the Layer 1 blockchain. When consensus is reached on the transaction data, the data will be posted back to the main chain and secured by the security mechanism of the underlying blockchain.
Specifically, there are two types of rollups in terms of security models: optimistic rollups and zero-knowledge (ZK) rollups \cite{Smith:23}.
An optimistic rollup is an approach used to scale the Ethereum network by moving some computations and state storage off-chain.  
In ZK rollups, transactions are bundled into batches and executed outside the Layer 1 blockchain. 
The summary of changes will be submitted to the main blockchain, rather than submitting each transaction individually. 
To guarantee the correctness of the changes, they provide proof of validity by utilizing zero-knowledge proofs.   
The rollup paradigm is based on a final settlement on the Layer 1 blockchain. It requires rollups to post a copy of every transaction to the Layer 1 blockchain. In order to address the data availability bottleneck, it is necessary to create dedicated space for rollups.  
For example, Danksharding scales Ethereum for high throughput by scaling the number of binary large objects (a.k.a. blobs) attached to blocks from 1 to 64 \cite{EthereumDanksharding:23}. 

\subsubsection{State channels}
State channels enable participants to securely transact off-chain by utilizing multi-signature contracts. Then, two on-chain transactions that can open and close the channel are submitted for final settlement with the main network \cite{Smith:23}.  
State channels represent a more generalized form of payment channels. They can be utilized not only for payments but also for any state updates on the blockchain, such as changes within a SC.
The most well-known examples are Bitcoin’s Lightning network and Ethereum’s Raiden network. 
The Lightning Network is a decentralized payments network that runs on top of the Bitcoin blockchain. It greatly improves the scalability of the Bitcoin blockchain by allowing users to make multiple transactions off-chain without broadcasting each transaction to the entire network.
Near-instant and low-cost Bitcoin settlements can be achieved between participants. 
The Lightning Network uses a two-party, multi-signature Bitcoin address (channel) to store funds. It requires both parties to agree on the new balance to spend funds from the channel.  
In this case, the network allows dynamic participation so that payments can be made through a network of channels \cite{JT:16}.  
Similarly, the Raiden network is Ethereum's version of Bitcoin's Lightning Network. It enables the transfer of tokens that are compliant with the ERC20 standard on the Ethereum blockchain. 
This is achieved through the use of digital signatures and hash-lock (i.e., balance proof).
Digital signatures ensure that neither party can exit any value transfers contained therein, as long as at least one participant decides to submit it to the blockchain.
The Raiden balance proof is a protocol executed by the Ethereum blockchain. Since no one other than these two participants can access the tokens stored in the payment channel SC, the Raiden balance proof is as binding as on-chain transactions \cite{Raiden}.

\subsubsection{Sidechains}
A sidechain is an independent blockchain that is linked to the main blockchain. It allows assets to move between the sidechain and the main chain. The purpose is to solve scalability issues by offloading some of the validation and transaction processing to the sidechain \cite{Smith:23}.   
Sidechains interact with Layer 1 blockchains in two primary ways: the first way is to provide a mechanism (i.e., a cross-chain bridge) for bridging assets from Layer 1 blockchain to their respective sidechain; the second method involves periodically publishing its state snapshots (i.e., highly compressed summaries of the balances of all accounts on its network) to Layer 1 blockchain network \cite{Cahill:22}.
Examples of sidechains include the Liquid network and Polygon Proof of Stake (PoS). 
The Liquid Network is a sidechain of the Bitcoin blockchain. 
It facilitates fast, secure, and private settlement of digital assets.  
The Liquid-version Bitcoins are backed by an equal amount of Bitcoins on the main chain, ensuring verifiable 1:1 backing. This allows users to trade using the speed and confidentiality of the Liquid network \cite{LiquidNetwork}.  
Polygon PoS is a 3-layer architecture sidechain of the Ethereum blockchain to connect Ethereum-compatible blockchain networks.  
The Ethereum layer consists of a set of staking SCs on the main chain, allowing users to stake tokens to join the system.
The Heimdall layer is a validation layer, consisting of PoS Heimdall nodes that run in parallel to the main chain. These nodes monitor the staking SCs and commit checkpoints from Polygon to the Ethereum main chain.  
The Bor layer is a layer for producing sidechain blocks.  
It is used to aggregate transactions into blocks for periodic verification by Heimdall nodes \cite{Tripathi:23}.

\subsection{Practical applications of blockchain in Web 3.0}

\begin{figure}[]
    \centering
    \includegraphics[scale=0.063]{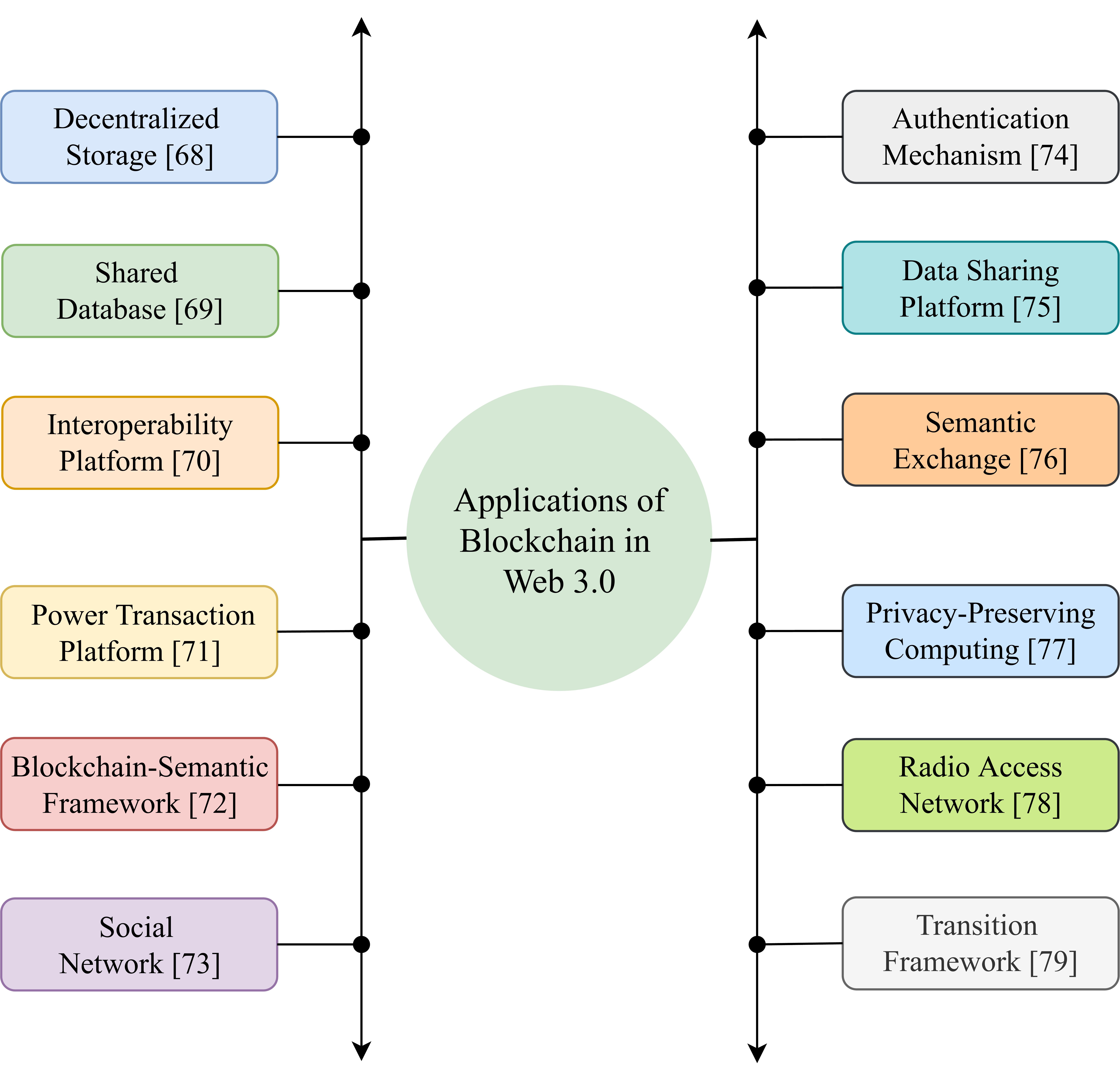}   
    \caption{Applications of blockchain in Web 3.0.}
    \label{fig: ApplicationsBlockchain}
\end{figure}

In this section, we further explore the applications of blockchain in Web 3.0, as shown in Fig.~\ref{fig: ApplicationsBlockchain}, to provide intuitions on how to leverage blockchain in the Web 3.0 ecosystem. 
Benet in \cite{Benet:14} designed a P2P distributed file storage system called InterPlanetary File System (IPFS).
It is a modular suite of protocols for storing and sharing data, aiming to store files by connecting computing devices around the world to the same file system. It plays a crucial role as the file storage solution underlying the decentralized vision of Web 3.0. Compared to traditional centralized storage solutions, IPFS uses a global P2P network, allowing for permanent and immutable data storage without single points of failure. Its content addressing and distributed hash table technology ensure fast and reliable file retrieval.
Drakato \textit{et al.} in \cite{Drakatos:21} proposed Triastore, a blockchain database system that can store and retrieve ML models from the blockchain. To this end, Triastore introduced Proof of Federated Learning for a global model, and Blockchain Consensus for committing the generated model data to a blockchain database. In the context of Web 3.0, the authors claimed that Triastore has the potential for big data analytics in telecommunications and smart city applications.
Liu \textit{et al.} in \cite{Liu:21} proposed an interoperability platform, HyperService, that provides interoperability and programmability between blockchains in order to make the Web 3.0 ecosystem connected. In particular, HyperService is powered by a unified programming framework for developers and a secure cryptographic protocol for blockchain.
A Web 3.0-based P2P platform, VA3, was developed by Chopra \textit{et al.} in \cite{Chopra:22} for electricity settlement at an individual level.
Specifically, VA3 automates the measurement of power consumption and production by feeding this data into a SC to modify the home router. Then, the SC could automatically manage electricity settlements.
Lin \textit{et al.} in \cite{Lin:2023} proposed a blockchain-semantic framework for Web 3.0, aiming to address the problem of unsustainable resource consumption for computation and storage due to the explosive growth of on-chain content and the growing user base. Specifically, an Oracle-based proof of semantic mechanism was introduced to facilitate on-chain and off-chain interactions while maintaining system security.  
Additionally, a DL-based sharding mechanism was designed to improve interaction efficiency.
Palanikkumar \textit{et al.} in \cite{Palanikkumar:23} proposed a decentralized social network system implemented using the Web 3.0 Library, which is a collection of Ethereum JavaScript application programming interfaces (APIs).
This library provides functionalities to interact with the Ethereum blockchain. In this way, an Online Social Network (OSN) service was created in a decentralized manner for democratic self-management.
Petcu \textit{et al.} \cite{Petcu:23} proposed a novel authentication mechanism utilizing Ethereum blockchain technologies, enabling the browser to interact with the user's software and hardware wallets to implement user authentication. With this approach, Web 3.0 authentication could provide enhanced security, privacy, and ownership of user data compared to existing authentication methods that rely on third-party authentication service providers.
Razzaq \textit{et al.} in \cite{Razzaq:23} proposed a Web 3.0 Internet of Things (IoT) data sharing framework based on IPFS. Specifically, blockchain and SCs are used to provide data security. Hybrid storage is used to achieve secure data exchange. Additionally, access control policies are stored on-chain to ensure policy integrity and allow for public auditing of any policy changes.
Lin \textit{et al.} in \cite{Lin:23} proposed a blockchain-based framework for semantic exchange in Web 3.0, which aims to achieve fair and efficient interactions. Specifically, it first tokenized semantic data as non-fungible tokens (NFTs). Trading strategies were then optimized via the Stackelberg game. Afterward, ZK proof was leveraged to allow the sharing of authentic semantic information. 
Guo \textit{et al.} in \cite{Guo:23} proposed a privacy-preserving computing architecture in Web 3.0. The main building blocks are state channel and computing sandbox,  
used to ensure secure and reliable computation. In addition, the onion routing technology was used to preserve user privacy.
Qiu \textit{et al.} in \cite{Qiu:23} proposed a framework, called FogBC-RAN, to establish a secure and decentralized communication system in Web 3.0. To this end, a cross-chain information transmission process was introduced for efficient cost-sharing. Additionally, a computational offloading strategy using matching game theory was used to minimize the system cost of computationally intensive Web 3.0 applications.
Yu \textit{et al.} in \cite{Guang:2022} proposed a framework called WebttCom that allows for a transition from Web 2.0 to Web 3.0. The proposed framework can build a connection between traditional Web 2.0 applications and Web 3.0 platforms, ensuring data privacy and governance while improving development efficiency.
Specifically, an interpreter mechanism was used to aggregate and process requests between the Web 2.0 and Web 3.0 domains.

\subsection{Summary and insights} 
Off-chain, cross-chain, and layer 2 solutions play a vital role in the blockchain-based Web 3.0 system. 
Off-chain technology greatly expands Web 3.0's ability to provide solutions to the real world by connecting on-chain and off-chain resources.
Cross-chain technology provides interoperability for isolated blockchain platforms, jointly promoting the development of the Web 3.0 ecosystem.
Layer 2 solutions improve the scalability of blockchain-based Web 3.0 by offloading computationally intensive operations. 
These technologies demonstrate effectiveness in their respective focuses while taking into account other aspects, such as decentralization, security, and privacy.
However, it is challenging to make effective tradeoffs. It is important to first identify key concerns and then develop a logical strategy. For example, scalability may take precedence over decentralization and security in DeFi, whereas security and decentralization are more important than scalability in the NFT market.

\section{AI for Web 3.0: an intelligent and semantic web} \label{sec:Artificial}
AI plays a crucial role in realizing a more decentralized, secure, and user-centered Web 3.0. By effectively integrating AI technology into various areas of the Web 3.0 ecosystem, it is promising to bring about an era of more intelligent, efficient, and personalized digital experiences.
However, the dominance of centralization has been a longstanding characteristic of AI-based solutions due to their heavy reliance on centralized massive datasets and computing resources. Traditional AI techniques typically require aggregating large amounts of data and performing computationally intensive training processes. This has inevitably led to a centralization of both the data and infrastructure that AI advances have been built upon up to this point. In contrast, Web 3.0 aims to build a decentralized architecture with no single point of control, raising novel challenges for integrating AI in a manner that is distributed, privacy-preserving, and aligned with the vision of decentralization.
As we explore the decentralized landscape of Web 3.0, it is necessary to consider how AI can adapt its centralization tendencies in order to flourish in this emerging environment.

In this section, we first show the relevance of AI and Web 3.0. Then we illustrate how generative AI and Web 3.0 complement each other through an in-depth study of generative AI. Afterward, we will introduce the practical applications of AI in Web 3.0. 
Finally, the summary and insights are provided.

\subsection{Relevance to Web 3.0}
\begin{figure}[]
    \centering
    \includegraphics[scale=0.065]{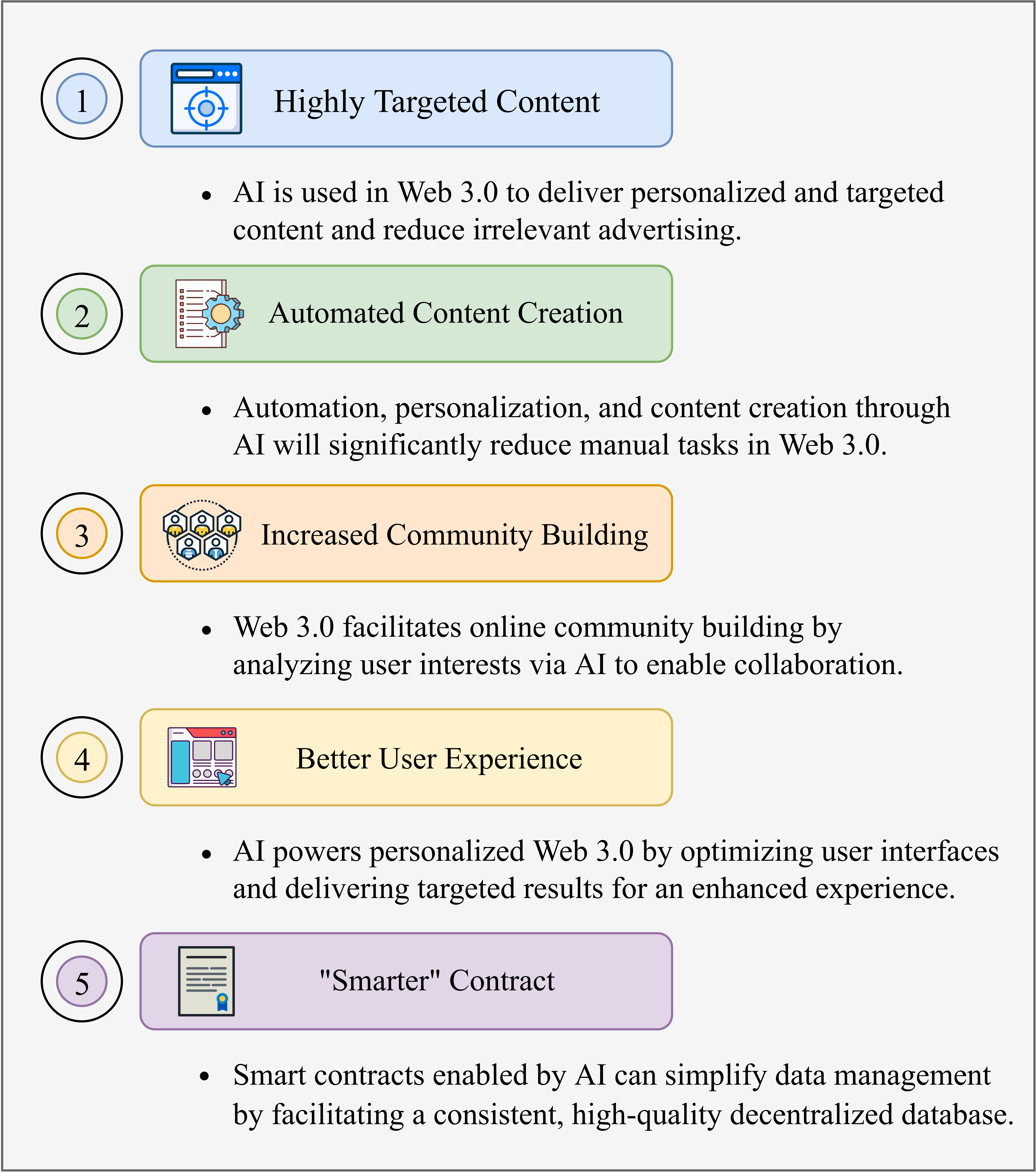} 
    \caption{Relevance of AI to Web 3.0.}
    \label{fig: aiRelevance}
\end{figure}

Web 3.0 represents an intelligent and personalized Web with the goal of providing users with a more seamless experience. AI can drive the applications of Web 3.0 to handle more complex tasks due to its ability to process and analyze large amounts of data. The high correlation between AI and Web 3.0 can be seen from the following key aspects as shown in Fig.~\ref{fig: aiRelevance}.

\subsubsection{Highly targeted content}
Web 3.0 is powered by AI technology to provide consumers with more personalized and targeted advertising. With the ability to understand user intent and preferences, Web 3.0 allows for more effective and efficient marketing campaigns. As a result, annoying and irrelevant advertisements can be eliminated. Only the most relevant content is available to customers. For example, in e-commerce, recommender systems make personalized product recommendations by analyzing users' browsing behavior, search history, and purchase history. 
\subsubsection{Automated content creation}

Web 3.0 allows for the creation of smarter, connected, and interactive web experiences that automate more complex tasks and reduce the need for human intervention. In Web 3.0, automated content creation will become more prevalent through the use of AI technologies. High-quality and dynamic content can be generated based on user queries and preferences. For example, ChatGPT not only helps users generate personalized product descriptions and marketing materials but also optimizes them in a real-time interactive manner. 

\subsubsection{Increased community building}
Web 3.0 is expected to foster community building through the utilization of AI technologies. AI helps connect people with similar interests, skills, and goals by analyzing user behavior and preferences, thereby creating more meaningful discussions and collaborations in online communities to increase engagement and build a sense of community. In this way, the Web will be more interactive, intelligent, and connected.

\subsubsection{Better user experiences} 
The convergence of Web 3.0 and AI holds significant potential in enhancing the user experience. AI can deliver more precise and relevant results to Web 3.0 users, as well as personalize their interfaces to improve the usability and accessibility of Web 3.0.  In order to accomplish this goal, AI-enabled websites need to classify data and present information that is deemed useful to individual users, providing a personalized and improved navigation experience. Users will search and find what they need more easily and precisely, making Web 3.0 applications more user-friendly. From this point of view, prioritizing user experience throughout the entire development of Web 3.0 applications will be more important than ever.

\subsubsection{``Smarter" contract}
In the era of Web 3.0, data management will be more important than ever as data becomes more complex and time-consuming to manage. Additionally, the nature of data ownership means that data will not be managed on a central server. As a decentralized solution, SCs can produce a clean version of data by connecting multiple data sources. In particular, when combined with AI, SCs have great potential to improve and simplify data management and avoid duplicate aggregation. If the underlying P2P network of Web 3.0 is regarded as a unified database, AI-powered SCs will facilitate the establishment of a consistent and high-quality database.

\begin{figure*}[]
    \centering
    \includegraphics[scale=0.052]{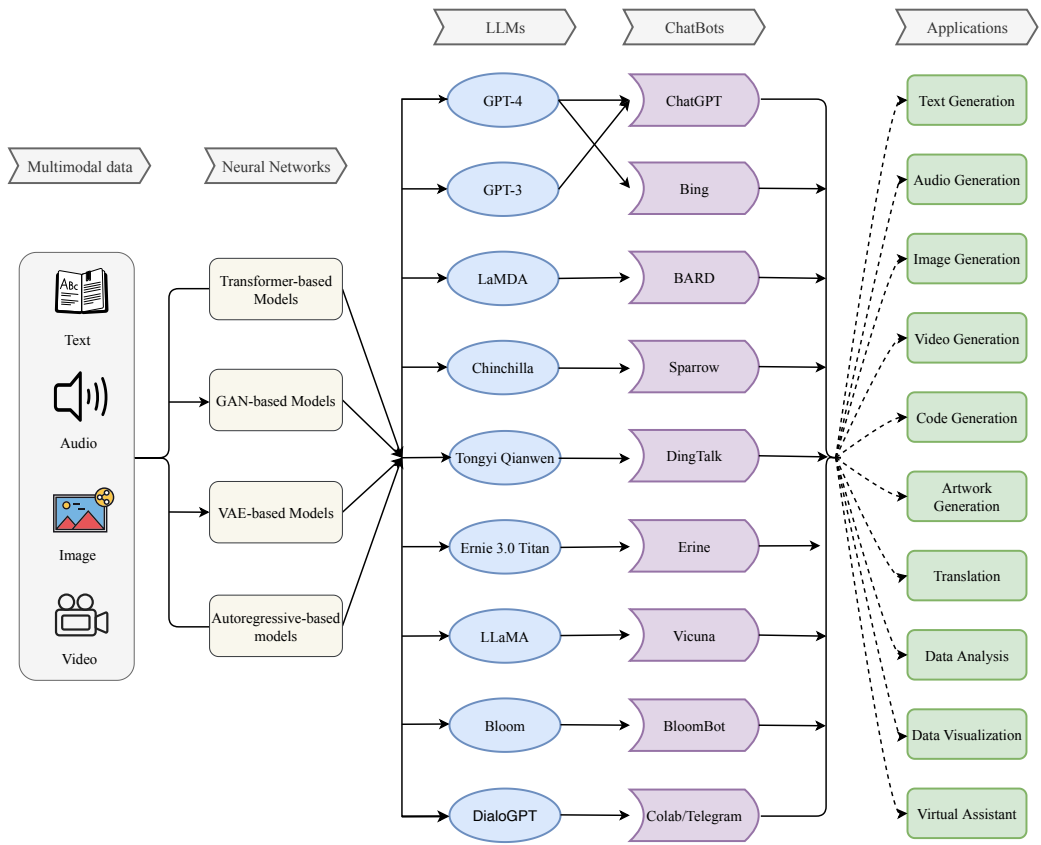} 
    \caption{Generative AI.}
    \label{fig: generativeAI}
\end{figure*}

\subsection{Generative AI}
Generative AI has emerged as the most exciting technology of AI advancement within two years. 
It empowers various applications by creating new data that is similar to human-generated data as shown in Fig.~\ref{fig: generativeAI}.
Notably, generative AI has gained significant interest in semantic communication \cite{LinAIGC:23, LinAIGC:2023, LiuAIGC:23, XiaAIGC:23, Cheng:23} and edge network \cite{DuAIGC:2023, LiuLifecycle:2023, XuAIGC:2023, DuGenerativeAIGC:2023, WangGenerativeAIGC:2023}. 
The rapid growth is attributed to the creation of large language models (LLMs) that may have billions or even trillions of parameters \cite{cao:23, Zhang:23}. 
Unlike traditional AI, which focuses on analyzing and processing existing data to accomplish tasks such as classification and clustering, generative AI creates new and original data by learning patterns and features from existing datasets.
With generative AI, Web 3.0 will be more creative and engaging than ever before.
The development cycle of decentralized applications (dApps) will be dramatically shortened to help companies stay ahead of the competition. In addition, generative AI has the potential to fundamentally change the way a wide range of traditional industries operates due to the highly automated creative process. In the entertainment industry, generative AI has brought huge disruption to the film industry.

\subsubsection{Underlying technologies}
Natural Language Processing (NLP) and Computer Vision (CV) are two distinct subfields of AI. The former focuses on the interaction between computers and human language. The latter focuses on enabling machines to interpret and understand visual information.
Both are important components of generative AI.

\begin{itemize}
\item \emph{NLP}:
NLP is dedicated to empowering computers to comprehend and react to text or speech input in the same manner that humans respond to their own text or speech. NLP makes it possible for computer algorithms to effectively summarize massive volumes of information and translate text across languages and spoken commands \cite{IBMNLP:20}.  
This is critical in Web 3.0 as natural language is ambiguous, making it difficult for algorithms to accurately recognize and process text or audio data.
With the increase in computing power and a large amount of decentralized data available in Web 3.0, 
computers will be intelligent enough to interpret information to provide faster and more precise results, 
which makes machines virtually indistinguishable from human users. Imagine that a voice assistant in Web 3.0 is able to process all the unstructured data on the network. They will comprehend the meaning of anything on the web and deliver a thorough answer rather than merely replying with Wikipedia information and reading Web 2.0 articles \cite{Treviso:23, Qin:23, KKKS:23}.

\item \emph{CV}:
CV aims to use computers to extract features from a large number of visual inputs and then provide recommendations. The computer will develop the ability to distinguish between images if enough data is supplied into the model. This will be greatly met in Web 3.0 with the explosive growth of decentralized data volume. Convolutional neural networks (CNN)-type algorithms and recurrent neural networks (RNN)-type algorithms are the pillars of computer vision \cite{Le:22, Bi:22}.  
A CNN aids a computer's vision by breaking down an image into labeled pixels which will be further used to perform convolutional operations.
Similarly, RNN techniques are applied in video applications to assist computers in understanding the connections between images within a sequence of frames. The development of self-driving cars relies on computer vision to interpret visual inputs from car cameras and other sensors, in order to understand the environment. It is crucial for distinguishing other objects on the road, such as various automobiles, traffic signs, pedestrians, and all other visual information.

\end{itemize}

\subsubsection{Types of generative AI}

Generative AI is a form of unsupervised learning, which means that the model learns to generate new data samples without being explicitly told what the correct output should be. Instead, the model is trained on a large dataset of examples and learns to capture the underlying patterns and structures in the data.
There are four main types of generative models as shown in Fig.~\ref{fig: generativeAI},   
including transformers, generative adversarial networks (GANs), variational autoencoders (VAEs), and autoregressive models.

\begin{itemize}
\item \emph{Transformers}: A transformer model is a neural network that understands contextual meaning by analyzing relationships and patterns in sequential data \cite{Vaswani:17}. It is based on an attention mechanism to selectively prioritize different parts of the input sequence to produce relevant outputs.  
It consists of an encoder and a decoder. The encoder processes input sequences to generate latent representations that capture semantic information.
One of the key advantages is its ability to process input sequences in parallel, which makes it much faster than traditional RNNs for long sequences. This, together with its effectiveness in capturing long-range dependencies (a subtle way of detecting the interactions and interdependencies of even distant data elements in a series of data), makes the transformer model a fundamental model driving a paradigm shift in AI.
\item \emph{GANs}: One of the key breakthroughs in the development of generative AI was the introduction of GANs \cite{Goodfellow:14}. A GAN involves a generator and a discriminator. These two networks oppose each other, using a two-player game-like approach to generate new data. 
The generator generates new data based on patterns it learns from the training dataset while the discriminator evaluates the authenticity of the generated data.  
This adversarial training approach allows the generator to generate data that is indistinguishable from real data, while the discriminator has enhanced capabilities in identifying the generated data.
\item \emph{VAEs}: A VAE is a neural network for unsupervised learning of complex data distributions \cite{KW:13}. It involves two sub-processes: the encoder maps input data into a latent space; the decoder then draws samples from the data distribution in this latent space to generate the output. 
Unlike traditional autoencoders, VAEs introduce randomness into the encoding process, which allows them to be used in generative AI for generating new data with similar patterns to the input data. Furthermore, they could be combined with other generative models to create more advanced and powerful generative models. 
\item \emph{Autoregressive models}: An autoregressive model is a type of statistical model used for forecasting future values in time series data based on prior observations \cite{Gregor:14}.  
It is assumed that there is an auto-correlated structure in the data where the current value of a time series can be modeled as a linear combination of prior values in the series. 
The term ``autoregressive" comes from the fact that these models involve regressing a time series data to its own past values. However, autoregressive models are primarily used for stationary time series with constant mean and variance over time. Non-stationary time series may require transformation before applying an autoregressive model.
\end{itemize}

\subsection{Generative AI for Web 3.0} 

Web 3.0 is envisioned to transform the Internet into a semantic, intelligent, and user-centric platform where information is interconnected through semantic understanding. Generative AI plays a crucial role in realizing this vision. 
There are four key drivers for integrating generative AI into Web 3.0, i.e.,
truly bringing semantics to Semantic Web 3.0, efficiently developing Web 3.0, easily performing data analysis, and proactively providing security assistance.

\subsubsection{Semantic understanding}

Web 3.0 proposes a digital realm where machines can interact and communicate with both other machines and human users. 
However, in order for machines to precisely and effectively communicate, they must first understand the meaning and subtle differences of digital information. This is why generative AI will be the cognitive layer of Web 3.0 as illustrated in Fig.~\ref{fig:stack}, driving machines to comprehend various types of content through DL algorithms. For instance, text, audio, images, and video as shown in Fig.~\ref{fig: generativeAI}.
An increasing number of practical applications also highlight the advantages of incorporating semantic capabilities into the Web 3.0 ecosystem. For example, 
Alice is the first intelligent non-fungible token powered by GPT-3, allowing it to adjust how it interacts with people based on each new interaction \cite{Alethea:2021}. 
Pregelj in \cite{Pregelj:2023} introduces a ChatGPT-based Web 3.0 plugin that enables wallet creation, on-chain queries, and on-chain operations directly from prompts.
SuperCool AI is a digital marketplace based on generative AI that generates and trades NFTs via prompts \cite{Sathavara:2023}.

\subsubsection{Efficient development}
To realize the decentralized and intelligent vision of Web 3.0, SCs are indispensable because most of the core applications and services in Web 3.0 are built on SCs. These include the creation and trading of NFTs, the development of dApps in the DeFi field, and the formulation of DAO rules, etc.
Generative AI can revolutionize the way SCs are created as they are essentially self-executing contracts with terms and conditions programmed directly into the codes.
By understanding context and expected outcomes, generative AI can advance the Web 3.0 ecosystem by efficiently developing SC code that ensures compliance with predefined rules.
For example, 
Web3-GPT is a chat assistant based on GPT4 that combines LLMs and AI agents, aiming to revolutionize the development and deployment process of SCs \cite{Sokoli:2023}. 
ETHGPT is a development toolkit that supports semantic search to provide professional tools and assistance to support the rapid development of the Web 3.0 ecosystem \cite{Savage:2023}. 
FlashGPT can efficiently generate and deploy secure and reliable Solidity SCs to a variety of Layer 1 and Layer 2 solutions through simple interactions \cite{Yang:2023}.

\subsubsection{Data analysis}
In the user-centric Web 3.0 ecosystem, generative AI will play an important role in simplifying data analysis.
Through natural language interfaces, generative AI can quickly and accurately analyze complex datasets based on simple prompts from users, providing users with valuable insights.
In this way, generative AI eliminates the need for Web 3.0 users to master relevant programming languages or advanced data analysis knowledge by automating data processing behind the scenes, lowering the barrier to participating in the Web 3.0 economic ecosystem.
For example, 
TokenGPT aims to simplify complex Web 3.0 investing by using generative AI to review SCs to conduct a comprehensive analysis of the market \cite{TokenGPT:2023}. 
CoinGPT is a generative AI-driven data analysis tool that allows users to connect their crypto wallets to analyze the transaction history of NFTs and other cryptocurrencies to improve transaction performance \cite{Zhang:2023}. 
Defi-Companion is a ChatGPT-based bot that assists Web 3.0 users in querying data from endpoints, performing data analysis, and identifying DeFi opportunities \cite{MG:2023}.

\subsubsection{Security assistance}
Security is critical to the Web 3.0 ecosystem. The P2P architecture, consensus mechanism, and cryptographic protocol of blockchain technology provide the first level of security, preventing 51\% attacks, double spends, Sybil attacks, etc.
However, these guarantees are not sufficient for complex Web 3.0 systems. Various security and reliability issues may still arise. 
Generative AI is a promising solution that provides additional protection for Web 3.0 users and their data through continuous monitoring of SCs. 
For example, Quantstamp can secure transactions and build user trust by automatically auditing SCs through generative AI to uncover vulnerabilities that may be missed by traditional methods \cite{Quantstamp:2017}.
ChainSecurity uses a combination of generative AI and formal verification methods to detect vulnerabilities in SCs \cite{ChainSecurity:2017}. 
Secure Semantic Snap, a ChatGPT-based MetaMask Snap, can semantically understand the target contract to protect users from malicious SCs \cite{Gupta:2022}.

\subsection{Web 3.0 for generative AI}
\subsubsection{Current solution of generative AI}

Data and computing power are considered to be the two major elements that promote the development of AI. The development of these two elements has also become a booster for the explosion of AI technology.
Based on the generative models, generative AI typically requires a significant amount of computing power and data. This is due to the fact that generative AI models are built on intricate mathematical algorithms, which need to analyze enormous datasets to recognize patterns and generate new content. In addition, training these models is computationally intensive, particularly for models that deal with high-resolution images or real-time processing. Today, LLMs are booming. However,   
the scale of LLMs has dramatically increased which could be a major obstacle for the majority of organizations. 
The high computational demands also restrict its use in computing environments with constrained resources, such as mobile devices or edge computing systems.

The current solution is cloud computing, as it provides the necessary computing resources and infrastructure required to develop and deploy chatbots at scale. Different chatbots can be deployed on different cloud computing platforms, depending on their specific needs and requirements. For example, ChatGPT is running on Microsoft Azure while Bard is executing on the Google Cloud platform. The cost of training such models is growing exponentially,  
which is unacceptable for many organizations. Ultimately, the tech giants will continue to dominate the market for generative AI, meaning that the value generated by this phenomenal field will be drawn by these large companies. This is still the typical way of operating in the Web 2.0 era, which is large-scale and highly centralized.

\subsubsection{Web 3.0 as a solution} 
\begin{figure*}[]
    \centering
    \includegraphics[scale=0.052]{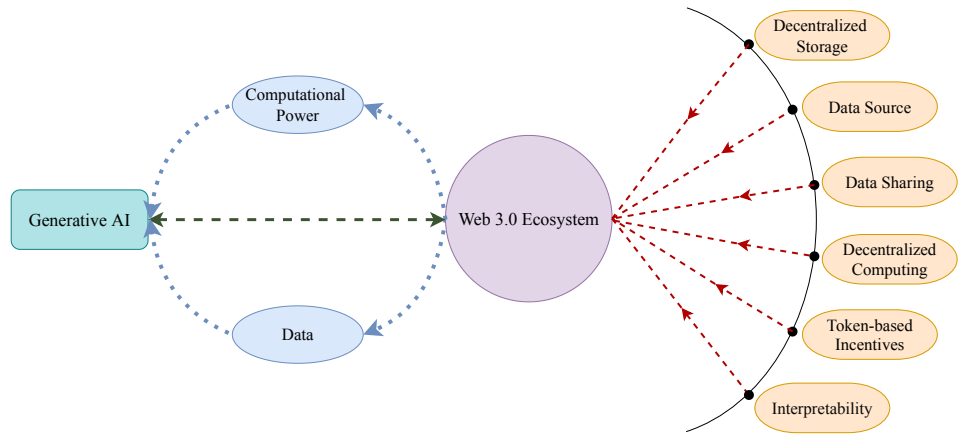}   
    \caption{Web 3.0 reduces the requirements for generative AI development.}
    \label{fig: Web3GeneraiveAI}
\end{figure*}

In the trend of decentralization, cloud computing solutions represent a compromise for the emerging industry. Web 3.0 provides a decentralized coordination platform that will facilitate unprecedented innovation and the adoption of generative AI.  
The enormous amount of data available for research, development, and industrial use is one of the key factors enabling the rapid development of generative AI. 
Accordingly, relevant solutions in this area have also received wide attention.
For example, MedDAO is an innovative decentralized autonomous organization (DAO) dedicated to addressing the critical issue of the shortage of medical images in training AI models within the global healthcare field.
Specifically, MedDAO creates aggregated and decentralized datasets by providing an anonymous, encrypted, and secure healthcare platform that incentivizes patients to contribute personal data \cite{Alam:2023}. 
A data-driven economy makes data the new gold. Correspondingly, computing power will be an important tool for contemporary ``gold diggers". Under the status quo of the technological monopoly of tech giants, the Web 3.0 ecosystem will effectively promote the broader development of generative AI. Data-driven industries are no longer just limited to large technology companies. Small-scale institutions and individuals will also benefit greatly, which is consistent with the vision of Web 3.0. 
Generally, how Web 3.0 can solve the dual challenges of data and computing power faced by generative AI can be elaborated from several main aspects as shown in Fig.~\ref{fig: Web3GeneraiveAI}.

\begin{itemize}
\item \emph{Decentralized storage}: Web 3.0 infrastructure can provide a more effective data storage solution. Instead of centralized servers, data can be stored on computer networks. In addition, there are technologies that allow for decentralized data storage such as the IPFS. In this way, generative AI models can access data from various sources, reducing the burden on individual computers and thereby assisting in solving the issues of data accessibility and availability for generative AI that the existing system confronts.
\item \emph{Data provenance}: Data provenance is a fundamental consideration in generative AI as it creates new data from existing data. Therefore, understanding the provenance of the training data is critical to assessing the quality and trustworthiness of the generated data. Web 3.0 enables the creation of a tamper-proof record of data provenance and data integrity. By ensuring that the data used to train AI models is trustworthy and comes from a reliable source, the accuracy and quality of generative language models can be improved.
\item \emph{Data sharing}:  
Web 3.0 can effectively facilitate data sharing as it provides accountability and transparency regarding data access. Users will have absolute control over their data, as they possess ownership of their personal data and digital identity through private keys that are exclusively under their control. 
Through protocols like IPFS, users can choose which data to share and with whom. Additionally, the network allows for secure data sharing through technologies such as threshold secret sharing \cite{Shamir:79, Das:2023, DasSS:2023} and revocable data sharing \cite{Ge:2019, Zhang:2021, ZhangIOT:2021, YLF:23}.  
This means individuals and organizations can share data to jointly train generative AI models through collaborative learning and secure multi-party computing schemes without compromising their privacy.  
\item \emph{Decentralized computing}: Web 3.0 enables decentralized computing through the use of blockchain networks. By leveraging the computing power of a network of decentralized computers, Web 3.0 can provide a more efficient and scalable computing environment for generative AI. In this case, instead of relying on a single, centralized server or data center to perform computational tasks, blockchain-based networks can distribute computational tasks across decentralized nodes coordinated by the Web 3.0 platform.
\item \emph{Token-based incentives}: Web 3.0 allows users to monetize their data using SCs. This is a key feature of Web 3.0, where users can sell their data directly for profit without the involvement of third parties. Small companies and individuals will benefit from such a marketplace platform as it removes barriers, levels the playing field, and promotes innovation.
Additionally, idle computing power could also be sold. GPUs used for gaming are typically utilized only a fraction of the time. Gamers can bid and receive payment for their idle computing power using SCs. In this case, AI developers can utilize this computing power to train and deploy their models.
\item \emph{Interpretability}: The interpretability of deep learning has long been a bottleneck. Deep learning-based generative AI inherits this. Due to the dramatic increase in the size of language models, interpretability has become more important than ever.
Web 3.0 allows all data processing and decisions to be tracked via blockchain. In turn, the generative paradigm of the data is analyzed in depth to achieve a constant understanding of generative AI and achieve effective control over it.
\end{itemize}

\subsection{Practical applications of AI in Web 3.0} \label{sec:AIApplication}
\begin{figure}[]
    \centering
    \includegraphics[scale=0.063]{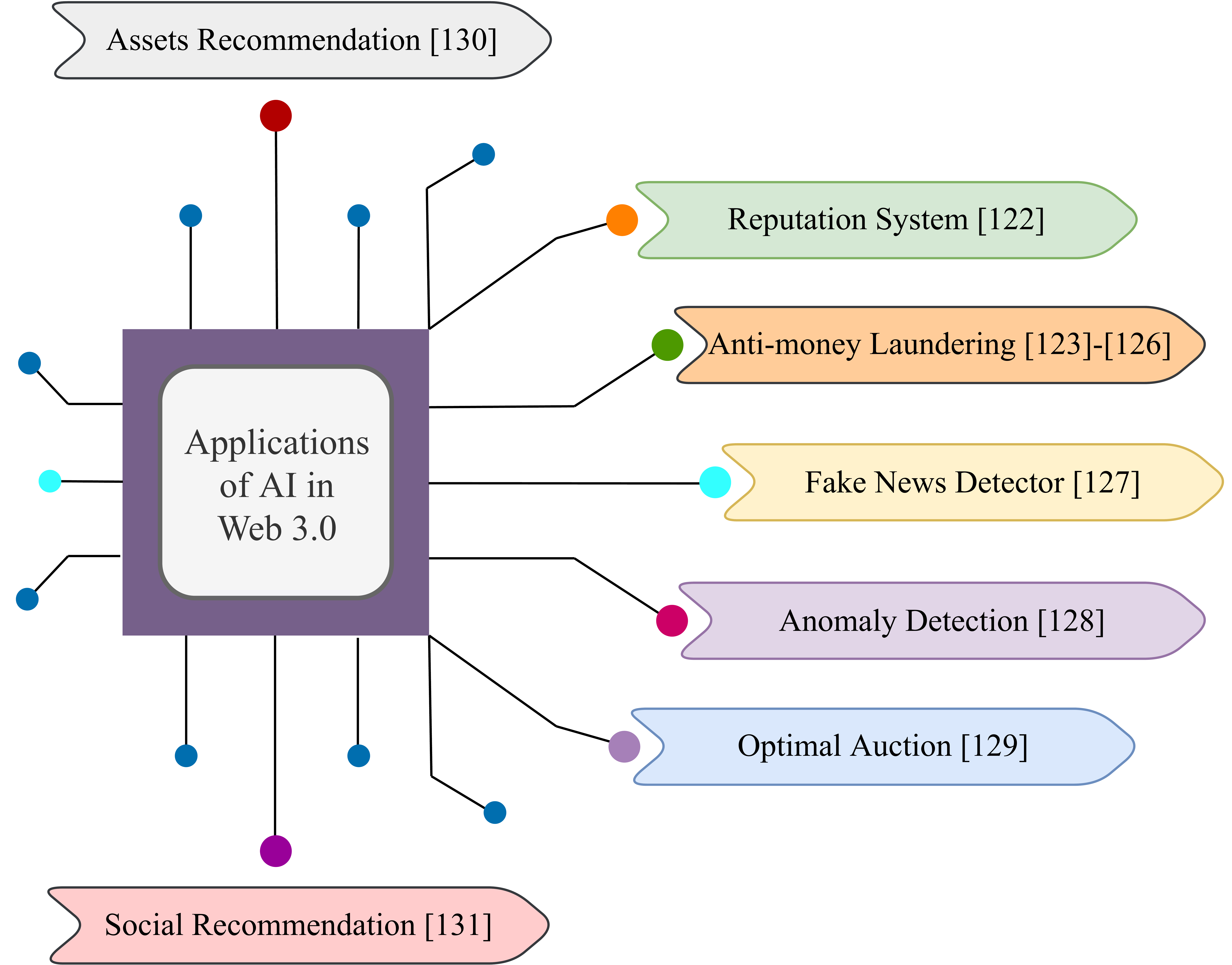}   
    \caption{Applications of AI in Web 3.0.}
    \label{fig: ApplicationsAI}
\end{figure}

In this section, we further explore the applications of AI in Web 3.0 as shown in Fig.~\ref{fig: ApplicationsAI}, to provide intuitions on how to leverage AI in the Web 3.0 ecosystem.
Keizer \textit{et al.} in \cite{Keizer:21} introduced the need for a decentralized trust and reputation system on the Web 3.0 platform by discussing the trust issues caused by Web 3.0's distributed shared services. Specifically, the paper proposed a framework based on deep reinforcement learning that allows reputation scores to be calculated in a decentralized manner while still being personalized for each user.
Lorenz \textit{et al.} in \cite{Lorenz:2021} proposed an active learning solution to address the challenging problem of detecting money laundering activities in cryptocurrency transactions when there is minimal labeled data available. Specifically, active learning was applied to develop an efficient classifier to reduce the number of labels required. This is achieved through an iterative sampling strategy of the most informative, yet unlabeled examples from the pool of data points.
Weber \textit{et al.} in \cite{Weber:19} explored the potential of using ML for the anti-money laundering of cryptocurrencies. The goal was to enable financial forensics by analyzing open data on blockchains despite the challenges posed by the anonymity of cryptocurrencies. The main contribution of this article was to open source the Elliptic dataset and provide benchmark methods to predict the binary classification of illegal transactions.
Alarab and Prakoonwit in \cite{AP:2023} developed a classification model that utilizes a combination of long-short-term memory (LSTM) and graph convolutional network (GNN) to classify illicit transactions in the Elliptic Bitcoin dataset based only on transaction features.
By studying different acquisition functions under the same experimental settings, the proposed model could achieve an accuracy of 97.77\%.
Lo \textit{et al.} in \cite{Lo:2023} proposed Inspection-L, a novel GNN framework for detecting money laundering activities in the Bitcoin network. At its core, Inspection-L combined a self-supervised Deep Graph Infomax and a supervised Random Forest to learn topological information and node characteristics within the transaction graph for the purpose of identifying illegal transactions.
The framework showed the potential for identifying suspicious financial activities.
Unzeelah \textit{et al.} in \cite{Unzeelah:22} proposed an AI-powered method to build a secure, reliable, and efficient platform in Web 3.0 to address the issues of misleading content and fake news spreading on current platforms. To this end, NLP technologies and ML models were implemented. For example, LSTM with Word2Vec and GloVe were used for word embeddings. Additionally, the combination of the Ethereum blockchain with the IPFS technique was utilized to decentralize the system and enable off-chain storage.
Kim \textit{et al.} in \cite{Kim:2022} presented a novel security mechanism using blockchain network traffic statistics as a metric for identifying malicious events.   
Specifically, a data collection engine periodically generated multi-dimensional, real-time data streams by monitoring underlying blockchain activities. Afterward, an anomaly detection engine was used to detect anomalies from the generated data instances based on One-class Support Vector Machine or AutoEncoder.
Xu \textit{et al.} in \cite{Minrui:23} introduced a quantum blockchain-powered Web 3.0 framework to provide information-theoretic security for decentralized data transmission and payment to cope with the situation that quantum computing subverts the conventional cryptosystems. 
In particular, an optimal auction for NFT transactions based on quantum deep learning is proposed to maximize revenue with sufficient liquidity in Web 3.0.
Yu \textit{et al.} in \cite{Yu:2023} proposed a framework for classifying referable non-fungible tokens (rNFTs) using GNN. The goal was to provide an effective recommendation system for Web 3.0 assets. First, the authors converted rNFT reference relationships into direct acyclic graphs (DAGs). Then,  the node and edge characteristics were modeled based on rNFT metadata and token transactions. Afterward, GraphSage was  modeled to contain the characteristics collected during the learning process. In this way, the model combined considerations of graph topology and attribute characteristics to enable supervised classification of existing and incoming NFT nodes.
Madhwal and Pouwelse in \cite{Madhwal:23} implemented a decentralized social recommendation system, Web3Recommend, that aimed to generate balanced recommendations for trust and relevance on Web 3.0 platforms. It addressed the challenges of generating recommendations in decentralized networks that lacked a central authority and were vulnerable to Sybil attacks. Web3Recommend combined MeritRank (a decentralized reputation scheme that provides Sybil resistance) and SALSA (a personalized graph algorithm). Specifically, MeritRank added decay parameters to SALSA to theoretically guarantee protection against Sybil attacks. By integrating with Music-DAO, an open-source Web 3.0 music-sharing platform, the proposed system was shown to generate personalized real-time recommendations.

\subsection{Summary and insights} 
Web 3.0 is a user-centric web where users can create and trade their digital assets. Generative AI and Web 3.0 have great potential to reinforce each other. On one hand, Generative AI lowers the barriers for ordinary users to enter the Web 3.0 world through its powerful API. Its semantic understanding capabilities allow users to easily create exclusive NFTs, providing a solid foundation for participating in the digital economy.
On the other hand, Web 3.0 can alleviate Generative AI’s huge demand for data and computing power. In this way, lightweight, personalized, and even decentralized generative AI will become possible.
Despite these promising aspects,
for generative AI to be effectively and securely integrated into Web 3.0 environments, challenges around bias and lack of explainability need to be addressed.
Moreover, there is an urgent need for effective storage and computing solutions to support the sustainable development of Web 3.0.

\section{Edge computing for Web 3.0: an interconnected and ubiquitous web} \label{sec:edgecomputing}
Web 3.0 is expected to be an interconnected and ubiquitous web that is accessible to everyone and anywhere at any time. With the rise of IoT devices and the increasing need for low-latency and real-time data processing, edge computing has become an important part of the Web 3.0 ecosystem. Edge computing is essentially a distributed computing paradigm in which computing services occur at the edge of the network as opposed to being performed in centralized data servers. Users can benefit from faster service via edge computing by bringing computing resources and data storage closer to where users actually consume the data.

In this section, we will first illustrate the close correlation between edge computing and Web 3.0.
Then we will explore the integration of edge computing and other cutting-edge technologies to promote the development of Web 3.0. Specifically, we proposed two solutions for decentralized storage and computing.
Afterward, we will introduce the practical applications of edge computing in Web 3.0.
Finally, the summary and insights are provided.

\subsection{Relevance to Web 3.0} 
\begin{figure}[]
    \centering
    \includegraphics[scale=0.031]{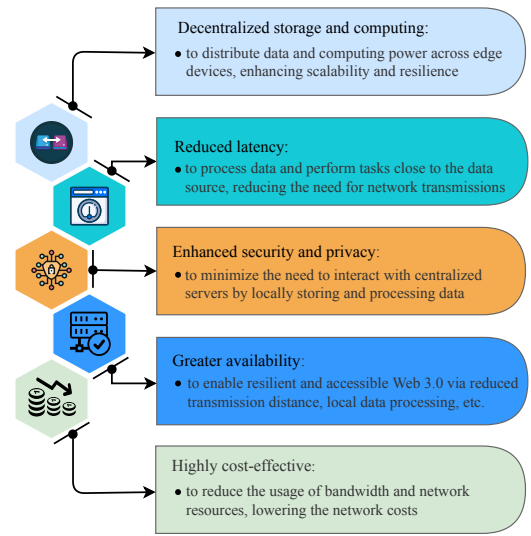} 
    \caption{Relevance of edge computing to Web 3.0.}
    \label{fig: edgeRelevance}
\end{figure}

Web 3.0 is a network of ubiquitous connectivity that aims to create a collaborative platform that is accessible to anyone, anywhere, and anytime.
The decentralized infrastructure of edge computing is highly compatible with this vision of Web 3.0.
Specifically, edge computing can support Web 3.0 in the following aspects as shown in Fig.~\ref{fig: edgeRelevance}.

\subsubsection{Decentralized storage and computing}
Edge computing plays a crucial role in facilitating decentralized storage and computing in the Web 3.0 ecosystem. Users are able to store and process data in edge devices without relying on centralized servers. This increases users' control over their data. In addition, edge computing can distribute storage and computing tasks across edge devices, making Web 3.0 more scalable and more resilient \cite{Xiong:23}. In this way, users can also share or monetize their idle storage space and computing power on the Web 3.0 platform, which contributes to the realization of data ownership and fair incentives in Web 3.0.

\subsubsection{Reduced latency}
Edge computing can significantly reduce the latency of Web 3.0 applications, improving the overall performance of the Web 3.0 ecosystem. Edge devices can process data and perform tasks close to the data source, reducing the need for network transmissions \cite{Singh:2021}. This is important for Web 3.0 applications as they typically require real-time interactions. With the proliferation of 5G, fast communication with edge devices has been greatly improved. Users can get fast responses from Web 3.0 applications running on edge devices. Therefore, with the help of edge computing, the user experience can be greatly enhanced, driving the widespread adoption of Web 3.0.
\subsubsection{Enhanced security and privacy}
In the architecture of edge computing, the natural isolation of edge devices improves the overall security and privacy of the Web 3.0 ecosystem. In edge computing, data is stored and processed locally on edge devices, minimizing the need for data to travel to and from centralized data servers. This reduces the risk of data being compromised during transmission. In addition, distributed data storage and processing across a wide range of edge devices protects data from a single point of failure. As data is fragmented and stored on multiple edge devices, an attacker would need to simultaneously compromise a significant number of devices to access or tamper with the information, which is challenging. 
For example, in the well-designed Web 3.0 ecosystem with Byzantine fault tolerance, an adversary might need to corrupt more than $33\%$ of the devices to attack the system. Otherwise the Web 3.0 ecosystem can continuously provide services.  Moreover,  since data is stored and processed on edge devices, without being collected by the centralized data servers, data privacy can be enhanced.

\subsubsection{Greater availability} 
The Web 3.0 ecosystem needs to function regardless of connectivity. The architecture of edge computing can provide the resilience to prevent a single point of system failure. In addition, edge computing allows data to be stored and processed locally on edge devices close to where the data is generated. This can significantly shorten the distance of data transfer. Moreover, edge computing allows local edge devices to perform data preprocessing, content caching, and load balancing. This reduces the amount of data that needs to be transferred over a constrained network. These three aspects improve the availability of the Web 3.0 ecosystem as well as the accessibility of its applications.

\subsubsection{Highly cost-effective}
Mass adoption of Web 3.0 will benefit from a variety of applications. Edge computing provides a more cost-effective solution for application development and deployment by allowing businesses or individuals to utilize edge devices. 
Specifically, edge computing enables the utilization of existing client devices, thereby eliminating the requirement for costly infrastructure. As such, the development of edge applications requires less upfront equipment investments and deployment time. 
Moreover, with edge computing, businesses can optimize their IT costs by processing data locally rather than in the cloud or large data centers \cite{Qiu:2020, Lin:2019}.  
In the Web 3.0 ecosystem, enterprises will benefit from utilizing edge computing infrastructure to reduce dependence on cloud providers, resulting in lower workloads and faster content delivery.
For example, MadeiraMadeira, a leading Brazilian retailtech, has significantly reduced its operating costs by offloading up to 90\% of transmitted data to edge computing resources \cite{Silva:2022}.

\subsection{Integration of edge computing and blockchain: a storage solution for Web 3.0}

Edge computing and blockchain are separate but interdependent technologies in Web 3.0. 
Edge computing can provide the infrastructure for blockchain nodes to store and verify transactions. Blockchain, on the other hand, can be truly decentralized by creating an open and secure computing environment.
Since both edge computing and blockchain are developed based on the concept of decentralized and distributed networks, they can become a powerful combination by complementing each other \cite{Liu:2018, Zhu:2021}.
The main benefit of this combination is that it enables secure communication and data processing, including data storage and computation, without the need for centralized servers. In this section, we will discuss in-depth how the integration of edge computing and blockchain can provide a decentralized storage solution for Web 3.0 applications that require high performance, low latency, secure, and decentralized storage.
Specifically, we will first introduce the main building blocks. Then, we will parse their functionalities to propose an edge storage solution.

\subsubsection{Building blocks}
The proposed decentralized storage solution consists of four main functional modules: network architecture, incentive mechanism, data integrity, and access control. These modules play a vital role in ensuring the effectiveness and security of the storage system.

\begin{itemize}
\item Network architecture:
Edge nodes act as storage nodes while edge devices act as data centers. These edge nodes are registered on the blockchain network in a specific way to provide excess storage space that can be used by the storage network. For example, a deposit is required to complete registration on the Ethereum network. When data needs to be stored, the data is distributed to available edge nodes for redundant storage. The reliability of storage is guaranteed by the blockchain-based token system.
Blockchain networks run in parallel to act as administrators. Data is hash-mapped to corresponding edge node locations for tracking and maintenance. In addition, the blockchain strictly enforces the access control mechanism and verifies the integrity of the data. Incentives are made by tracking and monitoring the amount of storage provided by each edge node and the integrity of the data.
Users and dApps interact with the storage network through APIs that communicate with the blockchain network.  
Specifically, edge nodes provide a decentralized storage layer. Blockchain manages the storage network and enforces policies. The API acts as the interface between users and dApps. The key is that this architecture combines the advantages of blockchain with the distributed resources of edge computing.

\item Incentive mechanism: 
An important component of this storage solution is the incentive mechanism that rewards edge nodes for providing storage. 
An effective incentive mechanism can solve the information asymmetry problem between users and the network \cite{Doe:2023}.
When an edge node initially registers to join the network, it needs to deposit a certain amount of tokens. The deposited amount depends on the amount of storage that the edge node wishes to provide. This ensures that edge nodes are committed to providing reliable and accessible storage. If an edge node does not do as promised, it will lose the deposited tokens. On the other hand, edge nodes can obtain corresponding rewards by meeting certain conditions. For example, they provide the amount of storage promised at registration and keep that storage accessible. They also need to regularly verify data integrity.

\item Data integrity: 
Blockchain verifies the integrity of data by maintaining hash values. When data is uploaded to the storage network, SC is responsible for distributing it to available edge nodes. Specifically, edge nodes first hash the data to generate a unique hash value. Then, the hash value and corresponding metadata will be recorded on the blockchain. Notably, edge nodes must periodically provide hash values of the data to prove that they have the correct version of the data. In this way, it can be ensured that the data stored on the edge nodes matches the original data uploaded to the storage network. If an edge node fails to provide the correct hash value, it will be tagged as an invalid storage node. Users can choose whether to download the required data according to the latest tag. The downloaded data is hashed locally using the same algorithm (i.e., SHA256). This local hash value is then compared to the hash value stored on the blockchain. If the hashes do not match, the data has been corrupted or modified. Users will need to re-download the required data from other edge nodes. In this way, even if some edge nodes are offline, users can obtain the correct data with the help of the original hash value stored on the blockchain.

\item Access control: 
As a Web 3.0 storage solution, the blockchain does not exist as a database but is used as a decentralized and immutable ledger. Data ownership, access rights, and decryption keys are all recorded on the blockchain. A large number of edge devices serve as actual storage units. At this point, the encrypted data is stored on the edge device, while the decryption key is managed by the blockchain.
When a user requests access to the data, the blockchain is first responsible for checking whether the user has the appropriate permissions. Access rights can also be further divided and enforced by the blockchain. For example, a small number of designated users can read and write data, while other users can only read data. In addition, access rights can be restricted to access only specific parts of the data, rather than the entire dataset. Authenticated users will be given the appropriate decryption key to access the data. The blockchain distributes the decryption keys only to users with appropriate access rights.
In this case, only authorized users with the correct decryption key can access the data. This ensures that even if an edge node is compromised, data remains secure without proper keys.
Moreover, activity logs of access attempts, key distribution, and successful access are recorded in the immutable ledger so that any unauthorized access attempts can be detected.

\end{itemize}

\subsubsection{A decentralized storage solution}

\begin{figure} []
	\centering
	\includegraphics[scale=0.055]{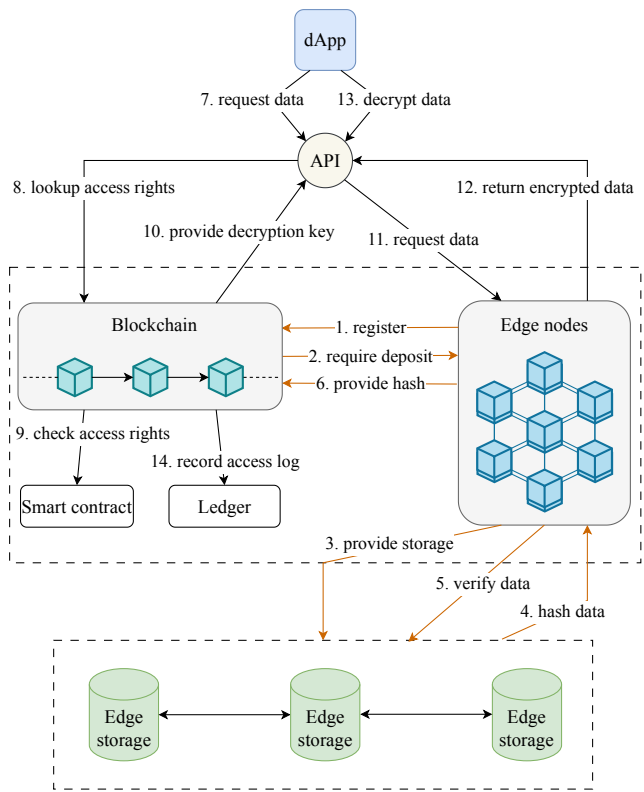}
	\caption{A decentralized storage solution in Web 3.0.}
        \label{fig:decentralizedStorage}
\end{figure}

The proposed framework (see Fig.~\ref{fig:decentralizedStorage}) aims to provide decentralized, secure, reliable, and low-latency storage by utilizing the integration of distributed edge resources with blockchain technology.
Edge nodes first register on the blockchain network. As part of the registration process, edge nodes must post a deposit of a certain value to signify their commitment to providing storage resource services.
A blockchain-based token incentive system is used to reward edge nodes for reliably contributing their storage. 
When storing data, edge nodes initially hash the data through a cryptographic hash function, such as SHA256, to generate a unique identifier. Edge nodes are then responsible for periodically verifying data integrity by comparing hash values. The latest hash is submitted to the blockchain network for review and modification, in order to ensure only valid data hashes are maintained. This allows data reliability and integrity to be guaranteed even if some edge nodes become unavailable.
Additionally, the blockchain network is used to manage access control, decryption keys, and access logging in a decentralized manner. Specifically, when a dApp requests data through the API, the blockchain uses predefined SCs to check whether it has the appropriate access right. If authorized, the blockchain will provide the decryption key to the API. The API then requests data from the specified edge node which will return encrypted data. At this point, the API can use the key to decrypt the encrypted data and provide the plaintext to the user. Afterward, the blockchain records an immutable log of the access.

\subsection{Integration of edge computing and AI: a computing solution for Web 3.0}
The convergence of edge computing and AI is known as edge AI, where AI models are deployed and executed at the edge of the network, close to where the data is generated \cite{Wang:2020, Deng:20, Cao:22}.
Edge AI is a promising combination for enhancing the functionalities of Web 3.0.  
With edge AI, edge devices can perform data analysis locally, without relying on traditional centralized servers. In this way, Web 3.0 applications that require high-speed data processing can give fast responses to complex environments. For example, edge AI can power edge devices such as robots, drones, and self-driving cars that require real-time analysis of sensor data. Edge AI can also improve the privacy and security of Web 3.0 applications by processing data at the edge device. In addition, the architecture of edge computing is distributed and redundant, the Web 3.0 system can continue to operate even if some edge nodes are damaged or offline. 
All of these characteristics are highly compatible with the Web 3.0 vision.
In this section, we will proceed to explain how edge AI can provide a decentralized computing solution for the Web 3.0 ecosystem.
{Specifically, we will first introduce the main building blocks. Then, we will use a sequence diagram to demonstrate the proposed edge AI-based decentralized computing solution.

\begin{figure*}[]
    \centering
    \includegraphics[scale=0.066]{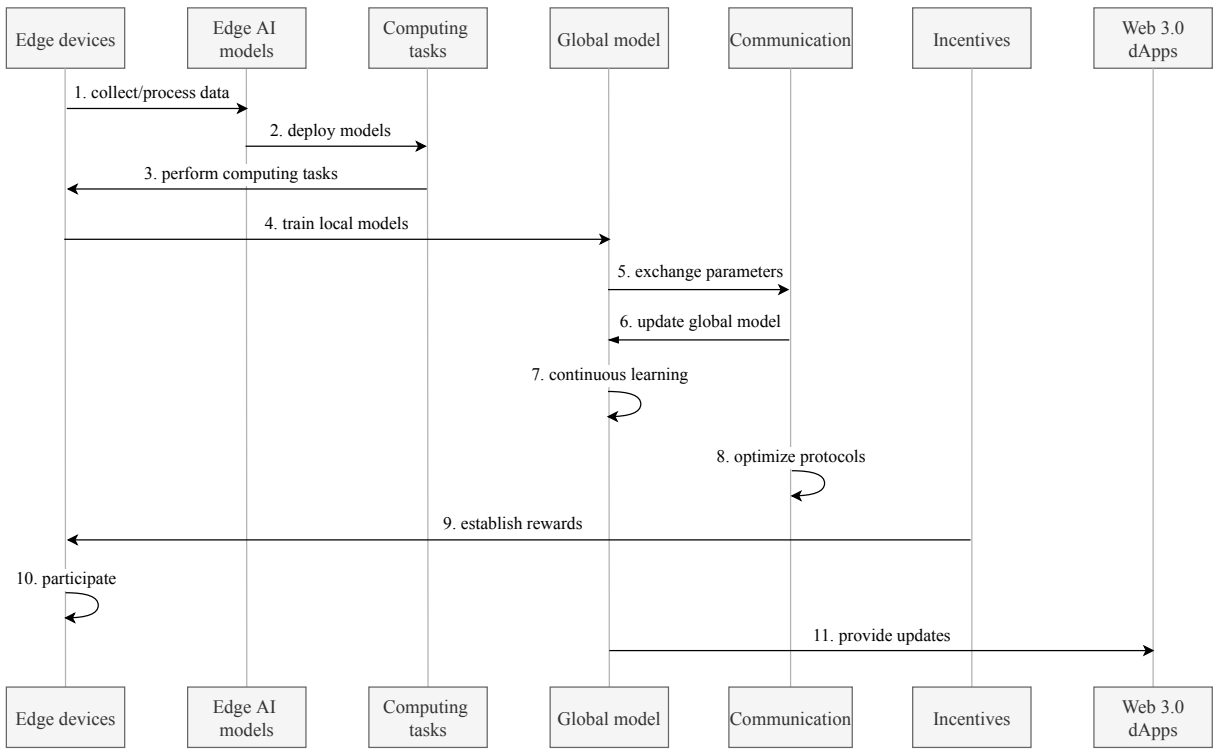}   
    \caption{Workflow of a decentralized computing solution in Web 3.0.}
    \label{fig: decentralizedComputing}
\end{figure*}

\subsubsection{Building blocks}
The proposed decentralized computing solution consists of five core components: network deployment, model deployment, incentive mechanism, collaborative learning, and communication-efficient protocol. 
They are designed to function in an integrated manner to facilitate efficient and effective edge operations. 

\begin{itemize}
\item Network deployment:
The first step is to deploy an edge network consisting of a large number of edge devices. Edge devices refer to any machines at the edge of the network that can perform specific tasks. For example, various types of IoT sensors, edge servers, smart devices, and so on. Specifically, IoT sensors are responsible for collecting data from the surrounding environment. Edge servers are designed to aggregate data from multiple IoT sensors and perform computations accordingly. Smart devices can both collect and process data.

\item Model deployment:
Through the edge network, AI models can be deployed to edge devices as needed by Web 3.0 applications. These AI models can perform various tasks such as CV, NLP, and predictive analytics. Specifically, CV models can be deployed on edge devices such as cameras and IoT sensors for object detection and recognition. NLP models can be used to perform language-related tasks on smart devices. Predictive models can analyze real-time data streams from IoT sensors for anomaly detection and predictive maintenance.

\item Incentive mechanism:
A proper incentive mechanism is significant for developing decentralized computing solutions in the Web 3.0 ecosystem. Owners of edge devices will have an incentive to participate and share their data and computing resources, driving more effective edge AI-based computing solutions. Rewards can be offered in a variety of ways. For example, owners of edge devices and data contributors can be rewarded with tokens for sharing data and resources. In the decentralized community, they can also gain a higher reputation, such as badges and rankings, for the quantity and quality of data and resources they provide, thereby unlocking more community services.

\item Collaborative learning:
Collaborative learning is fundamentally a decentralized form of learning, which matches the decentralized nature of Web 3.0.
In the context of edge AI, each edge device can train a local model based on the local data. Then honest edge devices within the edge network collaborate and share parameters to create a more general global model. Since there is no central authority, edge nodes can exchange learning parameters through peer-to-peer communication to optimize the global model. Furthermore, collaborative learning can help improve the global model by combining data from multiple data resources without sharing the original data.

\item Communication-efficient protocol:
Communication overhead is often the bottleneck in distributed systems. The performance of a distributed system is usually measured by the communication complexity. Communication complexity refers to the amount of data that needs to be communicated in the form of bits or messages between edge nodes in the network. Reducing communication complexity is critical to the scalability, bandwidth requirements, and network latency of the Web 3.0 ecosystem. There are several ways to significantly reduce communication overhead. First, the most fundamental way is to develop communication-efficient decentralized algorithms.  
Second, priorities can be assessed based on the incentive mechanism. In this way, the model parameters provided by edge nodes or devices with high priority will be adopted. Finally, limiting the frequency of model updates is also a straightforward way to reduce communication overhead. For example, changes in model parameters of edge nodes or devices only trigger communication when a certain threshold is reached.

\end{itemize}

\subsubsection{A decentralized computing solution}

As shown in Fig.~\ref{fig: decentralizedComputing}, deploying edge AI models can process data efficiently and effectively. This eliminates the need for extensive data exchange with centralized servers and provides real-time feedback to perform tasks such as object detection using computer vision or malware analysis using LLMs.
Edge devices train local models using their own data and share parameters with the global model. This sharing enables the global model to be continuously improved through collaboration, creating a positive loop ecosystem.
Optimization of communication protocols is critical to creating decentralized ecosystems where data and resources may be limited, as they allow efficient sharing of parameters between edge devices and global models. By optimizing communication, the communication overhead of edge devices is mitigated, making it easier for them to participate and contribute to the growth and improvement of the global model.
The incentive mechanism creates a mutually beneficial relationship that can further encourage edge devices to share resources and strengthen the global model to support various applications in the Web 3.0 ecosystem.

\subsection{Practical applications of edge computing in Web 3.0} 

\begin{figure}[]
    \centering
    \includegraphics[scale=0.063]{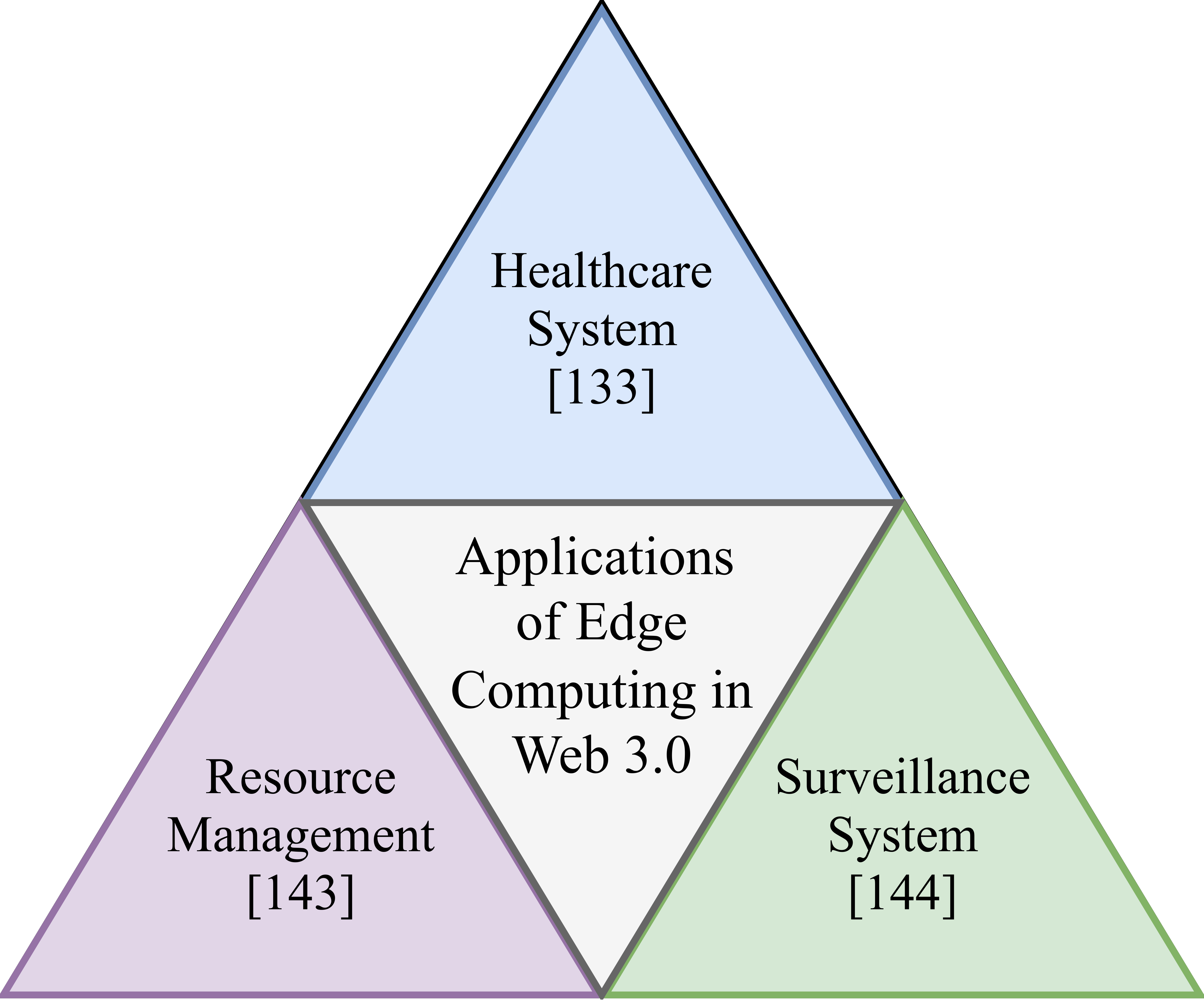}   
    \caption{Applications of edge computing in Web 3.0.}
    \label{fig: ApplicationsEdgeComputing}
\end{figure}

In this section, we further explore the applications of edge computing in Web 3.0, as shown in Fig.~\ref{fig: ApplicationsEdgeComputing}, to provide intuitions on how edge computing can be used in the Web 3.0 ecosystem.
Singh and Chatterjee in \cite{Singh:2021} proposed a smart healthcare system based on edge computing, aiming to address the challenges posed by the growing number of sensitive patient data faced by modern healthcare systems. 
The proposed system incorporated an intermediary edge computing layer tasked with preserving low latency and protecting patient privacy. This edge computing layer was used to encrypt and handle patient data privacy by applying Privacy-Preserving-Searchable-Encryption techniques. Additionally, an access control mechanism was also implemented by the layer to restrict unauthorized access to patient data stored remotely. 
Compared with traditional methods, the proposed approach shows improvements in performance, security, lower latency, transmission time, power consumption, and energy consumption. 
In the Web 3.0 ecosystem, no single entity can control large volumes of resources. Therefore, resource collaboration and management have become particularly important. As a promising solution, edge computing has the potential to maximize individual benefits through resource allocation mechanisms. 
Luong \textit{et al.} in \cite{Luong:18} presented a DL-based approach for deriving an optimal auction mechanism to coordinate edge resources in a decentralized environment.  
Specifically, miners’ valuations were used as training data to model deep neural networks for the purpose of performing monotonic transformations on miners’ bids. 
Wang \textit{et al.} in \cite{Wang:2019} proposed a video surveillance system based on the integration of edge computing, permissioned blockchain, IPFS technology, and CNNs. The goal was to address challenges that commonly arise within large-scale video surveillance, such as massive device access, high bandwidth requirements, vulnerabilities to attack, and real-time monitoring. Respectively, the system used edge computing to facilitate extensive wireless sensor information gathering and data processing in a distributed manner. Permissioned blockchain underlain the framework to ensure reliability and robustness. IPFS technology was used for massive video data storage to reduce bandwidth usage. CNN technology permitted object recognition capabilities within video streams, achieving real-time surveillance.

\subsection{Summary and insights} 
The exponential growth of the IoT has resulted in billions of devices being deployed around us. These edge devices are increasingly becoming the foundational carrier for supporting Web 3.0, generating massive amounts of heterogeneous and confidential data. In traditional centralized architectures, it is easy to unify and coordinate real-time data processing, computation, and decision-making. However, in a decentralized Web 3.0, this task becomes extremely challenging.
Especially as centralized solutions mature, the efficient allocation of resources for storage and computation in the Web 3.0 ecosystem needs to be further optimized.
In addition, the development of lightweight multimodal learning algorithms for edge devices is essential for edge resource management and optimization in Web 3.0 scenarios.

\section{Use cases for Web 3.0} \label{sec: applications}

\begin{table}[]
\centering
\small 
\setlength{\tabcolsep}{3pt} 
\renewcommand{\arraystretch}{1.8} 
\caption{Main differences between Web 3.0 and metaverse. }
\begin{tabular}{c|c|c}
\hline \hline
\multicolumn{1}{c|}{\textbf{Attributes}}             & \multicolumn{1}{c|}{\textbf{Web 3.0}} & \multicolumn{1}{c}{\textbf{Metaverse}} \\ \hline \hline
\textbf{Definition}      &     a new version of Web      &    digital reality             \\ \hline
\textbf{Focus}      & ownership          & experience                \\ \hline
\textbf{Interface}      &     front-end      &   VR/AR              \\ \hline
\textbf{Architecture}      &     fully decentralized      &   centralized/decentralized              \\ \hline
\textbf{Key technology}        & blockchain/AI       & extended reality   \\ \hline \hline
\end{tabular}
\label{tab:metaverse}
\end{table}

Although Web 3.0 is still in its emerging stage, it has a solid technological foundation. Blockchain, AI, and edge computing will enable Web 3.0 to play an important role in various fields.  
Notably, when it comes to Web 3.0 applications, Metaverse may come up in the discussion.  
However, Metaverse is considered to be a different kind of web state than Web 3.0. 
The main differences between Web 3.0 and Metaverse are provided in Table~\ref{tab:metaverse}.

Metaverse is a gigantic and shared virtual space created when the physical world converges with the virtual world.
As opposed to Web 3.0, which is primarily concerned with who will own and govern the Web, Metaverse focuses on how people will interact with it. Moreover, people may still browse the Web using the front ends of various end devices for Web 3.0. However, Metaverse users will access the Web using virtual reality headsets while navigating a digital avatar across the virtual environment.
If Metaverse manifests, it may do so in a centralized manner (e.g., Meta) or decentralized manner (e.g., Decentraland), or in any combination of the two \cite{Kuznetsov:21, Jeffries:21, Lodge:22}. 
Although there is a lot of overlap in the technology support for Web 3.0 and Metaverse, Web 3.0 relies more on blockchain, AI, and some other emerging technologies while Metaverse relies on extended reality technologies such as virtual reality (VR), augmented reality (AR), and mixed reality (MR). 
Therefore, even though the two concepts of Web 3.0 and Metaverse are highly related, the overlap does not necessarily mean that either is an application of the other. Metaverse aims to correlate the digital and physical worlds so that social life, the real economy, and physical identity can all find their counterparts in the virtual world. Web 3.0 incorporates many similar features and characteristics but is distinguished by its focus on decentralization, trustless, permissionless, and individual data ownership.

\begin{figure} []
	\centering
	\includegraphics[scale=0.061]{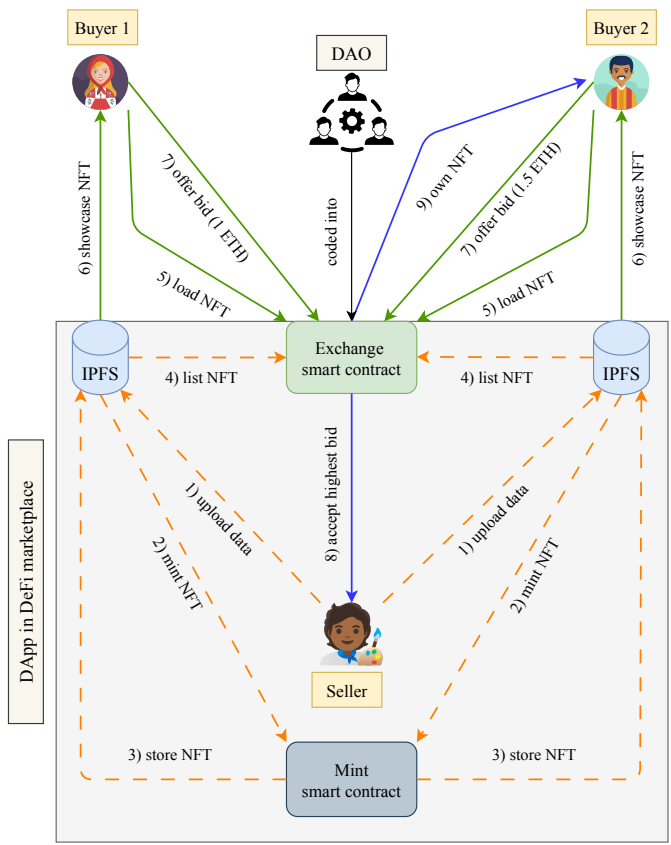}
	\caption{NFT trading process monitored by the DAO in the DeFi market.}
        \label{fig:NFTTrading}
\end{figure}

In this section, we will introduce three use cases in Web 3.0, i.e., NFTs, DeFi, and DAO.
Firstly, we will show how they can work together through a specific NFT trading process as shown in Fig.~\ref{fig:NFTTrading}.  
Next, we will discuss these three main use-cases in terms of definition, relationship to blockchain and AI, applications, and corresponding issues.

A key criterion for successful NFT trading is decentralization, which enables trustlessness and security. 
To achieve this, the entire trading process must be carried out by SCs deployed on the blockchain.
Furthermore, exchange contracts are designed to interact with all other NFT exchange contracts,
which implies a widely adopted and recognized standard interface (e.g., ERC-721) should be used.
Additionally, a DAO should be coded into the exchange SC such that users can collectively manage the trading process.
Notably, the exchange SC does not store NFT data itself, rather it only maintains information required to perform the transfer of ownership (e.g., NFT token ID, seller address, and buyer address). 
The actual NFT data is retrieved from a separate SC dedicated to minting the NFT.
Specifically, the seller first uploads data to a decentralized database (e.g., IPFS) in order to subsequently mint the corresponding NFT.
Mint SC then retrieves data and uses it to mint NFT.
Minted NFT will also be stored in the decentralized database and listed through the Exchange SC. 
On the buyer side, they send requests to Exchange SC to load NFTs and provide offers. 
Typically, the seller will accept the offer with highest bid. To accomplish this, Exchange SC will transfer the bid to the seller and NFT ownership to the buyer.

\subsection{NFTs}
\emph{1) What is an NFT?} An NFT is a token that represents a unique digital asset. It can be used to represent ownership and authenticity. 
The  assets cannot be exchanged with one another since each person has a digital signature, making them distinct and non-interchangeable \cite{Ethereum:NFT}. 

\emph{2) Blockchain for NFTs:}  NFTs are minted through SCs deployed on the blockchain, which determines the characteristics of the NFTs. 
The main features are listed below \cite{WLWC:21, RZIB:21, Chohan:21, Yang:23}.

\begin{itemize}
\item \emph{Ownership}: Ownership depends on where the private key associated with the NFT is stored. The transfer of the private key is the replacement of ownership, which realizes the trading of NFTs.
\item \emph{Scarcity}: Each NFT has a unique ID, making it a scarce and non-fungible asset. Therefore, the certificate of ownership can be used across the network, enabling the owner to be effectively verified.
\item \emph{Interoperability}: Users are allowed to seamlessly trade or share NFTs across multiple blockchain-based ecosystems, increasing the liquidity and reach of NFTs. 
\item \emph{Immutability}: The creation of an NFT means that the ownership and provenance of the NFT, as well as any related data or metadata, are permanently stored on the blockchain and cannot be changed.
\item \emph{Programmability}: The ownership and logic of NFTs can be programmed through SCs. This allows more complex functionality, such as transfer rules and scarcity, to be encoded directly into SCs.
\item \emph{Transparency}: Blockchain provides an immutable and transparent record of ownership, attributes, and SCs associated with NFTs. In this case, the integrity and provenance of unique digital assets can be ensured.
\end{itemize}

It is worth noting that an NFT can only have one owner at a time. This owner can add more attributes related to the asset in the NFT. The public ledger can be viewed by anybody, making it simple to verify and trace an NFT's ownership. In this way, creators can monetize their work, trade it globally, and have indisputable rights over their creations.

\emph{3) AI for NFTs: } 
AI and NFTs have a symbiotic relationship. AI greatly enhances NFTs through scarcity, personalization, and market optimization. In turn, NFTs provide a way to record, verify, and potentially monetize AI's contributions. Together, they are transforming collecting, creativity, identity, work, and value. Specifically, there are a variety of ways that AI can be used in the context of NFTs.

\begin{itemize}
\item \emph{Originality}: In art, generative AI models can be used to generate new ideas and designs by studying large databases of human-created artwork. This is especially useful for creators who want to issue large numbers of NFTs quickly.
\item \emph{Verification}: Blockchain is used to establish proof of ownership of NFTs. AI can play a complementary role by using techniques, such as fingerprinting, metadata analysis, and content matching, to further verify NFT ownership claims on blockchains. 
\item \emph{Enhanced interactivity}: AI-powered NFT can leverage deep learning methods to make the tokenized assets more dynamic and interactive.
The owner of NFTs can program to respond to certain commands or generate new designs, allowing the output of tokenized assets to evolve. 
\item \emph{Optimization}: AI can provide valuable insights into pricing, improving liquidity, and maximizing the value of NFT by analyzing historical sales data and attributes of successful NFT, especially for items, collections, and assets with limited supply.
\item \emph{Personalization}: AI provides personalized recommendations about purchasing or collecting new NFTs by analyzing a user's collection, interests, and preferences, leading to a more curated and valuable collection.
\item \emph{Wider ecosystem}: AI can be used to analyze the NFT market and forecast trends and demands, helping investors and collectors make more informed decisions when trading NFTs.
\end{itemize}

\emph{4) Applications for NFT:}
In this section, we will explore various notable applications of NFTs,  that showcase how blockchain and AI technologies can drive the advancement of NFTs.

\begin{itemize}
\item \emph{Securechain}:  
Securechain is a hybrid verification system designed to protect NFTs stored in hot wallets from wallet-draining attacks \cite{Lopez:2023}.
The main idea is that with a hybrid system, transfer verification can be created on-chain and accepted or rejected off-chain, unlike current solutions such as transferring NFTs to cold wallets.
It needs to be made clear that cold wallets make it inconvenient to use NFTs. In addition, since the private keys of hot wallets cannot be changed, once the key is compromised, users will be responsible for all resulting losses. In terms of specific implementation, there are various security measures such as authentication controller, contract controller, authority controller, user controller, and verification controller. Additionally, the event listening module allows the backend to monitor SC for any on-chain changes.
All these functionalities are designed to ensure that the backend only accepts verifications, but strictly prohibits any changes to the verified information, as this information is permanently recorded in the blockchain ledger.

\item \emph{NFTool}:  
NFTool is an NFT platform powered by a suite of tools including a chatbot, an NFT minting tool, SC auditor, NFT search, and built-in NFT deployment \cite{Zhu:2023}. Specifically,
a ChatGPT-style chatbot that serves as an intelligent guide capable of handling related cross-chain functions and issues.
To avoid creative concepts remaining un-minted, users can follow the detailed instructions of Mint NFT to launch their own NFTs on different chains.
However, errors in SCs can be easily exploited by malicious actors. It is crucial to utilize the SC analysis tool to ensure SCs are tested thoroughly and ready for deployment.
In addition, the NFT Search tool enables the exploration of NFT collections across the blockchain by utilizing Covalent, Zora, and Graph.
Users can convert fiat currencies into corresponding cryptocurrencies to help deploy SCs.

\item \emph{StoryChain}:   
StoryChain is a new interpretation of story editing, leveraging Artificial Intelligence Generated Content (AIGC) and blockchain technology to enable collaborative story creation \cite{Adiloglu:2023}.
With StoryChain, users can collaboratively craft stories comprising distinctive chapters and artistic works.
Notably, each page is an NFT minted for the corresponding user.
During the story creation process, the contract first verifies the appropriateness of the user prompt, then ChatGPT generates the text and illustrates it with an AI image generator. The entire story, along with the resulting images, will then be uploaded to IPFS.
The generated hash is submitted to the contract to mint the NFT for the author. This way, authors will have a permanent record of ownership and chapter details accessible on-chain via IPFS.
Alternatively, a voting mechanism allows the story community to democratically select the direction of future entries, treating each story like a DAO. The proposed chapter is subject to consensus among NFT owners to determine the direction of the story.

\end{itemize}

\begin{table*}[]
\centering
\small 
\setlength{\tabcolsep}{6pt} 
\renewcommand{\arraystretch}{1.8} 
\caption{Issues of NFTs.}
\begin{tabular}{c|l|l}
\hline \hline
& \multicolumn{1}{c|}{\textbf{Types}}    & \multicolumn{1}{c}{\textbf{Description}} \\ \hline \hline
\multirow{9}{*}{\textbf{Technical}}    
				       &  \multirow{1}{*}{\hfil Complexity}  & \multicolumn{1}{l}{The complex development of NFTs has not yet been simplified by high-quality tools.} \\ \cline{2-3} 
                                        & \multirow{1}{*}{\hfil Storage} & \multicolumn{1}{l}{The URL where the artwork is stored makes the artwork itself vulnerable to link damage.} \\ \cline{2-3} 
                                        &  Fees   & Artists pay more on average in the NFT market than they earn in sales. \\ \cline{2-3} 
                                        & \multirow{1}{*}{\hfil SC risks} & \multicolumn{1}{l}{SC vulnerability caused a massive attack leading to the NFT theft incident.} \\ \cline{2-3} 
                                        & \multirow{1}{*}{\hfil Rapid innovation}  &  \multicolumn{1}{l}{This creates a challenge of continuous change for those who adopt the technology.}  \\ \cline{2-3} 
                                        &  \multirow{1}{*}{\hfil Usability}    & {Slow confirmation and high gas prices will limit the rapid growth of NFTs.} \\ \cline{2-3} 
                                        &  \multirow{1}{*}{\hfil Extensibility} & \multicolumn{1}{l}{It is difficult to interact with other ecosystems.} \\ \cline{2-3} 
                                        & \multirow{1}{*}{\hfil Cybersecurity}  & \multicolumn{1}{l}{It is challenging to identify fake NFT stores and malicious proxies.} \\ \cline{2-3} 
                                        &             Security and privacy          &  NFT data is at risk of being lost or misused by malicious parties. \\ \cline{2-3} 
 \hline
\multicolumn{1}{l|}{\multirow{5}{*}{\textbf{Others}}} 

					       & \multirow{1}{*}{\hfil Regulatory}  & \multicolumn{1}{l}{New technologies bring unique regulatory and legal considerations.}  \\ \cline{2-3} 
\multicolumn{1}{l|}{}                  &  \multirow{1}{*}{\hfil Environmental concerns}  & \multicolumn{1}{l}{Transactions result in high energy consumption and consequent greenhouse gas emissions.}  \\ \cline{2-3}  
\multicolumn{1}{l|}{}                  & \multirow{1}{*}{\hfil Anti-money laundering}   & \multicolumn{1}{l}{Decentralized transactions can lead to money laundering.}  \\ \cline{2-3} 
\multicolumn{1}{l|}{}                  & \multirow{1}{*}{\hfil Copyright}  & \multicolumn{1}{l}{The public nature makes it easy for anyone to copy the referenced documents.} \\ \cline{2-3} 
\multicolumn{1}{l|}{}                  & \multirow{1}{*}{\hfil Ponzi scheme}  & \multicolumn{1}{l}{Critics have likened NFT to a Ponzi scheme.}  
\\ \hline \hline
\end{tabular}
\label{tab:NFT}
\end{table*}

\emph{5) Issues:}  
Items that typically end up lost on the Web can now be monetized through NFT technology. However, both the NFT technology and the NFT market are currently in their infancy. Critical infrastructure, including technology platforms and trading platforms, will continue to exist in a centralized form.   
We list the main issues from both technical and non-technical perspectives in Table~\ref{tab:NFT} 
\cite{Busch:22, Wang:2021, RZIB:21, WLWC:21, KT:22}.

\subsection{DeFi}
\emph{1) What is DeFi?} DeFi manages financial services primarily using blockchain technology and a number of cryptocurrencies. It is an alternative to the global financial system of the Web 3.0 era and aims to democratize finance, which is difficult to achieve in the traditional financial system \cite{ZAB:20, Werner:21, WLSL:21, Jiang:2023}.  
The biggest difference between DeFi and traditional financial institutions is that DeFi is decentralized and does not use a third party to carry out financial operations. DeFi is open and transparent and performs a range of functions based on the issuance code, which cannot be modified by any node and can only be updated if necessary with the consent of most nodes on the chain. A more detailed comparison of several key aspects is shown in Table~\ref{tab:diffDeFi} \cite{Kaihua:21, Ethereum:DeFi}.

\begin{table}[!b]
\centering
\small 
\setlength{\tabcolsep}{3pt} 
\renewcommand{\arraystretch}{1.8} 
\caption{Differences between DeFi and traditional finance.}
\begin{tabular}{c|c|c}
\hline \hline
\multicolumn{1}{c|}{\textbf{Attributes}}             & \multicolumn{1}{c|}{\textbf{DeFi}} & \multicolumn{1}{c}{\textbf{Traditional finance}} \\ \hline \hline
\bf{Infrastructure}      &     Decentralization      &    Centralization             \\ \hline
\bf{Custody}    & Users          & Companies                \\ \hline
\bf{Currency}     &     Digital asset      &   Fiat currency              \\ \hline
\bf{Speed}       & Minutes       & Depends (manual processing)  \\ \hline
\bf{Auditability}       &    Anyone    & Authenticated organizations  \\ \hline
\bf{Collateral}       &    Required    &  Optional \\ \hline
\bf{Anonymity}    & Yes           & No                           \\ \hline
\bf{Permission}   & No            & Yes                          \\ \hline
\bf{Availability} & Yes           & No                           \\ \hline
\bf{Transparency} & Yes           & No                           \\ \hline \hline
\end{tabular}
\label{tab:diffDeFi}
\end{table}

\emph{2) Blockchain for DeFi:}  Blockchain is the core technology that replaces traditional centralized institutions and enables decentralized financial services as shown in Fig.~\ref{fig: Defi} \cite{Schar:21, CB:20}.  
Cryptocurrencies, SCs, stablecoins, and dApps are the four components that comprise DeFi and are all based on blockchain technology \cite{HRS:21}.

\begin{itemize}
\item \emph{Cryptocurrency}: As one of the earliest and most successful deployments of blockchain, cryptocurrencies are ubiquitous in the decentralized world. It has underpinned the rise and continued success of DeFi. DeFi's successful operation requires the support of cryptocurrencies because they enable many core functions. For example, cryptocurrencies are used to represent and transfer value. Without cryptocurrencies, basic lending and borrowing functions would not be possible.
\item \emph{SCs}: As a blockchain-based program, a SC can be activated automatically when certain criteria are met. It eliminates third parties or central authorities typically required in traditional financial transactions. It also allows users to program any transaction into code, decentralizing the financial process. When a SC goes live, no one can change it. As a result, many of the business terms found in traditional financial industry agreements can be transferred to SCs and enforced through code.
\item \emph{Stablecoins}: One of the main drawbacks of many cryptocurrencies is their excessive volatility, which greatly reduces the incentive to participate for those users who do not have sufficient risk tolerance. To solve this problem, the concept of stablecoins was created. A stablecoin is a cryptocurrency with a fiat stable price. It is designed to maintain price parity with a stable asset, such as the US dollar or gold, to provide the necessary stability.
\item \emph{dApps}: A dApp is essentially an application that runs on the blockchain. Unlike traditional applications that run on large servers, dApps are created and deployed on the blockchain through SCs. The main benefits are their permissionless nature and resistance to censorship.  
\end{itemize}

\begin{figure}[]
    \centering
    \includegraphics[scale=0.58]{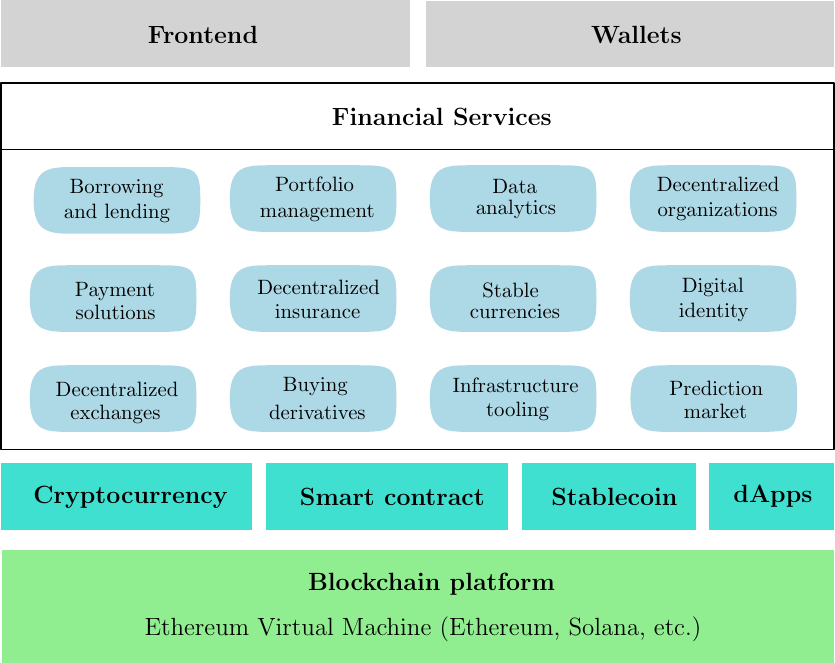}
    \caption{DeFi stack.}
    \label{fig: Defi}
\end{figure}

\emph{3) AI for DeFi:}
DeFi aims to democratize finance to bridge the limitations of traditional finance. It can provide financial services to anyone, anywhere, regardless of location, income, or background \cite{Zhao:19, Cao:20, Sadman:22}.  
With the rapid and widespread deployment of AI in multiple fields, AI is promising to provide novel solutions to enhance the DeFi ecosystem \cite{EasyFi:23, BinanceDeFi:23}. 
Specifically, it can be reflected in the following aspects.
\begin{itemize}
\item \emph{Fraud Detection}: 
AI can analyze transaction data, user behavior, and transaction patterns to detect anomalies 
that may indicate fraud, scams, or money laundering 
in DeFi's platform and protocols.
Take money laundering as an example, various techniques have been proposed to detect money laundering activities (see Section~\ref{sec:AIApplication}). 
\item \emph{Risk Analysis}: AI can assess the risks of different DeFi projects based on technology, market, operations, regulations, management team, and other factors. This can help DeFi users make informed decisions about which projects to invest in and which to avoid.
\item \emph{Automated Trading}: AI bots can analyze real-time market data, trends, and opportunities on the DeFi platform to automatically trade digital assets and provide liquidity to generate maximum profits.
\item \emph{Credit Scoring}: AI can analyze DeFi users' transaction history, loan data, and collateral information to generate credit scores for them. These scores can then be used to determine their eligibility for loans and credit limits under the DeFi loan agreement.
\item \emph{Personalization}: By understanding the user profile, investment goals, risk appetite, and portfolio details, AI can provide customized recommendations about interest rates and lending options that are best suited for each DeFi user.
\item \emph{Product Development}: Generative AI allows for rapid understanding of DeFi users' needs and issues, developing new products and expanding DeFi's ecosystem.
\end{itemize}
\begin{table*}[]
\centering
\small 
\setlength{\tabcolsep}{8pt} 
\renewcommand{\arraystretch}{1.8} 
\caption{Issues of DeFi.}
\begin{tabular}{c|l|l}
\hline \hline
& \multicolumn{1}{c|}{\textbf{Types}}    & \multicolumn{1}{c}{\textbf{Description}} \\ \hline \hline
\multirow{7}{*}{\textbf{Technical}}    
                                        & \multirow{1}{*}{\hfil Scalability} & \multicolumn{1}{l}{The slow-transaction nature of the consensus mechanism limits large-scale adoption.}  \\ \cline{2-3} 
                                        &  Oracle   & It is difficult for off-chain data to be reported securely on-chain. \\ \cline{2-3} 
                                        & \multirow{1}{*}{\hfil Operational security} & \multicolumn{1}{l}{A malicious third party could break the SC once they get the key.} \\ \cline{2-3} 
                                        & \multirow{1}{*}{\hfil Custodial risk}  &  \multicolumn{1}{l}{The theft of private keys can be catastrophic for users.} \\ \cline{2-3} 
                                        & \multirow{1}{*}{\hfil Bias}  &  \multicolumn{1}{l}{Data-driven decisions on DeFi protocols may reflect or even amplify biases.} \\ \cline{2-3} 
                                        & \multirow{1}{*}{\hfil Limited data}  &  \multicolumn{1}{l}{Limited transaction data makes it difficult to train accurate AI models.} \\ \cline{2-3} 
                                        & \multirow{1}{*}{\hfil Integration}  &  \multicolumn{1}{l}{Integrating centralized AI models and systems into a decentralized DeFi protocol is difficult.} \\ \cline{2-3} 
 \hline
\multirow{4}{*}{\textbf{Others}}                   &  \multirow{1}{*}{\hfil Regulatory}  & \multicolumn{1}{l}{Finding the right regulatory balance will be tricky.} \\ \cline{2-3} 
\multicolumn{1}{l|}{}                  & \multirow{1}{*}{\hfil Collateralization} & \multicolumn{1}{l}{Over-collateralization will limit DeFi lending business.} \\ \cline{2-3} 
\multicolumn{1}{l|}{}                  & \multirow{1}{*}{\hfil Dependencies}  & \multicolumn{1}{l}{The openness and composability of DeFi introduce dependencies to the system.} \\ \cline{2-3} 
\multicolumn{1}{l|}{}                  & \multirow{1}{*}{\hfil Responsibility}  & \multicolumn{1}{l}{DeFi shifts the responsibility from the third party to the user.}  \\ \hline \hline
\end{tabular}
\label{tab:riskDeFi}
\end{table*}

\emph{4) Applications for DeFi:}
In this section, we will show how blockchain and AI technologies can improve the DeFi ecosystem by exploring several applications of DeFi.

\begin{itemize}
\item \emph{Agrosurance}: 
Agrosurance aims to revolutionize how agricultural insurance and liquidity are managed by developing an innovative platform that delivers transparent, reliable, and decentralized solutions \cite{Goyal:2023}.
This platform consists of five SCs, each with a unique purpose.  
Specifically, AgroCoin contract allows the transfer and management of AgroCoins between users.
AgroSuranceLand contract represents and manages land assets by minting them into NFTs.
StakeManager contract enables users to earn AgroCoins through staking to incentivize liquidity and participation within agricultural insurance.
InsuranceManage uses Chainlink DONs to obtain off-chain real-time data to trigger predefined rules to calculate insurance premiums and verify claim eligibility.
FundManager contract acts as a repository for managed funds, storing tokens received from the InsuranceManager and StakeManager contracts.

\item \emph{Prompt DeFi}:   
Prompt DeFi simplifies the interaction process with DeFi into simple voice commands, thereby attracting more users without relevant backgrounds to enter the DeFi world \cite{KO:2023}.
This platform uses ChatGPT to execute transactions based on text prompts. In addition, the use of the web3auth library greatly simplifies the account creation process, making new users interacting with Web 3.0 as easy as with Web 2.0.
For example, Prompt DeFi integrates prominent DeFi platforms, including Uniswap, Lido, and Aave. This allows users to easily send and exchange tokens through Uniswap, deposit funds into Lido, and manage loans in Aave.
Furthermore, the platform utilizes Chainlink Automations to incorporate additional portfolio triggers, enhancing the overall user experience.

\item \emph{RoboFI}:   
RoboFi is a robotic DeFi ecosystem that enables robotic entities to generate and trade Energy Attribute Certificates (EACs) via a decentralized infrastructure powered by blockchain \cite{Dubyk:2023}. 
In RoboFi, there are two robotic entities that provide commercial services for humans and other robots.  
One of the robots, called EAC provider, owns a solar power plant that produces green electricity. 
When 1 MWh of electricity is generated, 1 EAC NFT is created to confirm the fact and origin of this electricity. 
The other robot, called EAC consumer, is charged from the common power grid of the RoboFi ecosystem. 
To be sustainable, the EAC consumer needs to confirm the origin and track of consumed electricity by connecting to the RoboFi NFT EACs trading platform.
The EAC consumer can showcase that it has the certificate to use the corresponding amount of green energy, confirming the creation of sustainable Value.

\end{itemize}

\emph{5) Issues:}  While DeFi can minimize transaction risks by eliminating third parties and enabling the exchange of financial assets on a trustless basis, innovations always come with a new set of issues. In particular, DeFi has not been stress tested for long-term or widespread use. As shown in Table~\ref{tab:riskDeFi}, a number of key issues must be addressed in order to offer customers and institutions a reliable, fault-tolerant ecosystem that can handle new financial applications at scales \cite{HRS:21, Iredale:21, Chen:2022, Ma:2023}.

\subsection{DAOs}

\emph{1) What is a DAO?} A DAO is an organization that operates fully autonomously on a blockchain, conforming to rules encoded through SCs and their underlying consensus mechanisms. It is designed to reduce or bypass the need for cascading human intervention or centralized coordination \cite{Qin:2022}. For this reason, a DAO is often referred to as a trustless system, which differs from the traditional model of management by a small group of people. 
All the rules are set up in advance in a SC and executed by P2P computing nodes. A more detailed comparison between DAOs and traditional organizations is shown in Table~\ref{tab:diffDAO} \cite{Wang:19, Qin:22, Ethereum:DAOs}.

\begin{table}[!b]
\centering
\small 
\setlength{\tabcolsep}{4pt} 
\renewcommand{\arraystretch}{1.9} 
\caption{Differences between DAOs and traditional organizations.}
\begin{tabular}{c|c|c}
\hline \hline
\multicolumn{1}{c|}{\textbf{Attributes}}             & \multicolumn{1}{c|}{\textbf{DAOs}} & \multicolumn{1}{c}{\textbf{Traditional organizations}} \\ \hline \hline
\bf{Trust}    &     Blockchain      &    Experience             \\ \hline
\bf{Governance}      & Community          & Main stakeholders                \\ \hline
\bf{Structure}     &     Democratic      &   Hierarchical              \\ \hline
\bf{Transparency}      & Complete       & Restricted  \\ \hline
\bf{Pattern}       &    Collaboration    & Competition  \\ \hline
\bf{Service}        &    Fully automated    &  Centralized automation \\ \hline
\bf{Access}       &    Open    &  Closed \\ \hline
\bf{Affiliation}       &    Multiple    &  One \\ \hline
\bf{Cost}   & Low           & High       \\ \hline \hline
\end{tabular}
\label{tab:diffDAO}
\end{table}

\emph{2) Blockchain for DAO:}  The underlying technology of DAO is the blockchain. It relies heavily on SCs, which are transparent, verifiable, autonomous, and publicly auditable. The workflow of DAO is shown in Fig.~\ref{fig: DAO}. 
SCs are used to establish the rules of DAO, which are set by a core team of community members. 
Due to the tamper-proof nature of the blockchain, once the contract is in effect on the blockchain, no one can change the code without a consensus reached through a vote of the members. The rules and logic in the code strictly limit its functionality \cite{FAH:20, ZhaoXi:22, SA:22}. 
Financing is usually achieved through a token offering, in which case the tokens are sold to raise funds and replenish the coffers of DAO. 
Token holders receive voting rights in exchange for their money. The voting rights are usually proportional to their holdings. Once fundraising is complete, DAO is ready for deployment.  
No particular organization has the right to change the code in the SC. It is entirely up to the token holders to decide. Based on these features, DAO can provide significant support for many applications.

\begin{figure}[]
    \centering
    \includegraphics[scale=0.52]{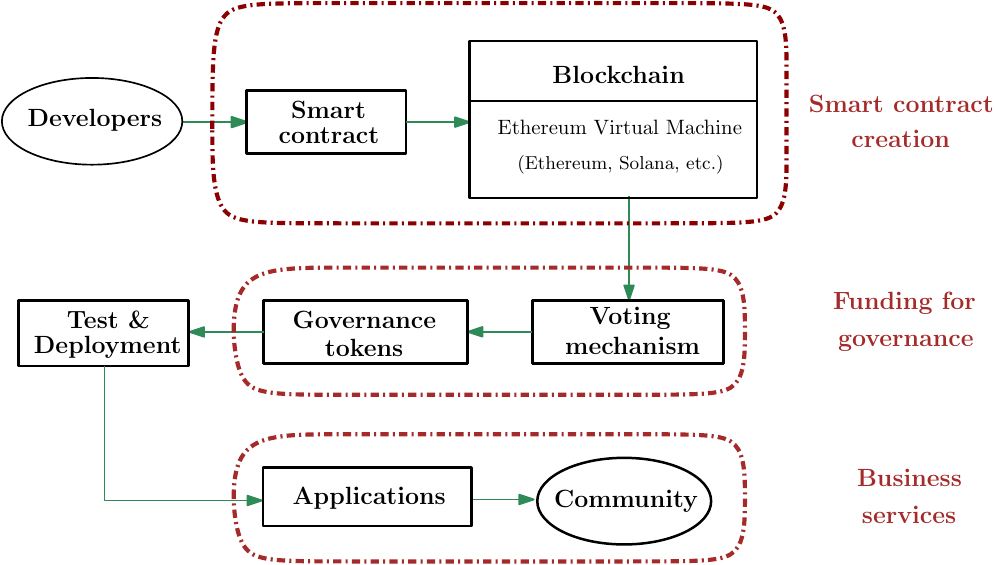}
    \caption{DAO workflow.}
    \label{fig: DAO}
\end{figure}

\emph{3) AI for DAO:}
The combination of DAO and AI can be complementary. AI plays a pivotal role in bringing more functionality and greater efficiency to DAOs. AI gains the resources it lacks; on the other hand, DAO gets critical autonomous decision-making capabilities \cite{HHH:22}. In the context of Web 3.0, AI can enhance DAOs to fit their mission in numerous ways.  
\begin{itemize}
\item \emph{Autonomous functions}: AI can autonomously perform certain automated tasks, such as computation and data analysis, based on constraints, goals, and rules defined by a DAO.
\item \emph{Enhanced control}: AI integrated into a DAO has no centralized control by the nature of DAOs. Its knowledge and capabilities will be designed to serve the overall goals and tasks of a DAO in a transparent and decentralized manner, thus enhancing decentralized control.
\item \emph{Facilitate collaboration}: AI integrated into DAOs can help facilitate collaboration among different stakeholders by providing shared information, coordination tools, and alignment metrics.
\item \emph{SCs}: Generative AI can aid in the rapid development of SCs by precisely coding the goals and rules of the DAO so that the SC enforces these terms in a fair manner.
\item \emph{Monitoring}: AI can help continuously monitor key metrics, risks, and processes within a DAO, providing comprehensive reporting for greater transparency and accountability.
\end{itemize}

\emph{4) Applications for DAO:}
In this section, we will show the promise of DAO powered by blockchain and AI technologies in decentralized governance through specific applications.

\begin{itemize}

\item \emph{OmniGovern DAO}:  
OmniGovern is a decentralized governance system deployed on a layer 2 blockchain, designed to simplify on-chain governance \cite{Nayan:2023}.
To enable seamless governance across blockchains, Layerzero acts as an interoperable middleware that connects multiple blockchains with Omnichain Fungible Token, facilitating gas-less transactions through a relay mechanism.
Worldcoin’s authentication mechanism is used to eliminate potential bot activity to ensure the integrity and authenticity of votes and proposals.
OmniGovern implements robust account abstraction on the Base Gorelli chain to simplify the complexity of user interaction with the platform by masking complex blockchain details.
Additionally, Covalent’s integration ensures transparency of voting results, proposal history, and token analysis, promoting trust and accountability.

\item \emph{Rooster DAO}:  
Rooster DAO is an investment fund managed in the form of DAO, focusing solely on investments in the Dotsama ecosystem \cite{Duportail:2022}.
To incentivize active participation, every member of Rooster Dao owns an evolving NFT that represents their level of engagement in investment proposals and voting. As members actively contribute, the rooster image gains color and levels up. 
Moreover, when a member proposes a highly profitable investment, all members collectively transfer a portion of the profits to the proposer. 
Specifically, there are two types of SCs. 
The NFT management contract is responsible for creating collections owned by the contract itself, minting and transferring NFTs, validating ownership, and adding resources to or removing resources from  NFTs.
The governance contract interacts with the NFT management contract to track proposals, authorizations, and votes.

\item \emph{DAOasis}:  
DAOasis provides a social networking platform facilitating the creation and management of multisigniture authorities \cite{KAL:2023}. This empowers users to execute transactions to manage their digital assets,  handle customized transactions, and perform secure cross-chain interoperability. Specifically,
Polybase is a decentralized database that provides a secure and decentralized method for data interaction.
The Gnosis Safe wallet offers users the ability to create accounts and safes. The Safe is a multi-signature wallet that requires a minimum number of signatures to perform a transaction, ensuring maximum security for all involved parties.
Connext facilitates seamless interactions across multiple blockchains by enabling cross-chain transactions and fund transfers.

\end{itemize}

\begin{table*}[]
\centering
\small 
\setlength{\tabcolsep}{10pt} 
\renewcommand{\arraystretch}{1.9} 
\caption{Issues of DAOs.}
\begin{tabular}{c|l|l}
\hline \hline
& \multicolumn{1}{c|}{\textbf{Types}}    & \multicolumn{1}{c}{\textbf{Description}} \\ \hline \hline
\multirow{8}{*}{\textbf{Technical}}    
                                        & \multirow{1}{*}{\hfil Security} & \multicolumn{1}{l}{The code is extremely difficult to fix, thus leaving a vulnerability.}  \\ \cline{2-3} 
                                        &  Speed   & Every user is given an opportunity to vote, which requires a longer voting period.  \\ \cline{2-3} 
                                        & \multirow{1}{*}{\hfil Engagement} & \multicolumn{1}{l}{Abandoning participation in governance will lead to a re-centralization of power.} \\ \cline{2-3} 
                                        & \multirow{1}{*}{\hfil  Pseudonym}  &  \multicolumn{1}{l}{Pseudonyms may hinder efforts to combat financial crime. } \\ \cline{2-3} 
                                        & \multirow{1}{*}{\hfil  Unfairness}  &  \multicolumn{1}{l}{Autonomous attributes may be undermined by a few users who have a larger voice.}  \\ \cline{2-3} 
                                        & \multirow{1}{*}{\hfil  Consistency}  &  \multicolumn{1}{l}{Data bias makes it challenging to fully align AI's mission with DAO's goals.}  \\ \cline{2-3} 
                                        & \multirow{1}{*}{\hfil  Autonomy}  &  \multicolumn{1}{l}{The integration of AI into DAOs brings some centralization of control and power.}  \\ \cline{2-3} 
                                        & \multirow{1}{*}{\hfil  Transparency}  &  \multicolumn{1}{l}{The lack of interpretability of AI can pose a challenge to DAO's decision-making process.}  \\ \cline{2-3} 
 \hline
\multirow{4}{*}{\textbf{Others}}                   &  \multirow{1}{*}{\hfil Regulatory}  & \multicolumn{1}{l}{The legal status is typically ambiguous and may change depending on the jurisdiction.} \\ \cline{2-3} 
\multicolumn{1}{l|}{}                  & \multirow{1}{*}{\hfil Inefficiency} & \multicolumn{1}{l}{ It is easy for a DAO to spend much more time discussing than implementing.} \\ \cline{2-3} 
\multicolumn{1}{l|}{}                  & \multirow{1}{*}{\hfil Social}  & \multicolumn{1}{l}{Inactive or non-voting shareholders can cause disruptions in an organization's operations.} \\ \cline{2-3} 
\multicolumn{1}{l|}{}                  & \multirow{1}{*}{\hfil Education}  & \multicolumn{1}{l}{It is difficult for people with different backgrounds to develop strategies together.}  \\ \hline \hline
\end{tabular}
\label{tab:riskDAO}
\end{table*}

\emph{5) Issues:} Although DAOs are still in the early stages of development, it has attracted widespread attention. However, improperly building or maintaining a DAO can have serious consequences. Since a DAO directly controls assets, vulnerabilities always can cause catastrophic damage. In addition, due to their infrastructure, DAOs suffer from many of the same limitations and security issues as the blockchains on which they operate. Beyond that, some of the major limitations, as well as issues, are listed in Table~\ref{tab:riskDAO} \cite{Ding:2022, Yu:23, GKSW:22, Tan:2023}.

\section{Challenges and Future Research Directions} \label{sec:challenges}

Advanced technologies, represented by blockchain, AI, and edge computing are driving the rapid development of the Web 3.0 ecosystem, which is expected to revolutionize the way people interact with the Web. However, the fast iteration of these technologies also presents many challenges and limitations to Web 3.0. 
The miscreants are well aware of this and try to exploit this iterative gap for their own nefarious purposes. 
As a result, some major challenges and open issues need to be addressed urgently. In the following, we will discuss some of the major issues that Web 3.0 is facing \cite{RB:16, Sheridan:22, Fan:2023, BNZ:17}.  
Additionally, we also discuss the potential research directions for building future Web 3.0 ecosystem.

\subsection{Key challenge and research direction of Web 3.0 in the context of blockchain}
\subsubsection{Key challenge} 
Scalability has proven to be a major obstacle to the rapid development of the Web 3.0 ecosystem.   
Scalability refers to the ability of a blockchain network to develop and adapt to growing demands as more users are added to the network without compromising its security or effectiveness.
As a typical problem of the blockchain trilemma, there is no general solution as of today. As the underlying ecology of Web 3.0, the blockchain platform often needs to make a trade-off between scalability, decentralization, and security in different scenarios. For example, Blockchain 3.0 chooses to sacrifice security for high throughput and fast transactions. This results in lower fault tolerance than those in blockchain 1.0 and blockchain 2.0.  
This could make Web 3.0 solutions based on such blockchain platforms more vulnerable to attacks. 
Meanwhile, digital wallets, existing as Web 3.0 portals, typically as browser extensions, pose security issues since digital wallets are connected to the blockchain through a centralized platform at this stage. Once the centralized platform is compromised, users will suffer huge financial losses.
For example, according to the Twitter account, Solana Status, an unidentified hacker has stolen approximately \$8 million from approximately 8,000 wallets on the Solana network \cite{SolanaStatus:2022}.
Therefore, the resolution of the trilemma, or how to make effective trade-offs, will greatly affect the development and deployment of Web 3.0.

\subsubsection{Future direction}
To effectively solve the scalability challenges faced by decentralized technologies required by Web 3.0, sharding is a promising solution. Although the concept of sharding has been proposed for years as a solution to scalability challenges, mainstream Web 3.0 platforms have yet to fully integrate sharding technology. The development and implementation of practical sharding techniques for Web 3.0 applications remain an ongoing area of research. Specifically, cross-shard communication is an urgent problem that needs to be solved as it requires additional protocols and mechanisms to ensure the consistency and validity of data and transactions across shards. Sharding security is also important for the Web 3.0 ecosystem. It refers to the risk of malicious nodes taking over a shard and manipulating its operations. One attractive solution to strengthen shard security is by using randomization and incentives to assign nodes to shards while applying cryptographic proofs (e.g., ZK proofs) to detect malicious behavior. 
Additionally, layer 2 scaling solutions are another important research direction that could complement sharding. By combining sharding with layer 2 solutions, blockchain networks can dramatically increase their throughput, thus realizing the Web 3.0 decentralized vision.
For example, state channels allow high-frequency transactions to be processed off-chain within individual shards, while only settlement states are periodically committed on-chain. Cross-shard state channels are used to maintain interconnectivity. In this way, the scaling issue will be effectively mitigated by combining layer 1 sharding with a layer 2 solution, which allows blockchain-powered Web 3.0 to support massive data concurrency.

\subsection{Key challenge and research direction of Web 3.0 in the context of AI}
\subsubsection{Key challenge}
Unconscious bias refers to implicit attitudes and stereotypes that influence people's understanding, actions, and decisions in an unconscious manner.
Although Web 3.0 is still in its early stages, unconscious bias seems to be creeping in.
According to a Bloomberg report, in December 2021, digital avatars of popular CryptoPunks NFTs fluctuated in price based on race, gender, and skin color, with medium and dark-skinned avatars averaging less than light-skinned NFTs \cite {EG:21}.
The researchers at People of Crypto Lab, an organization aiming to promote diversity, equity, and inclusion in the Web 3.0 ecosystem,  examined the current state of gender diversity among Web 3.0 startup founders and investors. They used Crunchbase's extensive database of nearly 2,800 participants worldwide to classify and analyze participants based on gender.
Only 13\% of Web 3.0 startups include females. Additionally, only 7\% of the founders of Web 3.0 startups are female. Both numbers are worse than average for startups overall \cite{Apotheker:23}. 
The lack of gender balance in the Web 3.0 ecosystem has significant implications for how people self-present, conduct business, and interact with each other online. As a result, the impacts of current underrepresentation maybe greater than in previous generations of the Web. 
Additionally, AI algorithms also display unconscious bias due to the unbalanced datasets they are trained with. For instance, 
Amazon had to stop using an AI recruiting tool that was biased against women in 2018. The tool was trained on a dataset containing resumes of applicants dating back 10 years. This dataset consisted mostly of male candidates, causing the AI algorithm to degrade resumes containing keywords such as ``woman" and ``female" \cite{Reuters:18}.
In 2019, researchers discovered bias against black patients in a commercially accessible AI system designed to predict patient outcomes. The algorithm had primarily been trained on data from white patients, resulting in higher misdiagnosis rates for black patients \cite{Ledford:19}.
Utilizing the biased dataset can harm the Web 3.0 ecosystem.
Given the signs of these impending problems, unconscious bias will be one of the major challenges for the Web 3.0 ecosystem.

\subsubsection{Future direction}
The main reason for introducing unintentional bias is usually that the training data is unbalanced or unrepresentative. However, data diversity is not always fully achieved in the real world due to factors such as structural inequalities and historical patterns. When AI systems developed based on such data become widely used, unconscious bias can create a vicious cycle. This is particularly concerning as generative AI plays an increasingly important role in creating vast amounts of new digital content in the decentralized Web 3.0 ecosystem. It is crucial to ensure that such content is not biased.
Given the prevalence of generative AI, a promising research direction is to build an AIGC-driven bias-free Web 3.0 ecosystem. Specifically, unconscious bias in Web 3.0 can be effectively reduced by leveraging AIGC to generate training data that scales diversity and inclusion. Furthermore, to build a more inclusive and secure Web 3.0, it is necessary to rethink fairness as well as privacy and security in AIGC-driven Web 3.0. An unfair AIGC model may further exacerbate inequalities in the Web 3.0 ecosystem. It is critical to ensure that AIGC algorithms are designed to promote inclusivity and avoid reinforcing bias.
With the rise of AIGC, the digital economy driven by Web 3.0 has been greatly enriched. The classic issues of privacy and security will also be an important area of research.

\subsection{Key challenge and research direction of Web 3.0 in the context of edge computing}
\subsubsection{Key challenge} 

Managing a large number of interconnected and heterogeneous devices poses significant challenges to the Web 3.0 ecosystem. As devices become more connected through edge computing architecture, threats can now spread more easily between devices. A vulnerability in one device may affect many other devices or even paralyze the entire system \cite{Antonakakis:2017}.
The devices themselves also come in many different forms, with different hardware, software, operating systems, and use cases. For example, smartphones and IoT sensor nodes will have very different technical specifications and deployment environments. This high degree of heterogeneity means that universal security solutions must account for a large number of potential configurations and usage patterns. As the number of connected devices continues to grow exponentially, the ways in which these different devices interact with each other also grow dramatically. It becomes incredibly difficult to develop a security approach that perfectly addresses each unique scenario. Solutions designed for maximum generality may lack accuracy for certain device types or environments. However, an approach that is too specific will not be able to scale and cover the entire content of the device ecosystem. How to achieve the right trade-off between generality and accuracy is the key to solving the problem \cite{Jin:2022}.
Furthermore, coordinating edge computing resources in a decentralized manner is a challenging task. Traditional centralized management approaches rely on a central authority or control point and may not scale well in a decentralized network environment.
Decentralized coordination of edge computing resources involves distributing decision-making and control mechanisms among multiple nodes or entities within the network. This approach aligns with Web 3.0 principles, which aim to distribute power and control among network participants. This requires further exploration of edge computing and blockchain technologies (e.g., consensus protocols, SCs, etc.).

\subsubsection{Future direction}
Solving device management and resource allocation issues in the Web 3.0 ecosystem requires the development of adaptive and scalable approaches.
One potential research direction is the modular architecture and protocol design of edge computing-enabled Web 3.0 to adapt to different device types, configurations, and use cases.
These frameworks should enable seamless integration and interoperability between heterogeneous devices to facilitate communication and collaboration across the Web 3.0 ecosystem.
In addition, since edge devices may dynamically join or leave the network, the allocation of resources and tasks becomes more challenging. Therefore, developing decentralized algorithms and mechanisms for resource allocation, load balancing, and task offloading helps optimize resource utilization and ensure efficient coordination among edge devices. Specifically, considering the functionality and availability of edge devices is crucial for the design of intelligent resource allocation algorithms. These algorithms should efficiently distribute tasks among edge devices to maximize resource utilization while minimizing latency and energy consumption.
Second, designing a load balancing mechanism is key to achieving even workload distribution across edge devices. Load balancing algorithms should consider factors such as device capacity, network congestion, and task demands to ensure optimal resource utilization and prevent overloading of specific devices.
Third, effective task offloading strategies need to be developed to determine when and which tasks should be offloaded from edge devices to other edge devices or edge computing clusters for acceleration, load balancing, etc. These policies should consider aspects such as task characteristics, network conditions, and device capabilities to minimize latency and improve overall system performance, thereby facilitating the scheduling of task processing in a system-wide view.

\section{Conclusion} \label{sec:conclusion}
This paper conducted a thorough investigation of the impact of blockchain, AI, and edge computing on Web 3.0, the most promising technologies driving the next generation of the Web. By exploring the development of the Web at different stages, the necessity and timeliness of web evolution are clarified. 
Technically, Web 3.0 is a back-end evolution based on blockchain technology that allows for the distribution of power and trust, enabling users to have more control over their personal data and digital assets. AI can empower Web 3.0 with key features, such as intelligent automation, enhanced security, and improved governance. Conversely, Web 3.0 can provide AI with the two most important elements: data and computing power. Edge computing brings practical benefits such as low latency and cost-effective performance to the Web 3.0 ecosystem. 
Then, this article conducted an extensive literature review and discussed the practical applications of each technology in Web 3.0.
We also proposed decentralized storage and computing solutions by exploring the integration of technologies.
Furthermore, we had an in-depth discussion on the mainstream use cases (i.e., NFTs, DeFi, and DAO) from the aspects of  definition, key attributes, related applications, and potential issues.
Finally, the key challenges and potential research directions were raised to provide guidance for research in related fields. 
Overall, Web 3.0 is envisioned as an ecosystem, built on top of blockchain, powered by AI, and optimized via edge computing, with the potential to change the way people interact with information fundamentally.
We expect that this survey can facilitate a clearer understanding of Web 3.0 and inspire further innovative research within this emerging field.



\begin{thebibliography}{100}
\providecommand{\url}[1]{#1}
\csname url@samestyle\endcsname
\providecommand{\newblock}{\relax}
\providecommand{\bibinfo}[2]{#2}
\providecommand{\BIBentrySTDinterwordspacing}{\spaceskip=0pt\relax}
\providecommand{\BIBentryALTinterwordstretchfactor}{4}
\providecommand{\BIBentryALTinterwordspacing}{\spaceskip=\fontdimen2\font plus
\BIBentryALTinterwordstretchfactor\fontdimen3\font minus
  \fontdimen4\font\relax}
\providecommand{\BIBforeignlanguage}[2]{{%
\expandafter\ifx\csname l@#1\endcsname\relax
\typeout{** WARNING: IEEEtran.bst: No hyphenation pattern has been}%
\typeout{** loaded for the language `#1'. Using the pattern for}%
\typeout{** the default language instead.}%
\else
\language=\csname l@#1\endcsname
\fi
#2}}
\providecommand{\BIBdecl}{\relax}
\BIBdecl

\bibitem{Statista:22}
Statista, ``{Worldwide digital population July 2022},''
  \url{https://www.statista.com/statistics/617136/digital-population-worldwide/\#statisticContainer},
  Sep. 2022.

\bibitem{KS:2022}
G.~Korpal and D.~Scott, ``Decentralization and {Web3} technologies,'' 2022,
  techRxiv preprint.

\bibitem{Web3Foundation}
Web3Foundation, ``Web 3.0 technology stack,''
  \url{https://web3.foundation/about/}, 2022.

\bibitem{VK:21}
S.~Voj{\'\i}{\v{r}} and J.~Ku{\v{c}}era, ``Towards re-decentralized future of
  the web: Privacy, security and technology development,'' \emph{Acta
  Informatica Pragensia}, vol. 2021, no.~3, pp. 349--369, 2021.

\bibitem{Zarrin:21}
J.~Zarrin \emph{et~al.}, ``Blockchain for decentralization of {Internet}:
  prospects, trends, and challenges,'' \emph{Cluster Computing}, vol.~24,
  no.~4, pp. 2841--2866, May 2021.

\bibitem{Wang:22}
Q.~Wang \emph{et~al.}, ``Exploring {Web3} from the view of blockchain,'' 2022,
  arXiv preprint arXiv:2206.08821.

\bibitem{Ray:23}
P.~Ray, ``{Web3}: A comprehensive review on background, technologies,
  applications, zero-trust architectures, challenges and future directions,''
  \emph{Internet of Things and Cyber-Physical Systems}, 2023.

\bibitem{Gan:23}
W.~Gan \emph{et~al.}, ``{Web 3.0}: The future of {Internet},'' 2023, arXiv
  preprint arXiv:2304.06032.

\bibitem{Ren:23}
X.~Ren \emph{et~al.}, ``Building resilient {Web 3.0} with quantum information
  technologies and blockchain: An ambilateral view,'' 2023, arXiv preprint
  arXiv:2303.13050.

\bibitem{Huang:23}
R.~Huang \emph{et~al.}, ``An overview of {Web3. 0} technology: Infrastructure,
  applications, and popularity,'' 2023, arXiv preprint arXiv:2305.00427.

\bibitem{Shen:23}
M.~Shen \emph{et~al.}, ``Artificial intelligence for {Web 3.0}: A comprehensive
  survey,'' 2023, arXiv preprint arXiv:2309.09972.

\bibitem{Beniiche:2022}
A.~Beniiche \emph{et~al.}, ``Society 5.0: {Internet} as if people mattered,''
  \emph{IEEE Wireless Communications}, vol.~29, no.~6, pp. 160--168, 2022.

\bibitem{Park:2023}
A.~Park \emph{et~al.}, ``Interoperability: Our exciting and terrifying {Web3}
  future,'' \emph{Business Horizons}, vol.~66, no.~4, pp. 529--541, 2023.

\bibitem{O'Reilly:05}
T.~O'Reilly, ``What is {Web} 2.0? design patterns and business models for the
  next generation of software,''
  \url{https://www.oreilly.com/pub/a/web2/archive/what-is-web-20.html}, Sep.
  2005.

\bibitem{Ethereum:22}
Ethereum, ``{Introduction to Web3},'' \url{https://ethereum.org/en/web3/}, May.
  2022.

\bibitem{Berners-Lee:20}
T.~Berners-Lee \emph{et~al.}, ``Solid protocol,''
  \url{https://solidproject.org/TR/protocol#terminology}, Dec. 2020.

\bibitem{WebFoundation:17}
World-Wide-Web-Foundation, ``Three challenges for the web, according to its
  inventor,'' \url{https://webfoundation.org/2017/03/web-turns-28-letter/},
  Mar. 2017.

\bibitem{PEW:19}
PewResearchCenter, ``Americans and privacy: Concerned, confused and feeling
  lack of control over their personal information,''
  \url{https://www.pewresearch.org/internet/2019/11/15/americans-and-privacy-concerned-confused-and-feeling-lack-of-control-over-their-personal-information/pi_2019-11-14_privacy_0-01/},
  Nov. 2019.

\bibitem{Statista:23}
Statista, ``{Number of data records exposed worldwide from 1st quarter 2020 to
  1st quarter 2023},''
  \url{https://www.statista.com/statistics/1307426/number-of-data-breaches-worldwide/},
  Jun. 2023.

\bibitem{DRZ:18}
S.~Duan, M.~Reiter, and H.~Zhang, ``{BEAT}: Asynchronous {BFT} made
  practical,'' in \emph{Proceedings of ACM SIGSAC Conference on Computer and
  Communications Security}, Jan. 2018, pp. 2028--2041.

\bibitem{YMRGA:19}
M.~Yin, D.~Malkhi, M.~Reiter, G.~Gueta, and I.~Abraham, ``Hotstuff: {BFT}
  consensus with linearity and responsiveness,'' in \emph{Proceedings of the
  ACM Symposium on Principles of Distributed Computing (PODC)}, Jul. 2019, pp.
  347--356.

\bibitem{MXCSS:16}
A.~Miller, Y.~Xia, K.~Croman, E.~Shi, and D.~Song, ``The {Honey Badger} of
  {BFT} protocols,'' in \emph{ACM SIGSAC Conference on Computer and
  Communications Security}, Oct. 2016.

\bibitem{JBR:19}
M.~Jalalzai, C.~Busch, and G.~Richard, ``Proteus: A scalable {BFT} consensus
  protocol for blockchains,'' in \emph{IEEE International Conference on
  Blockchain (Blockchain)}, Jul. 2019.

\bibitem{GAGMPRSTT:19}
G.~Gueta, I.~Abraham, S.~Grossman, D.~Malkhi, B.~Pinkas, M.~Reiter,
  D.~Seredinschi, O.~Tamir, and A.~Tomescu, ``{SBFT}: A scalable and
  decentralized trust infrastructure,'' in \emph{Annual IEEE/IFIP International
  Conference on Dependable Systems and Networks (DSN)}, Jun. 2019.

\bibitem{LWCLX:20}
P.~Li, G.~Wang, X.~Chen, F.~Long, and W.~Xu, ``Gosig: a scalable and
  high-performance {Byzantine} consensus for consortium blockchains,'' in
  \emph{Proceedings of the 11th ACM Symposium on Cloud Computing}, 2020, pp.
  223--237.

\bibitem{Nak:08}
S.~Nakamoto, ``Bitcoin: A peer-to-peer electronic cash system,'' 2008.

\bibitem{Wood:16}
G.~Wood, ``Polkadot: Vision for a heterogeneous multi-chain framework,''
  \emph{White Paper}, vol.~21, pp. 2327--4662, 2016.

\bibitem{KRDO:17}
A.~Kiayias, A.~Russell, B.~David, and R.~Oliynykov, ``Ouroboros: A provably
  secure {Proof-of-Stake} blockchain protocol,'' in \emph{Advances in
  Cryptology -- CRYPTO. Lecture Notes in Computer Science}, vol. 10401, 2017.

\bibitem{GHMVZ:17}
Y.~Gilad, R.~Hemo, S.~Micali, G.~Vlachos, and N.~Zeldovich, ``Algorand: Scaling
  {Byzantine} agreements for cryptocurrencies,'' in \emph{Proceedings of
  Symposium on Operating Systems Principles}, Oct. 2017, pp. 51--68.

\bibitem{BLMR:14}
I.~Bentov, C.~Lee, A.~Mizrahi, and M.~Rosenfeld, ``{Proof of Activity}:
  Extending {Bitcoin's} {Proof of Work} via {Proof of Stake},'' \emph{ACM
  SIGMETRICS Performance Evaluation Review}, vol.~42, no.~3, pp. 34--37, 2014.

\bibitem{Yakovenko:18}
A.~Yakovenko, ``Solana: A new architecture for a high performance blockchain
  v0. 8.13,'' \emph{Whitepaper}, 2018.

\bibitem{KKZ:20}
K.~Karantias, A.~Kiayias, and D.~Zindros, ``{Proof-of-Burn},'' in
  \emph{International conference on financial cryptography and data
  security}.\hskip 1em plus 0.5em minus 0.4em\relax Springer, 2020, pp.
  523--540.

\bibitem{NEM:18}
NEM, ``Nem whitepaper,''
  \url{https://whitepaper.io/document/583/nem-whitepaper}, 2018.

\bibitem{Fisch:18}
B.~Fisch, ``Poreps: {Proofs of Space} on useful data,'' \emph{Cryptology ePrint
  Archive}, 2018.

\bibitem{DFKP:15}
S.~Dziembowski, S.~Faust, V.~Kolmogorov, and K.~Pietrzak, ``{Proofs of
  Space},'' in \emph{Annual Cryptology Conference}.\hskip 1em plus 0.5em minus
  0.4em\relax Springer, Nov. 2015, pp. 585--605.

\bibitem{Agbozo:23}
E.~Agbozo \emph{et~al.}, ``Applying multi-criteria decision making to
  prioritization of {Web 3.0} development factors,'' in \emph{E-business
  technologies conference proceedings}, vol.~3, no.~1, 2023, pp. 229--232.

\bibitem{YC:18}
N.~Chaudhry and M.~Yousaf, ``Consensus algorithms in blockchain: comparative
  analysis, challenges and opportunities,'' in \emph{International Conference
  on Open Source Systems and Technologies (ICOSST)}.\hskip 1em plus 0.5em minus
  0.4em\relax IEEE, 2018, pp. 54--63.

\bibitem{FCHC:20}
M.~Ferdous, M.~Chowdhury, M.~Hoque, and A.~Colman, ``Blockchain consensus
  algorithms: A survey,'' 2020, arXiv preprint arXiv:2001.07091.

\bibitem{BMB:20}
S.~Bamakan, A.~Motavali, and A.~Bondarti, ``A survey of blockchain consensus
  algorithms performance evaluation criteria,'' \emph{Expert Systems with
  Applications}, vol. 154, p. 113385, 2020.

\bibitem{IBMAI:20}
I.~C. Education, ``Artificial intelligence {(AI)},''
  \url{https://www.ibm.com/cloud/learn/what-is-artificial-intelligence}, Jun.
  2020.

\bibitem{HUA:23}
H.~Hua \emph{et~al.}, ``Edge computing with artificial intelligence: A machine
  learning perspective,'' \emph{ACM Computing Surveys}, vol.~55, no.~9, pp.
  1--35, Apr. 2023.

\bibitem{BP:22}
J.~Bambacht and J.~Pouwelse, ``{Web3}: A decentralized societal infrastructure
  for identity, trust, money, and data,'' 2022, arXiv preprint
  arXiv:2203.00398.

\bibitem{Huang:2023}
G.~Huang \emph{et~al.}, ``Efficient and low overhead website fingerprinting
  attacks and defenses based on {TCP/IP} traffic,'' in \emph{Proceedings of the
  ACM Web Conference 2023}, Apr. 2023, pp. 1991--1999.

\bibitem{Cadena:2020}
W.~D. la~Cadena \emph{et~al.}, ``Trafficsliver: Fighting website fingerprinting
  attacks with traffic splitting,'' in \emph{Proceedings of the 2020 ACM SIGSAC
  Conference on Computer and Communications Security}, Oct. 2020, pp.
  1971--1985.

\bibitem{Voshmgir:20}
S.~Voshmgir, \emph{Token Economy: How the {Web3} reinvents the Internet}.\hskip
  1em plus 0.5em minus 0.4em\relax Token Kitchen, 2020, vol.~2.

\bibitem{Kasireddy:21}
P.~Kasireddy, ``The architecture of a {Web 3.0} application,''
  \url{https://www.preethikasireddy.com/post/the-architecture-of-a-web-3-0-application},
  Sep. 2021.

\bibitem{deController:2023}
H.~Xu \emph{et~al.}, ``{deController}: a {Web3} native cyberspace
  infrastructure perspective,'' \emph{IEEE Communications Magazine}, 2023.

\bibitem{Bassi:21}
C.~Bassi, ``Sustainable blockchain: Estimating the carbon footprint of
  algorand’s pure {Proof-of-Stake},''
  \url{https://algorand.com/resources/blog/sustainable-blockchain-calculating-the-carbon-footprint},
  Apr. 2021.

\bibitem{Kohli:22}
V.~Kohli \emph{et~al.}, ``An analysis of energy consumption and carbon
  footprints of cryptocurrencies and possible solutions,'' \emph{Digital
  Communications and Networks}, vol.~9, no.~1, pp. 79--89, Jun 2022.

\bibitem{Digiconomist:23}
Digiconomist, ``Bitcoin energy consumption index,''
  \url{https://digiconomist.net/bitcoin-energy-consumption}, Nov. 2023.

\bibitem{Digiconomist:2023}
------, ``Ethereum energy consumption index,''
  \url{https://digiconomist.net/ethereum-energy-consumption}, Nov. 2023.

\bibitem{Breidenbach:2021}
L.~Breidenbach \emph{et~al.}, ``Chainlink 2.0: Next steps in the evolution of
  decentralized oracle networks,'' \emph{Chainlink Labs}, vol.~1, pp. 1--136,
  2021.

\bibitem{Buterin:2016}
V.~Buterin, ``Chain interoperability,'' \emph{R3 Research Paper}, vol.~9, 2016.

\bibitem{BVGC:21}
R.~Belchior, A.~Vasconcelos, S.~Guerreiro, and M.~Correia, ``A survey on
  blockchain interoperability: Past, present, and future trends,'' \emph{ACM
  Computing Surveys}, vol.~54, no.~8, pp. 1--41, 2021.

\bibitem{WWC:23}
G.~Wang, Q.~Wang, and S.~Chen, ``Exploring blockchains interoperability: A
  systematic survey,'' \emph{ACM Computing Surveys}, 2023.

\bibitem{BMRS:18}
M.~Borkowski, D.~McDonald, C.~Ritzer, and S.~Schulte, ``Towards atomic
  cross-chain token transfers: State of the art and open questions within
  tast,'' \emph{Distributed Systems Group TU Wien (Technische Universit at
  Wien), Report}, vol.~8, 2018.

\bibitem{Multichain}
Multichain, ``Multichain bridge,''
  \url{https://docs.multichain.org/getting-started/introduction}.

\bibitem{Threshold}
Threshold, ``{tBTC} bridge,'' \url{https://threshold.network/earn/btc/}.

\bibitem{Parity}
Parity, ``Bridging the dapp-scaling now with parity bridge,''
  \url{https://www.parity.io/blog/tag/parity-bridge}, Mar. 2018.

\bibitem{Solana:20}
Solana, ``Wormhole,'' \url{https://solana.com/ecosystem/wormhole}, Apr. 2020.

\bibitem{Smith:23}
C.~Smith \emph{et~al.}, ``Scaling,''
  \url{https://ethereum.org/en/developers/docs/scaling/}, Apr. 2023.

\bibitem{EthereumDanksharding:23}
Ethereum, ``Danksharding,''
  \url{https://ethereum.org/en/roadmap/danksharding/}, Nov. 2023.

\bibitem{JT:16}
P.~Joseph and D.~Thaddeus, ``The bitcoin lightning network: Scalable off-chain
  instant payments,'' 2016.

\bibitem{Raiden}
Raiden-{Network}, ``What is the raiden network?''
  \url{https://raiden.network/101.html}.

\bibitem{Cahill:22}
A.~Cahill and S.~Deshpande, ``Layer-2 scaling solutions: A framework for
  comparison,''
  \url{https://www.tbstat.com/wp/uploads/2022/05/20220505_Layer2ScalingSolutions_TheBlockResearch.pdf},
  May 2022.

\bibitem{LiquidNetwork}
Liquid-{Network}, ``The liquid network,'' \url{https://liquid.net/}.

\bibitem{Tripathi:23}
A.~Tripathi, ``{Introduction to Polygon PoS},''
  \url{https://wiki.polygon.technology/docs/pos/polygon-architecture/}, Jun
  2023.

\bibitem{Benet:14}
J.~Benet, ``{IPFS} - content addressed, versioned, {P2P} file system,'' 2014,
  arXiv preprint arXiv:1407.3561.

\bibitem{Drakatos:21}
P.~Drakatos \emph{et~al.}, ``{Triastore}: A {Web 3.0} blockchain datastore for
  massive {IoT} workloads,'' in \emph{2021 22nd IEEE International Conference
  on Mobile Data Management (MDM)}, 2021, pp. 187--192.

\bibitem{Liu:21}
Z.~Liu \emph{et~al.}, ``Make {Web 3.0} connected,'' \emph{IEEE Transactions on
  Dependable and Secure Computing}, vol.~19, pp. 2965--2981, 2022.

\bibitem{Chopra:22}
A.~Chopra \emph{et~al.}, ``Va3: A {Web 3.0} based {I2I} power transaction
  platform,'' in \emph{2022 7th IEEE Workshop on the Electronic Grid (eGRID)},
  2022, pp. 1--5.

\bibitem{Lin:2023}
Y.~Lin \emph{et~al.}, ``A unified blockchain-semantic framework for wireless
  edge intelligence enabled {Web 3.0},'' \emph{IEEE Wireless Communications},
  2023.

\bibitem{Palanikkumar:23}
D.~Palanikkumar \emph{et~al.}, ``An enhanced decentralized social network based
  on {Web3} and {IPFS} using blockchain,'' in \emph{2023 7th International
  Conference on Trends in Electronics and Informatics (ICOEI)}, 2023, pp.
  616--623.

\bibitem{Petcu:23}
A.~Petcu \emph{et~al.}, ``A secure and decentralized authentication mechanism
  based on {Web 3.0} and {Ethereum} blockchain technology,'' \emph{Applied
  Sciences}, vol.~13, no.~4, 2023.

\bibitem{Razzaq:23}
A.~Razzaq \emph{et~al.}, ``{IoT} data sharing platform in {Web 3.0} using
  blockchain technology,'' \emph{Electronics}, vol.~12, no.~5, 2023.

\bibitem{Lin:23}
Y.~Lin \emph{et~al.}, ``A blockchain-based semantic exchange framework for {Web
  3.0} toward participatory economy,'' \emph{IEEE Communications Magazine},
  Aug. 2023.

\bibitem{Guo:23}
S.~Guo \emph{et~al.}, ``Blockchain-assisted privacy-preserving data computing
  architecture for {Web3},'' \emph{IEEE Communications Magazine}, vol.~61,
  no.~8, pp. 28--34, Aug. 2023.

\bibitem{Qiu:23}
Y.~Qiu \emph{et~al.}, ``Fog-assisted blockchain radio access network for
  {Web3},'' \emph{IEEE Communications Magazine}, Aug. 2023.

\bibitem{Guang:2022}
G.~Yu \emph{et~al.}, ``Towards {Web3} applications: Easing the access and
  transition,'' 2022, arXiv preprint arXiv:2210.05903.

\bibitem{LinAIGC:23}
Y.~Lin \emph{et~al.}, ``Blockchain-aided secure semantic communication for
  {AI}-generated content in {Metaverse},'' \emph{IEEE Open Journal of the
  Computer Society}, vol.~4, pp. 72--83, 2023.

\bibitem{LinAIGC:2023}
------, ``A unified framework for integrating semantic communication and
  {AI}-generated content in {Metaverse},'' 2023, arXiv preprint
  arXiv:2305.11911.

\bibitem{LiuAIGC:23}
G.~Liu \emph{et~al.}, ``Semantic communications for artificial intelligence
  generated content ({AIGC}) toward effective content creation,'' 2023, arXiv
  preprint arXiv:2308.04942.

\bibitem{XiaAIGC:23}
L.~Xia \emph{et~al.}, ``{Generative AI} for semantic communication:
  Architecture, challenges, and outlook,'' 2023, arXiv preprint
  arXiv:2308.15483.

\bibitem{Cheng:23}
R.~Cheng \emph{et~al.}, ``A wireless {AI}-generated content ({AIGC})
  provisioning framework empowered by semantic communication,'' 2023, arXiv
  preprint arXiv:2310.17705.

\bibitem{DuAIGC:2023}
H.~Du \emph{et~al.}, ``Enabling {AI}-generated content ({AIGC}) services in
  wireless edge networks,'' 2023, arXiv preprint arXiv:2301.03220.

\bibitem{LiuLifecycle:2023}
Y.~Liu \emph{et~al.}, ``Blockchain-empowered lifecycle management for
  {AI}-generated content ({AIGC}) products in edge networks,'' 2023, arXiv
  preprint arXiv:2303.02836.

\bibitem{XuAIGC:2023}
M.~Xu \emph{et~al.}, ``Joint foundation model caching and inference of
  generative {AI} services for edge intelligence,'' 2023, arXiv preprint
  arXiv:2305.12130.

\bibitem{DuGenerativeAIGC:2023}
H.~Du \emph{et~al.}, ``Generative {AI}-aided optimization for {AI}-generated
  content ({AIGC}) services in edge networks,'' 2023, arXiv preprint
  arXiv:2303.13052.

\bibitem{WangGenerativeAIGC:2023}
J.~Wang \emph{et~al.}, ``A unified framework for guiding generative {AI} with
  wireless perception in resource constrained mobile edge networks,'' 2023,
  arXiv preprint arXiv:2309.01426.

\bibitem{cao:23}
Y.~Cao \emph{et~al.}, ``A comprehensive survey of {AI}-generated content
  ({AIGC}): A history of generative {AI} from {GAN to ChatGPT},'' 2023, arXiv
  preprint arXiv:2303.04226.

\bibitem{Zhang:23}
C.~Zhang \emph{et~al.}, ``A complete survey on generative {AI (AIGC)}: Is
  {ChatGPT} from {GPT-4 to GPT-5} all you need?'' 2023, arXiv preprint
  arXiv:2303.11717.

\bibitem{IBMNLP:20}
I.~C. Education, ``Natural language processing {(NLP)},''
  \url{https://www.ibm.com/cloud/learn/natural-language-processing}, Jul. 2020.

\bibitem{Treviso:23}
M.~Treviso \emph{et~al.}, ``Efficient methods for natural language processing:
  A survey,'' \emph{Transactions of the Association for Computational
  Linguistics}, vol.~11, pp. 826--860, 2023.

\bibitem{Qin:23}
C.~Qin \emph{et~al.}, ``Is {ChatGPT} a general-purpose natural language
  processing task solver?'' 2023, arXiv preprint arXiv:2302.06476.

\bibitem{KKKS:23}
D.~Khurana, A.~Koli, K.~Khatter, and S.~Singh, ``Natural language processing:
  State of the art, current trends and challenges,'' \emph{Multimedia tools and
  applications}, vol.~82, no.~3, pp. 3713--3744, 2023.

\bibitem{Le:22}
N.~Le \emph{et~al.}, ``Deep reinforcement learning in computer vision: A
  comprehensive survey,'' \emph{Artificial Intelligence Review}, pp. 1--87,
  2022.

\bibitem{Bi:22}
Y.~Bi \emph{et~al.}, ``A survey on evolutionary computation for computer vision
  and image analysis: Past, present, and future trends,'' \emph{IEEE
  Transactions on Evolutionary Computation}, 2022.

\bibitem{Vaswani:17}
A.~Vaswani \emph{et~al.}, ``{Attention is all you need},'' \emph{Advances in
  neural information processing systems}, vol.~30, 2017.

\bibitem{Goodfellow:14}
I.~Goodfellow \emph{et~al.}, ``Generative adversarial nets,'' in \emph{Advances
  in neural information processing systems}, 2014, pp. 2672--2680.

\bibitem{KW:13}
D.~Kingma \emph{et~al.}, ``Auto-encoding variational bayes,'' 2013, arXiv
  preprint arXiv:1312.6114.

\bibitem{Gregor:14}
K.~Gregor \emph{et~al.}, ``Deep autoregressive networks,'' in
  \emph{International Conference on Machine Learning}, 2014, pp. 1242--1250.

\bibitem{Alethea:2021}
AletheaAI, ``{Robert Alice},''
  \url{https://www.sothebys.com/en/buy/auction/2021/natively-digital-a-curated-nft-sale-2/to-the-young-artists-of-cyberspace},
  Jun. 2021.

\bibitem{Pregelj:2023}
J.~Pregelj, ``Web3 plugin for {ChatGPT},''
  \url{https://github.com/jernejpregelj/web3-chatgpt-plugin}, May 2023.

\bibitem{Sathavara:2023}
D.~Sathavara \emph{et~al.}, ``{SuperCool-AI},''
  \url{https://supercool.vercel.app/}, Jun 2023.

\bibitem{Sokoli:2023}
M.~Sokoli \emph{et~al.}, ``{Web3GPT},'' \url{https://w3gpt.ai/}, May 2023.

\bibitem{Savage:2023}
J.~Savage, ``{ETHGPT},'' \url{https://github.com/Jsavage1325/ETHGPT}, May 2023.

\bibitem{Yang:2023}
C.~Yang \emph{et~al.}, ``{FlashGPT},''
  \url{https://github.com/alt-research/flashGPT}, Mar. 2023.

\bibitem{TokenGPT:2023}
{TokenGPT}, ``Tokengpt,'' \url{https://ethglobal.com/showcase/tokengpt-ko8x6},
  May 2023.

\bibitem{Zhang:2023}
M.~Zhang, ``{CoinGPT},'' \url{https://github.com/mccowanzhang/coingpt}, Jun.
  2023.

\bibitem{MG:2023}
Memosys and Greg, ``Defi-companion-covalent,''
  \url{https://github.com/gregfromstl/defi-companion}, Aug. 2023.

\bibitem{Quantstamp:2017}
Quantstamp, ``Quantstamp,'' \url{https://quantstamp.com/}, 2017.

\bibitem{ChainSecurity:2017}
ChainSecurity, ``Chainsecurity,'' \url{https://chainsecurity.com/}, 2017.

\bibitem{Gupta:2022}
K.~Gupta \emph{et~al.}, ``{Secure Semantic Snap},''
  \url{https://nozk.kanavgupta.xyz/}, Dec 2022.

\bibitem{Alam:2023}
A.~Alam \emph{et~al.}, ``{MedDAO},''
  \url{https://github.com/Arsalaan-Alam/hackfs}, Jun. 2023.

\bibitem{Shamir:79}
A.~Shamir, ``How to share a secret,'' \emph{Communications of the ACM},
  vol.~22, no.~11, pp. 612--613, 1979.

\bibitem{Das:2023}
S.~Das \emph{et~al.}, ``Practical asynchronous high-threshold distributed key
  generation and distributed polynomial sampling,'' in \emph{32nd USENIX
  Security Symposium (USENIX Security 23)}, Aug. 2023, pp. 5359--5376.

\bibitem{DasSS:2023}
------, ``A new paradigm for verifiable secret sharing,'' 2023, cryptology
  ePrint Archive.

\bibitem{Ge:2019}
C.~Ge \emph{et~al.}, ``Revocable identity-based broadcast proxy re-encryption
  for data sharing in clouds,'' \emph{IEEE Transactions on Dependable and
  Secure Computing}, vol.~18, no.~3, pp. 1214--1226, 2019.

\bibitem{Zhang:2021}
W.~Zhang \emph{et~al.}, ``A secure revocable fine-grained access control and
  data sharing scheme for {SCADA in IIoT} systems,'' \emph{IEEE Internet of
  Things Journal}, vol.~9, no.~3, pp. 1976--1984, 2021.

\bibitem{ZhangIOT:2021}
J.~Zhang \emph{et~al.}, ``Revocable and privacy-preserving decentralized data
  sharing framework for fog-assisted {Internet of Things},'' \emph{IEEE
  Internet of Things Journal}, vol.~9, no.~13, pp. 10\,446--10\,463, 2021.

\bibitem{YLF:23}
X.~Yang, W.~Li, and K.~Fan, ``A revocable attribute-based encryption {EHR}
  sharing scheme with multiple authorities in blockchain,'' \emph{Peer-to-peer
  Networking and Applications}, vol.~16, no.~1, pp. 107--125, 2023.

\bibitem{Keizer:21}
N.~Keizer \emph{et~al.}, ``The case for {AI} based {Web3} reputation systems,''
  in \emph{2021 IFIP Networking Conference (IFIP Networking)}, 2021, pp. 1--2.

\bibitem{Lorenz:2021}
J.~Lorenz \emph{et~al.}, ``Machine learning methods to detect money laundering
  in the {Bitcoin} blockchain in the presence of label scarcity,'' in
  \emph{Proceedings of the first ACM international conference on AI in
  finance}, Oct. 2021, pp. 1--8.

\bibitem{Weber:19}
M.~Weber \emph{et~al.}, ``Anti-money laundering in {Bitcoin}: Experimenting
  with graph convolutional networks for financial forensics,'' 2019, arXiv
  preprint arXiv:1908.02591.

\bibitem{AP:2023}
I.~Alarab and S.~Prakoonwit, ``Graph-based {LSTM} for anti-money laundering:
  Experimenting temporal graph convolutional network with bitcoin data,''
  \emph{Neural Processing Letters}, vol.~55, no.~1, pp. 689--707, 2023.

\bibitem{Lo:2023}
W.~Lo \emph{et~al.}, ``Inspection-l: self-supervised {GNN} node embeddings for
  money laundering detection in bitcoin,'' \emph{Applied Intelligence},
  vol.~53, pp. 19\,406--19\,417, Aug 2023.

\bibitem{Unzeelah:22}
M.~Unzeelah and Z.~Memon, ``Fighting against fake news by connecting machine
  learning approaches with {Web3},'' in \emph{2022 International Conference on
  Emerging Trends in Smart Technologies (ICETST)}, 2022, pp. 1--6.

\bibitem{Kim:2022}
J.~Kim \emph{et~al.}, ``A machine learning approach to anomaly detection based
  on traffic monitoring for secure blockchain networking,'' \emph{IEEE
  Transactions on Network and Service Management}, vol.~19, no.~3, pp.
  3619--3632, 2022.

\bibitem{Minrui:23}
M.~Xu \emph{et~al.}, ``When quantum information technologies meet blockchain in
  {Web 3.0},'' \emph{IEEE Network}, 2023.

\bibitem{Yu:2023}
G.~Yu \emph{et~al.}, ``Predicting {NFT} classification with {GNN}: A
  recommender system for {Web3} assets,'' in \emph{2023 IEEE International
  Conference on Blockchain and Cryptocurrency (ICBC)}, 2023, pp. 1--5.

\bibitem{Madhwal:23}
R.~Madhwal and J.~Pouwelse, ``Web3recommend: Decentralised recommendations with
  trust and relevance,'' 2023, arXiv preprint arXiv:2307.01411.

\bibitem{Xiong:23}
Z.~Xiong \emph{et~al.}, ``When mobile blockchain meets edge computing,''
  \emph{IEEE Communications Magazine}, vol.~56, no.~8, pp. 33--39, 2018.

\bibitem{Singh:2021}
A.~Singh and K.~Chatterjee, ``Securing smart healthcare system with edge
  computing,'' \emph{Computers \& Security}, vol. 108, 2021.

\bibitem{Qiu:2020}
T.~Qiu \emph{et~al.}, ``Edge computing in industrial {Internet of Things}:
  Architecture, advances and challenges,'' \emph{IEEE Communications Surveys \&
  Tutorials}, vol.~22, no.~4, pp. 2462--2488, 2020.

\bibitem{Lin:2019}
L.~Lin \emph{et~al.}, ``Computation offloading toward edge computing,''
  \emph{Proceedings of the IEEE}, vol. 107, no.~8, pp. 1584--1607, 2019.

\bibitem{Silva:2022}
T.~Silva, ``Cloud computing or edge computing: Cost comparison,''
  \url{https://www.azion.com/en/blog/cloud-computing-or-edge-computing-cost/},
  Dec. 2022.

\bibitem{Liu:2018}
M.~Liu \emph{et~al.}, ``Computation offloading and content caching in wireless
  blockchain networks with mobile edge computing,'' \emph{IEEE Transactions on
  Vehicular Technology}, vol.~67, no.~11, pp. 11\,008--11\,021, 2018.

\bibitem{Zhu:2021}
Y.~Zhu \emph{et~al.}, ``Blockchain-enabled access management system for edge
  computing,'' \emph{Electronics}, vol.~10, no.~9, 2021.

\bibitem{Doe:2023}
D.~Doe \emph{et~al.}, ``Promoting the sustainability of blockchain in {Web 3.0}
  and the {Metaverse} through diversified incentive mechanism design,''
  \emph{IEEE Open Journal of the Computer Society}, 2023.

\bibitem{Wang:2020}
X.~Wang \emph{et~al.}, ``Convergence of edge computing and deep learning: A
  comprehensive survey,'' \emph{IEEE Communications Surveys \& Tutorials},
  vol.~22, no.~2, pp. 869--904, 2020.

\bibitem{Deng:20}
S.~Deng \emph{et~al.}, ``Edge intelligence: The confluence of edge computing
  and artificial intelligence,'' \emph{IEEE Internet of Things Journal},
  vol.~7, no.~8, pp. 7457--7469, 2020.

\bibitem{Cao:22}
L.~Cao, ``Decentralized {AI}: Edge intelligence and smart blockchain,
  metaverse, {Web3}, and desci,'' \emph{IEEE Intelligent Systems}, vol.~37,
  no.~3, pp. 6--19, 2022.

\bibitem{Luong:18}
N.~Luong \emph{et~al.}, ``Optimal auction for edge computing resource
  management in mobile blockchain networks: A deep learning approach,'' in
  \emph{2018 IEEE international conference on communications (ICC)}, 2018, pp.
  1--6.

\bibitem{Wang:2019}
R.~Wang \emph{et~al.}, ``A video surveillance system based on permissioned
  blockchains and edge computing,'' in \emph{2019 IEEE international conference
  on big data and smart computing (BigComp)}, 2019, pp. 1--6.

\bibitem{Kuznetsov:21}
N.~Kuznetsov, ``{Facebook’s centralized metaverse a threat to the
  decentralized ecosystem?}''
  \url{https://cointelegraph.com/news/facebook-s-centralized-metaverse-a-threat-to-the-decentralized-ecosystem},
  Nov. 2021.

\bibitem{Jeffries:21}
A.~Jeffries, ``{Can Facebook align with the values of the metaverse?}''
  \url{https://www.marketingdive.com/news/can-facebook-align-with-values-metaverse/608768/},
  Oct. 2021.

\bibitem{Lodge:22}
M.~Lodge, ``What is decentraland?''
  \url{https://www.investopedia.com/what-is-decentraland-6827259}, Nov. 2022.

\bibitem{Ethereum:NFT}
Ethereum, ``{Non-fungible tokens (NFT)},''
  \url{https://ethereum.org/en/nft/\#what-are-nfts}.

\bibitem{WLWC:21}
Q.~Wang, R.~Li, Q.~Wang, and S.~Chen, ``Non-fungible token {(NFT)}: Overview,
  evaluation, opportunities and challenges,'' 2021, arXiv preprint
  arXiv:2105.07447.

\bibitem{RZIB:21}
W.~Rehman, H.~Zainab, J.~Imran, and N.~Bawany, ``{NFTs}: Applications and
  challenges,'' in \emph{International Arab Conference on Information
  Technology (ACIT)}.\hskip 1em plus 0.5em minus 0.4em\relax IEEE, 2021, pp.
  1--7.

\bibitem{Chohan:21}
U.~Chohan, ``{Non-fungible tokens: Blockchains, scarcity, and value},''
  \emph{Critical Blockchain Research Initiative (CBRI) Working Papers}, 2021.

\bibitem{Yang:23}
L.~Yang \emph{et~al.}, ``{Generic-NFT}: A generic non-fungible token
  architecture for flexible value transfer in {Web3},'' 2023, techRxiv
  preprint.

\bibitem{Lopez:2023}
E.~Lopez, ``Securechain,'' \url{https://github.com/eduardfina/Securechain},
  Jun. 2023.

\bibitem{Zhu:2023}
J.~Zhu \emph{et~al.}, ``{NFTool},'' \url{https://github.com/Bobliuuu/NFTool},
  Aug. 2023.

\bibitem{Adiloglu:2023}
C.~Adiloglu, ``{StoryChain},'' \url{https://storychain.ai}, Mar. 2023.

\bibitem{Busch:22}
K.~Busch, ``Non-fungible tokens {(NFTs)},''
  \url{https://crsreports.congress.gov/product/pdf/R/R47189}, Jul. 2022.

\bibitem{Wang:2021}
Q.~Wang \emph{et~al.}, ``{Non-fungible token (NFT)}: Overview, evaluation,
  opportunities and challenges,'' 2021, arXiv preprint arXiv:2105.07447.

\bibitem{KT:22}
R.~Kr{\"a}ussl and A.~Tugnetti, ``{Non-fungible tokens (NFTs): A review of
  pricing determinants, applications and opportunities},'' \emph{Applications
  and Opportunities}, 2022.

\bibitem{ZAB:20}
D.~Zetzsche, D.~Arner, and R.~Buckley, ``{Decentralized finance},''
  \emph{Journal of Financial Regulation}, vol.~6, no.~2, pp. 172--203, 2020.

\bibitem{Werner:21}
S.~Werner \emph{et~al.}, ``{SoK}: Decentralized finance {(DeFi)},'' 2021, arXiv
  preprint arXiv:2101.08778.

\bibitem{WLSL:21}
P.~Winter, A.~Lorimer, P.~Snyder, and B.~Livshits, ``What's in your wallet?
  privacy and security issues in {Web 3.0},'' 2021, arXiv preprint
  arXiv:2109.06836.

\bibitem{Jiang:2023}
E.~Jiang \emph{et~al.}, ``Decentralized finance ({DeFi}): A survey,'' 2023,
  arXiv preprint arXiv:2308.05282.

\bibitem{Kaihua:21}
Q.~Kaihua \emph{et~al.}, ``{CeFi vs. DeFi}--comparing centralized to
  decentralized finance,'' 2021, arXiv preprint arXiv:2106.08157.

\bibitem{Ethereum:DeFi}
Ethereum, ``{Decentralized finance (DeFi)},''
  \url{https://ethereum.org/en/defi/}.

\bibitem{Schar:21}
F.~Sch{\"a}r, ``Decentralized finance: On blockchain-and smart contract-based
  financial markets,'' \emph{FRB of St. Louis Review}, 2021.

\bibitem{CB:20}
Y.~Chen and C.~Bellavitis, ``{Blockchain disruption and decentralized finance:
  The rise of decentralized business models},'' \emph{Journal of Business
  Venturing Insights}, vol.~13, p. e00151, 2020.

\bibitem{HRS:21}
C.~Harvey, A.~Ramachandran, and J.~Santoro, \emph{DeFi and the Future of
  Finance}.\hskip 1em plus 0.5em minus 0.4em\relax John Wiley \& Sons, 2021.

\bibitem{Zhao:19}
X.~Zhao \emph{et~al.}, ``{FinBrain: when finance meets AI 2.0},''
  \emph{Frontiers of Information Technology \& Electronic Engineering},
  vol.~20, no.~7, pp. 914--924, 2019.

\bibitem{Cao:20}
L.~Cao, ``{AI} in finance: A review,'' \emph{Available at SSRN 3647625}, 2020.

\bibitem{Sadman:22}
N.~Sadman \emph{et~al.}, ``Promise of {AI in DeFi}, {A systematic review},''
  \emph{Digital}, vol.~2, no.~1, pp. 88--103, 2022.

\bibitem{EasyFi:23}
EasyFi, ``Artificial intelligence {(AI)} \& decentralized finance {(DeFi)}: A
  match made in heaven,''
  \url{https://medium.com/easify-network/artificial-intelligence-ai-decentralized-finance-defi-a-match-made-in-heaven-483d24129481},
  Feb. 2023.

\bibitem{BinanceDeFi:23}
Binance, ``How {AI} will influence {DeFi}: Promises and delusions.''

\bibitem{Goyal:2023}
Y.~Goyal \emph{et~al.}, ``{AgroSurance},''
  \url{https://github.com/agrosurance}, Mar. 2023.

\bibitem{KO:2023}
A.~Kondaurova and K.~Orlov, ``Prompt {DeFi},''
  \url{https://github.com/Digberi/promptdefi-web}, Mar. 2023.

\bibitem{Dubyk:2023}
A.~Dubyk \emph{et~al.}, ``{RoboFI},'' \url{https://robofi.482.solutions/}, Jul.
  2023.

\bibitem{Iredale:21}
G.~Iredale, ``Pros and cons of decentralized finance {(DeFi)},''
  \url{https://101blockchains.com/pros-and-cons-of-decentralized-finance/},
  Jul. 2021.

\bibitem{Chen:2022}
C.~Chen \emph{et~al.}, ``When digital economy meets {Web 3.0}: Applications and
  challenges,'' \emph{IEEE Open Journal of the Computer Society}, 2022.

\bibitem{Ma:2023}
W.~Ma \emph{et~al.}, ``A comprehensive study of governance issues in
  decentralized finance applications,'' 2023, arXiv preprint arXiv:2311.01433.

\bibitem{Qin:2022}
R.~Qin \emph{et~al.}, ``{Web3}-based decentralized autonomous organizations and
  operations: Architectures, models, and mechanisms,'' \emph{IEEE Transactions
  on Systems, Man, and Cybernetics: Systems}, vol.~53, no.~4, pp. 2073--2082,
  2022.

\bibitem{Wang:19}
S.~Wang \emph{et~al.}, ``{Decentralized autonomous organizations: Concept,
  model, and applications},'' \emph{IEEE Transactions on Computational Social
  Systems}, vol.~6, no.~5, pp. 870--878, 2019.

\bibitem{Qin:22}
R.~Qin \emph{et~al.}, ``{Web3}-based decentralized autonomous organizations and
  operations: Architectures, models, and mechanisms,'' \emph{IEEE Transactions
  on Systems, Man, and Cybernetics: Systems}, 2022.

\bibitem{Ethereum:DAOs}
Ethereum, ``{Decentralized autonomous organizations (DAOs)},''
  \url{https://ethereum.org/en/dao/}.

\bibitem{FAH:20}
Y.~E. Faqir, J.~Arroyo, and S.~Hassan, ``An overview of decentralized
  autonomous organizations on the blockchain,'' in \emph{Proceedings of the
  16th international symposium on open collaboration}, 2020, pp. 1--8.

\bibitem{ZhaoXi:22}
X.~Zhao \emph{et~al.}, ``{Task management in decentralized autonomous
  organization},'' \emph{Journal of Operations Management}, vol.~68, no. 6-7,
  pp. 649--674, 2022.

\bibitem{SA:22}
C.~Santana and L.~Albareda, ``Blockchain and the emergence of decentralized
  autonomous organizations {(DAOs)}: An integrative model and research
  agenda,'' \emph{Technological Forecasting and Social Change}, vol. 182, p.
  121806, 2022.

\bibitem{HHH:22}
M.~Haque and M.~H. S.~Hossain, ``A comprehensive review and architecture of a
  decentralized automated direct government system using artificial
  intelligence and blockchain,'' \emph{International Journal of Scientific \&
  Engineering Research}, vol.~13, 2022.

\bibitem{Nayan:2023}
K.~Nayan, ``{OmniGovern DAO},''
  \url{https://github.com/kamalbuilds/OmniGovern-DAO/}, Aug. 2023.

\bibitem{Duportail:2022}
D.~Duportail, ``Rooster {DAO},'' \url{https://github.com/RoosterDao}, Jul.
  2022.

\bibitem{KAL:2023}
S.~Kapadia, A.~Aghadi, and N.~Lionis, ``{DAOasis},''
  \url{https://github.com/Suhel-Kap/DAOasis}, Mar. 2023.

\bibitem{Ding:2022}
W.~Ding \emph{et~al.}, ``Desci based on {Web3 and DAO}: A comprehensive
  overview and reference model,'' \emph{IEEE Transactions on Computational
  Social Systems}, vol.~9, no.~5, pp. 1563--1573, 2022.

\bibitem{Yu:23}
G.~Yu \emph{et~al.}, ``Leveraging architectural approaches in {Web3}
  applications-a {DAO} perspective focused,'' in \emph{2023 IEEE International
  Conference on Blockchain and Cryptocurrency (ICBC)}, 2023, pp. 1--6.

\bibitem{GKSW:22}
D.~Gogel, B.~Kremer, A.~Slavin, and K.~Werbach, ``Decentralized autonomous
  organizations: Beyond the hype,'' \emph{White Paper}, Jun. 2022.

\bibitem{Tan:2023}
J.~Tan \emph{et~al.}, ``Open problems in {DAOs},'' 2023, arXiv preprint
  arXiv:2310.19201.

\bibitem{RB:16}
R.~Rudman and R.~Bruwer, ``Defining {Web 3.0}: Opportunities and challenges,''
  \emph{The Electronic Library}, 2016.

\bibitem{Sheridan:22}
D.~Sheridan \emph{et~al.}, ``{Web3} challenges and opportunities for the
  market,'' 2022, arXiv preprint arXiv:2209.02446.

\bibitem{Fan:2023}
Y.~Fan \emph{et~al.}, ``The current opportunities and challenges of {Web
  3.0},'' 2023, arXiv preprint arXiv:2306.03351.

\bibitem{BNZ:17}
C.~Barabas, N.~Narula, and E.~Zuckerman, ``Defending {Internet} freedom through
  decentralization: Back to the future,'' \emph{The Center for Civic Media \&
  The Digital Currency Initiative MIT Media Lab}, 2017.

\bibitem{SolanaStatus:2022}
SolanaStatus, ``An incident resulted in approximately 8,000 wallets being
  drained,'' \url{https://twitter.com/SolanaStatus}, Aug. 2022.

\bibitem{EG:21}
M.~Egkolfopoulou and A.~Gardner, ``Even in the {Metaverse}, not all identities
  are created equal,''
  \url{https://www.bloomberg.com/news/features/2021-12-06/cryptopunk-nft-prices-suggest-a-diversity-problem-in-the-metaverse#xj4y7vzkg},
  Dec. 2021.

\bibitem{Apotheker:23}
J.~Apotheker \emph{et~al.}, ``Web3 already has a gender diversity problem,''
  \url{https://www.bcg.com/publications/2023/how-to-unravel-lack-of-gender-diversity-web3},
  Feb. 2023.

\bibitem{Reuters:18}
Reuters, ``Amazon scrapped a secret {AI} recruitment tool that showed bias
  against women,''
  \url{https://venturebeat.com/ai/amazon-scrapped-a-secret-ai-recruitment-tool-that-showed-bias-against-women/},
  Oct. 2018.

\bibitem{Ledford:19}
H.~Ledford, ``Millions of black people affected by racial bias in health-care
  algorithms,'' \url{https://www.nature.com/articles/d41586-019-03228-6}, Oct.
  2019.

\bibitem{Antonakakis:2017}
M.~Antonakakis \emph{et~al.}, ``Understanding the mirai botnet,'' in \emph{26th
  USENIX security symposium (USENIX Security 17)}, 2017, pp. 1093--1110.

\bibitem{Jin:2022}
X.~Jin \emph{et~al.}, ``Edge security: Challenges and issues,'' 2022, arXiv
  preprint arXiv:2206.07164.

\end{thebibliography}

\end{document}